\documentclass[pre,twocolumn,showpacs,showkeys,superscriptaddress,amsmath,amssymb,floatfix,eqsecnum]{revtex4}
\usepackage{graphicx}
\usepackage{bm,texdraw}

\begin{document}

\title{Excess free energy and Casimir forces in systems with
  long-range interactions of van-der-Waals type: General
  considerations and exact spherical-model results}
\author{Daniel Dantchev} \thanks{e-mail:
daniel@imbm.bas.bg}
\affiliation{%
Fachbereich Physik, Universit{\"a}t Duisburg-Essen, Campus Essen, 45117 Essen,
Germany
}%
\affiliation{%
Institute of Mechanics---Bulgarian Academy of Sciences, Academic Georgy Bonchev Street
building~4, 1113 Sofia, Bulgaria}
\author{H.~W. Diehl}
\affiliation{%
Fachbereich Physik, Universit{\"a}t Duisburg-Essen, Campus Essen, 45117 Essen,
    Germany
}%
\author{Daniel Gr{\"u}neberg}\thanks{e-mail:
danielg@theo-phys.uni-essen.de}
\affiliation{%
Fachbereich Physik, Universit{\"a}t Duisburg-Essen, Campus Essen, 45117 Essen,
Germany
}%

\date{\today}

\begin{abstract}
  We consider systems confined to a $d$-dimensional slab of
  macroscopic lateral extension and finite thickness $L$ that undergo
  a continuous bulk phase transition in the limit $L\to\infty$ and are
  describable by an $O(n)$ symmetrical Hamiltonian. Periodic boundary
  conditions are applied across the slab. We study the effects of
  long-range pair interactions whose potential decays as $b
  x^{-(d+\sigma)}$ as $x\to\infty$, with $2<\sigma<4$ and
  $2<d+\sigma\leq 6$, on the Casimir effect at and near the bulk
  critical temperature $T_{c,\infty}$, for $2<d<4$.  These
  interactions decay sufficiently fast to leave bulk critical
  exponents and other universal bulk quantities unchanged---i.e., they
  are irrelevant in the renormalization group (RG) sense. Yet they
  entail important modifications of the standard scaling behavior of
  the excess free energy and the Casimir force ${\mathcal F}_C$. We
  generalize the phenomenological scaling ans{\"a}tze for these
  quantities by incorporating these long-range interactions. For the
  scaled reduced Casimir force per unit cross-sectional area, we
  obtain the form $L^{d}\,{\mathcal F}_C/k_BT\approx
  \Xi_0\big(L/\xi_\infty\big)+ g_\omega\,
  L^{-\omega}\Xi_\omega\big(L/\xi_\infty\big)+g_\sigma\,
  L^{-\omega_\sigma}\, \Xi_\sigma\big(L/\xi_\infty\big)$.  Here
  $\Xi_0$, $\Xi_\omega$, and $\Xi_\sigma$ are universal scaling
  functions; $g_\omega$ and $g_\sigma$ are scaling fields associated
  with the leading corrections to scaling and those of the long-range
  interaction, respectively; $\omega$ and
  $\omega_\sigma=\sigma+\eta-2$ are the associated
  correction-to-scaling exponents, where $\eta$ denotes the standard
  bulk correlation exponent of the system without long-range
  interactions; $\xi_\infty$ is the (second-moment) bulk correlation
  length (which itself involves corrections to scaling). The
  contribution $\propto g_\sigma$ decays for $T\neq T_{c,\infty}$
  algebraically in $L$ rather than exponentially, and hence becomes
  dominant in an appropriate regime of temperatures and $L$.  We
  derive exact results for spherical and Gaussian models which confirm
  these findings. In the case $d+\sigma =6$, which includes that of
  nonretarded van-der-Waals interactions in $d=3$ dimensions, the
  power laws of the corrections to scaling $\propto b$ of the
  spherical model are found to get modified by logarithms. Using
  general RG ideas, we show that these logarithmic singularities
  originate from the degeneracy $\omega=\omega_\sigma=4-d$ that occurs
  for the spherical model when $d+\sigma=6$, in conjunction with the
  $b$~dependence of $g_\omega$.
\end{abstract}
\pacs{05.70.Jk, 68.35.Rh, 11.10.Hi, 64.65.-k, 75.40.-s}

\keywords{Casimir effect, fluctuation-induced forces, renormalization
  group, spherical model, long-range interactions}

\maketitle

\section{Introduction}
\label{sec:intro}

When macroscopic bodies are immersed into a medium, the forces acting
between them in its absence are usually altered. Moreover, additional
(effective) forces not present without the medium may be induced by
fluctuations occurring in it. A well-known example of
fluctuation-induced forces is the so-called \emph{Casimir} force
between metallic bodies, named after its discoverer H.~B.~G.\ Casimir
\cite{Cas48}, that is induced by vacuum fluctuations of the
electromagnetic field and was recently verified through high-precision
experiments \cite{Lam97,MR98}.

Although the Casimir effect was well received at the time of its
discovery, interest in it diminished soon afterwards, and for a long
time it did not attract much attention. Since approximately 1970 there
has been a resurge of interest in it, which has evolved into an
enormous research activity during the past decades
\cite{casiqmrev,PMG86,MT97,KG99,BMM01,Mil01,casitdrev,Kre94,Kre99,BDT00}.

There are a number of good reasons for this development. To begin
with, fluctuation-induced forces are ubiquitous in nature.  Casimir's
original work \cite{Cas48} was concerned with the force induced by
vacuum fluctuations of the electromagnetic field.  Subsequently
it has been realized that analogous forces exist that are not mediated
by massless particles such as photons, but are induced by low-energy
excitations such as spin-waves---or, more generally, Goldstone modes
in systems with a spontaneously broken continuous symmetry---or
thermal fluctuations.  Since Goldstone modes are massless, the
associated fluctuations are scale invariant and induce a long-ranged
Casimir force. The same applies to thermal fluctuations at critical
points because of the divergence of the correlation length.  The
upshot is that Casimir forces have turned out to be of interest for
many diverse fields of physics, such as  quantum field theories
\cite{PMG86,MT97,KG99,Mil01,BMM01}, condensed matter physics, the
physics of fluids and quantum fluids \cite{Kre94,Kre99,BDT00}, wetting
phenomena \cite{FdG78,NI85,Die88}, microfluidics, and nanostructured
materials \cite{CAKBC01}.

Second, owing to the progress in experimental techniques made in
recent years, detailed investigations of Casimir forces have become
possible \cite{Lam97,MR98,GC00a,GC00b,GC02,UBMCR03,BI05,IB05}.  Third,
a further important reason for the ongoing interest in Casimir forces
is that they exhibit \emph{universal} features: Microscopic details of
both the fluctuating medium as well as the immersed macroscopic bodies
do not normally matter, at least as long as long-range interactions
are absent or may be safely ignored.

Last but not least, an equally important reason has been the
theoretical progress in dealing with interacting field theories with
boundaries that has been achieved since the 1980s
\cite{DD80,DD81b,DD81a,DD83a,Sym81,Die86a,Die97}. This has led to
detailed investigations of the Casimir effect for interacting field
theories \cite{Sym81,KD92a,KD92b,KD92c,EKD93,KED95,casitdrev}.

In this paper we will be concerned with the \emph{thermodynamic
  Casimir effect}---i.e., the Casimir effect induced by \emph{thermal
  fluctuations}. Our aim is to study the effects of long-range
interactions of van-der-Waals type on the Casimir force in systems
undergoing a continuous bulk phase transition. To this end we shall
consider long-range two-body interactions with a pair potential
$v^{(\sigma)}(x)$ that behaves as
\begin{equation}
  \label{eq:vsigma}
  v^{(\sigma)}(x)\mathop{\approx}_{x\to\infty}\text{const}\,x^{-(d+\sigma)}
\end{equation}
in the large-distance limit. The familiar dispersion forces in fluids
belong to this category: Important examples are the nonretarded and
retarded van-der-Waals interactions of a ${d=3}$~dimensional fluid,
which correspond to the cases $\sigma=3$ and $\sigma=4$, respectively.

According to scaling considerations (to be recalled in
Sec.~\ref{sec:bg}), the leading infrared singularities at the critical
point of systems with short-range forces do not get modified by such
long-range interactions because the associated pair potentials decay
sufficiently fast at large distances. They are irrelevant in the
renormalization group (RG) sense, giving corrections to the leading
critical behavior. Long-range interactions of this kind have been
termed ``subleading long-range interactions''
\cite{DR01,CD02a,CD02b,DKD03}. They are generically present in fluids
\cite{remdispforces,MDB04,BGO05} but occur also in other, for example,
magnetic systems.

For typical three-dimensional systems of the $n$-vector type with
$n<\infty$, the associated correction-to-scaling exponent
$\omega_\sigma$ is \emph{larger} than the familiar exponent $\omega$
that governs the leading corrections-to-scaling (see, e.g.,
Ref.~\cite{ZJ96}). Hence such long-range interactions yield
\emph{next-to-leading} corrections-to-scaling.

Despite their irrelevance, they have important consequences, even for
the near-critical behavior of bulk systems. Since they involve pair
potentials that decrease as inverse powers of the distance $x$ in the
limit $x\to\infty$, the usual exponential large-$x$ decay of
correlations away from the critical point gets replaced by an
algebraic one.

Their consequences for the medium-induced force between two
macroscopic bodies immersed into the medium a distance $L$ apart is of
a similar kind and importance: They yield contributions that decay
quite generally as an inverse power of $L$, irrespective of whether or
not the temperature $T$ is close to the bulk critical temperature
$T_{c,\infty}$ of the medium. When $T\simeq T_{c,\infty}$, they compete with the
long-ranged Casimir force produced by critical or near-critical
fluctuations. As previous work \cite{DR01,CD02a,CD02b,DKD03,CD03}
suggests, and will be shown in detail below, they actually become the
\emph{dominant} part of the medium-induced force in a certain regime
of temperatures and $L$.

We will consider the case of a slab geometry of cross-sectional area
$A=L_\parallel^{d-1}$ and thickness $L$. Reliable results for this
geometry are important for the interpretation of Monte Carlo
simulations of appropriate models with subleading long-range
interactions.

In view of our above remarks, a most obvious system class to consider
would be fluids. To describe the long-distance physics of classical
fluids near their liquid-gas critical point, a one-component order
parameter is used. Instead of considering this case, we will focus our
attention on systems that involve an $n$-component order parameter and
can be modeled by an $O(n)$ symmetrical Hamiltonian, and investigate
them in the limit $n\to\infty$. For simplicity, we will restrict ourselves
to the case of \emph{periodic boundary conditions} along all---namely,
both the perpendicular as well as the $d-1$ principal
parallel---directions.  Under these conditions, the large-$n$ limit of
the $O(n)$ model is equivalent to the spherical model
\cite{Sta68,Sta69,KT77}. We will present \emph{exact} results for the
Casimir force at and above the bulk critical temperature $T_{c}$, both
for spherical and Gaussian models with subleading long-range
interactions.

Our motivation for considering spherical models is twofold. First,
studying the effects of such long-range interactions on the Casimir
force for such models is an interesting problem in its own right.
Second, the exact results obtained for these models provide nontrivial
checks for the results of perturbative field-theoretic renormalization
group approaches and are expected to give valuable guidance for
acceptable approximations, an issue we plan to take up in a subsequent
paper \cite{GDD05}.

A special feature of the spherical model with $2<d<4$ is that the
correction-to-scaling exponents $\omega_\sigma$ and $\omega$ become equal when
$d+\sigma=6$, a condition satisfied, for example, for nonretarded
van-der-Waals interactions in $d=3$ dimensions. As our exact results
show, the corrections-to-scaling induced by the long-range
interaction~(\ref{eq:vsigma}) then get modified by logarithms.

The remainder of this paper is organized as follows. In the next
section, we provide the required background on Casimir forces. We
begin by recalling the definition of the Casimir force. Then we
discuss its scaling form when all interactions are short ranged,
specify the form of the subleading long-range interactions to be
considered, and recapitulate the scaling arguments which show that
they do not modify the leading critical singularities. Next, we
generalize the finite-size scaling ansatz by incorporating them. In
Sec.~\ref{sec:SM} we introduce the spherical model with subleading
long-range interactions which we solve for $2<d<4$ to produce exact
large-$n$ results for the Casimir force. The finite-size behavior of
the equation of state is analyzed in Sec.~\ref{sec:ES}.
Section~\ref{sec:fsfe} deals with the finite-size behavior of the free
energy and the Casimir force. Sec.~\ref{sec:concl} contains a brief
summary and discussion. Finally, there are three appendixes in which
various technical details are explained.

\section{Background}
\label{sec:bg}

\subsection{Definition and scaling form of Casimir force}
\label{sec:casiforcedef}

We consider a statistical mechanical system, a model magnet or fluid,
whose shape is a $d$-dimensional slab of thickness $L$ and
hyperquadratic cross-section with area $A=L_\|^{d-1}$. As previously
mentioned, we choose periodic boundary conditions along all $d$
principal hypercubic axes, so that the system has the topology of a
$d$-torus. Unless stated otherwise, the dimensionality $d$ is presumed
to satisfy $2<d<4$.

Let $F_{L,A}(T)$ be the total free energy of the system.  Taking the
thermodynamic limit {$L_\parallel \to \infty$} at fixed $L<\infty $,
we denote the reduced free energy per cross-sectional area $A$ as
$f_L(T)\equiv \lim_{A\to\infty}F_{L,A}/Ak_BT$. For $L\to \infty$,
$f_L(T)/L$ approaches $f_{\text{bk}}(T)$, the reduced bulk free
energy density \cite{rem:additivity}.  We therefore introduce the
reduced excess free energy by
\begin{equation}
  \label{eq:fexd}
f _{\text{ex}}(T,L)=f_L(T)-L f_{\text{bk}}(T)\;.
\end{equation}

The limit $L\to \infty$ of this quantity exists, but depends on the
boundary conditions: for periodic boundary conditions and the film
geometry with boundary planes $\mathfrak{B}_1$ and $\mathfrak{B}_2$
introduced above, we have \cite{rem:additivity}
\begin{equation}
  \label{eq:fexinfty}
  f_{\text{ex}}(T,H,\infty)=\left\{
  \begin{array}[c]{ll}0\;,&\text{periodic bc,}\\
   f_{s,1}+f_{s,2}\;,&\text{film geometry,}
  \end{array}
\right.
\end{equation}
where $f_{s,i}$, $i=1,2$, are the surface excess free energy of the
respective semi-infinite systems bounded by $\mathfrak{B}_i$.

In either case, the thermodynamic Casimir force per unit area is
defined in terms of $f_{\text{ex}}$ as
\begin{equation}
   \label{eq:FCdef}
{\mathcal F}_C(T,L)=-k_BT\,\frac{\partial
  f_{\text{ex}}(T,L)}{\partial L}\,.
\end{equation}

According to this definition, this quantity is a generalized force
conjugate to the thickness $L$ of the slab, which approaches zero as
$L\to \infty$. We are interested in its behavior for $L \gg a$, where
$a$ is a typical microscopic length scale (we henceforth set to
unity). Suppose for the moment that all interactions are short-ranged.
Then finite-size scaling theory should be applicable in this limit.
According to it, the Casimir force takes the scaling form
\cite{BDT00,Pri90}
\begin{equation}
{\mathcal F}_C(T,L)/k_BT=L^{-d}\,\Xi_0(L/\xi_\infty)\;,
\end{equation}
where $\xi_\infty$ is the bulk correlation length
\cite{rem:corrlength}, while $\Xi_0$ is a universal scaling function.
This holds up to eventual contributions from regular background terms
and irrelevant scaling fields, which we disregard for the moment but
will come back to later, in particular, in Sec.~\ref{sec:FSC}.

As the temperature $T$ approaches its bulk critical value $T_{c,\infty}$,
with $L$ fixed at a finite value, the correlation length $\xi_\infty$
diverges and $L/\xi_\infty\to 0$. The corresponding limiting value of the
scaling function $\Xi_0$ (which exists) is conventionally written as
\begin{equation}
  \label{eq:X0Delta}
  \Xi_0(0)=(d-1)\,\Delta_C\,,
\end{equation}
which defines the so-called Casimir amplitude $\Delta_C$ \cite{FdG78}.
This quantity is related to the critical Casimir force via
\begin{equation}
  \label{eq:cadef}
{\mathcal F}_C(T_{c,\infty},L)/k_BT_{c,\infty}=(d-1)\frac{\Delta_C}{L^d}\,.
\end{equation}
Just as the scaling function $\Xi_0$, it is a \emph{universal} quantity;
it is independent of microscopic details, but depends on the bulk
universality class considered and on other gross features such as
boundary conditions.

Let us be a bit more precise. Suppose that instead of choosing
periodic boundary conditions we considered a lattice model with free
boundary conditions along the perpendicular direction. Then the
topmost and lowest layers of the system would be free surfaces,
corresponding to macroscopic planar boundaries between which the
Casimir force acts. Provided (a) no symmetry-breaking boundary terms are
included in the Hamiltonian and (b) no long-range surface order is
possible for $T>T_{c,\infty}$, one expects the long-distance physics of the
system near the bulk critical point to be described by an $O(n)$
$\phi^4$ model with \emph{Dirichlet} boundary conditions. This is
because upon coarse graining, the lattice model with free boundary
conditions maps onto such a continuum field theory, albeit one
satisfying \emph{Robin boundary conditions} inside of averages
\cite{Die86a,Die97}.

If conditions (a) and (b) are satisfied, one has reason to believe
that the theory belongs to the basin of attraction of the fixed point
describing the so-called ordinary surface transition. This fixed point
is infrared-stable and corresponds to a Dirichlet boundary condition
on large scales. The analogs of the Casimir amplitude $\Delta_C$ and
the scaling function $\Xi_0$ for this case of Dirichlet boundary
conditions on both surface planes differ from their counterparts for
periodic boundary conditions. Details of the mesoscopic Robin boundary
condition---or microscopic details of the boundaries---do not matter
as long as the resulting continuum theory belongs to the basin of
attraction of the mentioned fixed point.

More generally, we have for a film geometry bounded in one direction
by a pair of parallel boundary planes $\mathfrak{B}_1$ and
$\mathfrak{B}_2$ the following situation. Universal quantities such as
the Casimir amplitude $\Delta_C$ or the scaling function $\Xi_0$ depend
(for given bulk universality class and short-range interactions) on
gross properties of \emph{both} boundary planes. Let SUC$_i$ denote
the universality class pertaining to the surface critical behavior of
the semi-infinite system with boundary plane $\mathfrak{B}_i$
(``surface universality class'' SUC), where $i=1$ or $2$. To specify
universal quantities like the Casimir amplitude, we can write
$\Delta_C^{\text{SUC}_1,\text{SUC}_2}$, where possible choices of
$\text{SUC}_1$ and $\text{SUC}_2$ are ``ord'', ``sp'', and ``norm'',
the SUC of the ordinary, special, and normal (or extraordinary
\cite{BD94}) transition, respectively. The above-mentioned case of
Dirichlet boundary conditions on $\mathfrak{B}_1$ and $\mathfrak{B}_2$
corresponds to the choices $\text{SUC}_1=\text{SUC}_2=\text{ord}$.

Systems with $O(n)$-symmetrical Hamiltonian and short-range
interactions have been studied in such film geometries for various
choices of $\text{SUC}_1$ and $\text{SUC}_2$ by means of the
$\epsilon$-expansion about the upper critical dimension $d^*=4$
\cite{KD92a,KD92b,KD92c,EKD93,KED95,Kre94}, Monte Carlo simulations
\cite{KL96,Kre99,DK04}, and other techniques \cite{BDT00}. A fairly
up-to-date survey of pertinent results may be found in the latter
reference. More recent results are contained in Ref.~\cite{DK04}.
Aside from these cases and the one of periodic boundary conditions,
also slabs with \emph{anti}periodic boundary conditions have been
considered for systems with short-range interactions
\cite{KD92a,KD92b}.

Going back to the case of periodic boundary conditions, we now turn to
the question of how to include subleading long-range interactions.

\subsection{Subleading long-range interactions}
\label{sec:slri}

We consider long-range two-body interactions with a pair potential
$v^{(\sigma)}(x)$ of the kind (\ref{eq:vsigma}). Let us begin by
recalling how the relevance or irrelevance of such interactions for
bulk critical behavior can be assessed.

\subsubsection{Relevance/Irrelevance criterion}
 \label{sec:relirrcrit}

Let ${\mathcal H}_{\text{sr}}$ be the standard $\phi^4$ Hamiltonian
representing the bulk universality class of the $n$-vector model with
short-range interactions for $d$ below $d^*=4$, its upper critical
dimension.  At the bulk critical point, the $n$-component order
parameter field $\bm{\phi}$ transforms as
$\bm{\phi}\to\ell^{-\Delta[\phi]}\bm{\phi}$ under changes
$\mu\to\mu\ell$ of the momentum scale, where the scaling dimension
$\Delta[\phi]$ is given by
\begin{equation}
  \label{eq:Deltaphi}
  \Delta[\phi]=(d-2+\eta)/2\;.
\end{equation}

Adding to ${\mathcal H}_{\text{sr}}$ a long-range interaction term with
pair potential $v^{(\sigma)}(x)$, we consider the Hamiltonian
\begin{equation}
  \label{eq:Hsrsigma}
  {\mathcal H}={\mathcal H}_{\text{sr}}+b\int
  {\mathcal O}^{(\sigma)}(\bm{x})\,d^dx\,,
\end{equation}
where ${\mathcal O}^{(\sigma)}(\bm{x})$ denotes the nonlocal operator
\begin{equation}
  \label{eq:Osigma}
  {\mathcal O}^{(\sigma)}(\bm{x})=\int d^dy\,v^{(\sigma)}(y)\,
  \bm{\phi}{\Big(\bm{x}-\frac{\bm{y}}{2}\Big)}\,
  \bm{\phi}{\Big(\bm{x}+\frac{\bm{y}}{2}\Big)}\,,
\end{equation}
and $b$ is the associated coupling constant.

We now ask under what conditions the short-range fixed point remains
infrared-stable with respect to this
${\mathcal O}^{(\sigma)}$~perturbation. Upon insertion of the limiting
form (\ref{eq:vsigma}) into it, we can use Eq.~(\ref{eq:Deltaphi}) to
conclude that the scaling dimension of the associated scaling
operator, at the short-range fixed point, is given by
\begin{equation}
  \label{eq:phiscinv}
  \Delta[{\mathcal O}^{(\sigma)}]=d-2+\eta+\sigma\;.
\end{equation}
The corresponding scaling field $g_\sigma\sim b$ varies as
$\ell^{-y_\sigma}$ in the infrared limit $\ell\to 0$, with the
RG~eigenexponent
\begin{equation}
  \label{eq:ysigma}
  y_\sigma\equiv-\omega_\sigma
  =d-\Delta[{\mathcal O}^{(\sigma)}]=2-\eta-\sigma\;.
\end{equation}
Depending on whether the correction-to-scaling exponent
$\omega_\sigma>0$ or $\omega_\sigma<0$, the short-range fixed point is
locally stable or unstable to such perturbations. Hence we arrive at
the following \emph{irrelevance/relevance criterion}: The long-range
perturbation $\propto b$ is irrelevant at the short-range fixed point
if
\begin{equation}
  \label{eq:irrcrit}
  \sigma>2-\eta\;,
\end{equation}
and relevant if $\sigma<2-\eta$. Note that here and elsewhere in this paper,
$\eta$ always means the correlation exponent of the short-range case.

The case when this criterion suggests these long-range interactions to
be relevant has been studied in the literature in the context of bulk
critical behavior. For $\sigma<2$, the upper critical dimension above
which Landau theory holds is lowered from $d^*=4$ to
$d^*_{\text{lr}}(\sigma)=2\sigma$. In the regime
$\sigma<d<d^*_{\text{lr}}(\sigma)$, the values of the critical
exponents depend on $\sigma$, where the analog of $\eta$ is given
exactly by $\eta_{\text{lr}}=2-\sigma$
\cite{FMN72,Sak73,BZJLG76,AHA76,FS82b,AF88}. For given $d$, a
crossover from the critical behavior characterized by these critical
exponents to one representative of systems with short-range
interactions is predicted to occur at $\sigma=2-\eta$
\cite{Sak73,HN89,Hon90,Car96,Jan98}. This crossover has recently been
reexamined for $d=2$ by numerical means \cite{LB02}.

Since we assume in our subsequent analysis that $2<\sigma<4$, the
irrelevance criterion~(\ref{eq:irrcrit}) is satisfied. Associated with
the long-range interaction~(\ref{eq:Osigma}) therefore is an
irrelevant scaling field $g_\sigma\sim b$ whose RG~eigenexponent is
given in Eq.~(\ref{eq:ysigma}). We next generalize the finite-size
scaling ansatz for the free energy by incorporating $g_\sigma$.

\subsubsection{Finite-size scaling}
  \label{sec:FSC}

Allowing a magnetic field $H$ to be present, we consider the reduced
free energy per unit cross-sectional area $A=L_\perp^{d-1}$ of the
previously specified slab with periodic boundary conditions, in the
thermodynamic limit {$L_\perp\to\infty$}. According to the
phenomenological theory of finite-size scaling
\cite{Fis71,Bar83,Pri90}, this quantity can be decomposed into a
regular background contribution $f_L^{\text{reg}}(T,H)$ and a
singular part $f_L^{\text{sing}}(T,H)$:
\begin{equation}
  \label{eq:fsing}
  f_L(T,H)=f_L^{\text{sing}}(T,H)+f_L^{\text{reg}}(T,H)\;.
\end{equation}
This decomposition entails analogous decompositions of the bulk and
excess free-energy densities $f_{\text{bk}}(T,H)$ and
$f_{\text{ex}}(T,H,L)$, respectively.

Before turning to the singular parts, let us briefly comment on the
regular background terms. For simple lattice systems with short-range
interactions it has been found that the regular background terms of
the excess free energy in the case of periodic boundary conditions
agree to high accuracy with those of the bulk free energy
\cite{Pri90}. This is understandable: Periodic boundary conditions
preclude surface and edge contributions to the total free energy and
hence terms of this kind that are analytic in temperature and magnetic
field. Yet, it must be remembered that free energies and their regular
background contributions are no universal properties, but depend on
microscopic details of the system considered. Suppose a given system
with periodic boundary conditions that belongs to the bulk
universality class of the $d$-dimensional, $n$-component $\phi^4$
model.  Then we can choose a simple lattice $n$-vector model with
nearest-neighbor interactions to investigate its universal critical
behavior. However, inclusion of any irrelevant interaction---in
particular, long-range interactions---that were dropped when making
the transition from the original system to the lattice model is
expected to modify the regular background contributions of the (bulk
and excess) free energy. In other words, the empirical fact that the
regular background contributions of the bulk and excess free energies
of simple lattice models with short-range interactions can be chosen
to be equal when periodic boundary conditions are applied, does not
imply that the same is true for microscopically more realistic model
with additional (irrelevant) interactions. In particular, this must be
kept in mind when adding irrelevant long-range interactions.

The singular parts $f^{\text{sing}}$,
$f_{\text{bk}}^{\text{sing}}$, and $f_{\text{ex}}^{\text{sing}}$
should have a scaling form.  Specifically,
$f^{\text{sing}}_{\text{ex}}(T,H,L)$ should take the finite-size
scaling form
\begin{eqnarray}\label{eq:freeenergyper}
\lefteqn{f^{\text{sing}}_{\rm ex}(T,h,L)}&&\nonumber\\
&=&L^{-(d-1)}X(g_tL^{1/\nu},g_hL^{\Delta/\nu};g_\sigma
L^{-\omega_\sigma},g_\omega L^{-\omega},\ldots)\nonumber\\
\end{eqnarray}
on sufficiently large length scales, where $\omega$ is the previously
mentioned standard correction-to-scaling exponent of short-range
systems. Further, $g_t$, $g_h$, $g_\omega$, and $g_\sigma$ denote
scaling fields.  The first two are the leading even and odd relevant
bulk scaling fields (namely, the ``thermal'' and ``magnetic'' scaling
fields). For simple magnetic systems they behave as
\begin{equation}
  \label{eq:gt}
  g_t\approx a_t\,t,\quad  t=(T-T_{c,\infty})/T_{c,\infty},
\end{equation}
and
\begin{equation}
\label{eq:gh}
g_h\approx
  a_h\,h,\quad  h=H/k_BT_{c,\infty},
\end{equation}
near the bulk critical point $(T,H)=(T_{c,\infty},0)$, where $a_t$ and $a_h$
are nonuniversal metric factors; for fluid systems, both become linear
combinations of $t$ and $\delta\mu$, the deviation of the chemical
potential from the critical point, because of ``mixing'' (see, e.g.,
Refs.~\cite{KG81,Nic81}). For simplicity, we will use magnetic
language and work with the above expressions henceforth.

Likewise, the previously introduced scaling field associated with the
long-range interaction (\ref{eq:Osigma}) is expected to vary as
\begin{equation}
  \label{eq:gsigma}
g_\sigma\approx a_\sigma b
\end{equation}
for small $b$. The ellipsis in Eq.~(\ref{eq:freeenergyper}) stands for
analogous expressions involving further scaling fields, all of which
we assume to be irrelevant; this means, in particular, that all
\emph{relevant} scaling fields other than $g_t$ and $g_h$ are taken to
vanish.  Moreover, we assume that none of the suppressed irrelevant
scaling fields is \emph{dangerous irrelevant} (see, e.g., Appendix D
of Ref.~\cite{Fis83}), so that all of them may be safely set to zero.

Current estimates of the correction-to-scaling exponent $\omega(n,d)$
of the $d$-dimensional $n$-vector model give $\omega(1,3)\simeq 0.81$
and somewhat smaller values for $n=2$ and $n=3$, such as
$\omega(3,3)\simeq 0.80$ \cite{GZJ98,PV00,KY05}. On the other hand,
the well-known exact spherical-model (SM) value is
\begin{equation}
  \label{eq:omegainfty}
  \omega_{\text{SM}}(2<d<4)=\omega(\infty,2<d<4)=4-d\;.
\end{equation}

Let us compare these numbers with the appropriate analogs for the
correction-to-scaling exponent $\omega_\sigma$ one can derive from
Eq.~(\ref{eq:ysigma}). The cases of nonretarded and retarded
van-der-Waals interactions in $d$ dimensions correspond to the choices
${\sigma=d}$ and ${\sigma=d+1}$, giving $\omega_d=d-2+\eta$ and
$\omega_{d+1}=d-1+\eta$, respectively. Both exponents are positive in
the regime of dimensions $2<d<4$ we are concerned with. For finite
$n$, where $\eta>0$, the latter remains larger than $\omega$ in this
whole regime, whereas $\omega_d$ would become smaller than $\omega$
slightly below $d=3$. In the spherical limit $n\to\infty$, this sign
change of $\omega_d-\omega$ occurs at $d=3$ where $\omega_d=\omega=1$.
In our analysis of the spherical model given below we shall first
assume that $d+\sigma<6$. Then the possibility that
$\omega>\omega_\sigma$ is ruled out. The borderline case $d+\sigma=6$
of the spherical model is special because $\omega=\omega_\sigma=4-d$.
Owing to this degeneracy, it requires special attention and will be
discussed separately.

For the time being we therefore take it for granted that the
irrelevant scaling fields $g_\omega $ and $g_\sigma$ yield leading and
next-to-leading corrections to scaling in the critical regime,
respectively. However, away from the bulk critical point, the
long-range interaction is expected to modify the large-$L$ behavior of
$f_{\text{ex}}$ and the Casimir in a \emph{qualitative} manner so that
they decay as inverse powers of $L$ rather than exponentially
\cite{CD03}. To see how this translates into properties of the scaling
function $X$, let us denote the scaling variables appearing in
Eq.~(\ref{eq:freeenergyper}) as
\begin{eqnarray}
  \label{eq:runvar}
  \check{t}&=&g_tL^{1/ \nu }\;,\nonumber\\
\check{h}&=&g_hL^{\Delta/ \nu}\;,\nonumber\\
\check{g}_\sigma&=&g_\sigma L^{-\omega_\sigma}\;,\nonumber \\
\check{g}_\omega&=&g_\omega L^{-\omega}\;,
\end{eqnarray}
and expand $X$ as
\begin{eqnarray}
  \label{eq:Xexp}
X(\check{t},\check{h};\check{g}_\sigma,\check{g}_\omega)
   &=& X_0(\check{t},\check{h})+\check{g}_\sigma\,
  X_\sigma(\check{t},\check{h})
\nonumber\\&&
\strut +\check{g}_\omega\,
  X_\omega(\check{t},\check{h})+\ldots\;,
\end{eqnarray}
where it is understood that all suppressed scaling fields have been
set to zero.

The scaling functions $X_0$ and $X_\omega$ obviously are
properties of the \emph{short-range} universality class. A similar,
though somewhat more restricted statement applies to $X_\sigma$:
Just as the other two, it may be viewed as the expectation value of a
quantity, computed at the infrared-stable fixed point of the $\phi^4$
model with short-range interactions in a periodic slab of thickness
$L=1$. However, it differs from those inasmuch as, in its case, this
quantity is the nonlocal operator ${\mathcal O}^{(\sigma)}(\bm{x})$
associated with the long-range interaction, whereas the other do not
involve this interaction at all.

At the bulk critical point $g_t=g_h=0$, all three of these scaling
functions are expected to take finite, nonzero values.  Specifically,
the critical value of $X_0$ yields the Casimir amplitude:
\begin{equation}
  \label{eq:X000}
  \Delta_C \equiv X_0(0,0)\;.
\end{equation}
We denote its analogs for $X_\omega$ and $X_\sigma$ as
\begin{eqnarray}
  \label{eq:X0sigma}
  \Delta_{\omega,C}&\equiv& X_\omega(0,0)\;,\\
  \label{eq:X0omega}
 \Delta_{\sigma,C}&\equiv& X_\sigma(0,0)\;.
\end{eqnarray}
The former controls the leading corrections to the asymptotic behavior
of the critical excess free energy, the latter its contribution linear
in $g_\sigma$ originating from the long-range
interaction~(\ref{eq:Osigma}).

Next, we turn to a discussion of the behavior as $L\to \infty$ when
$T>T_{c,\infty}$. In this limit, either the scaling variable
$\check{t}$, or both $\check{t}$ and $\check{h}$, tend to infinity. As
explained above, both functions $X_0$ and $X_\omega$ must
decrease as $\sim \exp[-L/ \xi^{(\text{sr})}(T,H)]$, where
$\xi^{(\text{sr})}(T,H)$ is the true correlation length of the system
with short-range interactions. Let us set $g_h=0$ for the sake of
simplicity. As $\check{t}\to\infty$ we then should have
\begin{equation}
  \label{eq:X0largeL}
  X_0(\check{t},0) \mathop{\sim}\limits_{\check{t}\to
    \infty}\exp\Big[-|\text{const}|\,\check{t}^{\nu}+O(\ln
    \check{t})\Big]\;,
\end{equation}
and similar asymptotic behavior for $X_\omega$. However, for the
function $ X_\sigma(\check{t},0)$ we anticipate the limiting
form
\begin{equation}
  \label{eq:X0sigmalargeL}
   X_\sigma(\check{t},0) \mathop{\approx}\limits_{\check{t}\to
    \infty}c_{\sigma}\,{\check{t}}^{-\nu\,\zeta}\;.
\end{equation}
The exponent $\zeta$ introduced here characterizes the asymptotic
dependence on $L/\xi$ via $X_\sigma\sim (L/\xi)^{-\zeta}$. Our
results for both the spherical (${n=\infty}$) and the Gaussian model
(GM) derived in the following sections yield
\begin{equation}
  \label{eq:zetaGM}
  \zeta_{\text{SM}}=\zeta(n=\infty)=\zeta_{\text{GM}}=2 \;,
\end{equation}
in conformity with Ref.~\cite{CD03}.

In the regime $L/\xi\gg 1$ where $X_0$ and $X_\omega$ are
exponentially small, the implied contribution $\sim g_\sigma$ to the
excess free energy should become dominant:
\begin{equation}
  \label{eq:fexlargeL}
  f_{\text{ex}}\big(Lg_t^\nu\gg 1\big)\approx
  g_\sigma\,c_\sigma\,L^{-(d+\sigma+\eta+\zeta-3)}
  g_t^{-\nu\zeta}\;.
\end{equation}
and imply a corresponding large-$L$ behavior
\begin{equation}
  \label{eq:FClargeL}
  {\mathcal F}_C\sim g_\sigma L^{-(d+\sigma+\eta+\zeta-2)}g_t^{-\nu\zeta}
\end{equation}
of the Casimir force.

In the cases of the spherical and Gaussian models, where $\eta=0$ and
$\zeta$ is given by Eq.~(\ref{eq:zetaGM}), the large-$L$ dependence of
$f_{\text{ex}}$ reduces to $\sim L^{-(d+\sigma-1)}$.  Our exact results for
the spherical and Gaussian models given below confirm these findings.
In fact, there are reasons to expect that the latter $L$~dependence
applies more generally even when $\eta>0$. As proven some time ago by
Iagolnitzer and Souillard \cite{IS77}, using the
Griffiths-Sherman-Kelly inequalties \cite{rem:GKS}, the two-point net
correlation function of a ferromagnetic system whose interactions
decay as $v^{(\sigma)}(x)$ in Eq.~(\ref{eq:vsigma}) cannot decay faster
than the potential. Although we are not aware of any rigorous proof
that they cannot decay slower than the potential either
\cite{rem:lbound,Dan01}, it seems most natural to us to assume that this
cumulant decays as $x\to\infty$ according to the same power law as the
interaction potential, barring eventual logarithmic corrections in
special cases.

Now, the correlation function
\begin{equation}
  \label{eq:Gofx}
  G(\bm{x})\equiv\langle S(\bm{x})\,S(\bm{0})\rangle -\langle S(\bm{x}\rangle
\langle S(\bm{0})\rangle
\end{equation}
of a slab of size $\infty^{d-1}\times L$ under periodic boundary
conditions (pbc) can be expressed in terms of its bulk counterpart
$G_\infty$ via
\begin{equation}
  \label{eq:GLpbc}
  G_L^{(\text{pbc})}(\bm{x})=\sum_{j=-\infty}^\infty
  G_\infty(\bm{x}-jL\hat{\bm{e}}_1)\;,
\end{equation}
where $\hat{\bm{e}}_1$ is a unit vector along the finite $1$-direction.
For fixed $\bm{x}$, the terms with $j\neq 0$ yield $L$-dependent
deviations from the $j=0$ bulk term that decay $\sim L^{-d-\sigma}$ as
$L\to\infty$, provided the large-distance behavior~(\ref{eq:vsigma})
of $v^{(\sigma)}$ carries over to $G_\infty$. This suggests that (away
from criticality) the excess contribution to the free energy of a
$d$-dimensional volume is down by a factor $L^{-\sigma}$, so that the
excess density $f_{\text{ex}}$ behaves as $L^{1-d-\sigma}$. To ensure
consistency with Eq.~(\ref{eq:fexlargeL}), we must therefore have
\begin{equation}
  \label{eq:zetaex}
  \zeta=2-\eta\;.
\end{equation}

Although our results for the spherical and Gaussian models described
below are in conformity with this prediction, they do not provide a
nontrivial check of it because $\eta$ vanishes. Such a check should in
principle be possible within the framework of the
$\epsilon$~expansion. To this end, one would have to compute the
scaling function $X_\sigma$ using RG~improved perturbation
theory to sufficiently high orders, verify its limiting
behavior~(\ref{eq:X0sigmalargeL}), and confirm its consistency with
Eq.~(\ref{eq:zetaex}).

\section{Spherical model}
\label{sec:SM}

\subsection{Definition of the model}
\label{sec:moddef}

Let $\mathfrak{L}\subset\mathbb{Z}^d$ be the set of sites $\bm{x}$ of
a simple hypercubic lattice of size $L_1{\times} L_2{\times}\cdots{\times} L_d$.
Imposing periodic boundary conditions along all $d$ principal
directions, we consider a spherical model with the Hamiltonian
\begin{eqnarray}
\label{eq:spmH}
\frac{{\mathcal H}}{k_BT}&=&-\frac{1}{2}\sum_{\bm{x}, \bm{x}'\in
\mathfrak{L}}\frac{J(\bm{x}-\bm{x}')}{k_BT}\, S(\bm{x}) S(\bm{x}')
\nonumber\\
&& \strut -h\sum_{\bm{x} \in {\mathfrak{L}}}S(\bm{x})+s\sum_{\bm{x}\in
  {\mathfrak{L}}} S^2(\bm{x})
\end{eqnarray}
whose spin variables $S(\bm{x})\in \mathbb{R}$ satisfy the mean
spherical constraint
\begin{equation}
  \label{eq:msc}
    \left \langle \sum_{\bm{x}\in {\mathfrak{L}}}S^2(\bm{x}) \right
    \rangle=|{\mathfrak{L}}|\;.
\end{equation}
Here $|{\mathfrak{L}}|$, the cardinality  of the set
$\mathfrak{L}$, is the total number of sites (or spins). Further, $s$
is a real positive variable, called spherical field, whose value is to
be determined from Eq.~(\ref{eq:msc}). For systems such as the one
considered here, whose spins are all equivalent by translational
invariance, the constraint~(\ref{eq:msc}) fixes all averages $\langle
S^2(\bm{x})\rangle$, $\forall\bm{x} \in \mathfrak{L}$, to be unity.

As before, $h=H/k_BT$ denotes a reduced magnetic field. The pair
interaction $J(\bm{x})$ consists of nearest-neighbor bonds and a
long-ranged contribution of the type $v^{(\sigma)}$ specified in
Eq.~(\ref{eq:vsigma}), with $2<\sigma<4$; we use the choice
\begin{equation}
\label{eq:sint}
J(\bm{x}) = J_1\,\delta_{x,1}+
\frac{J_2}{(\rho_0^2+x^2)^{(d+\sigma)/2}}\;,
\end{equation}
with $J_1\geq 0$ and $J_2>0$, where $\rho_0>0$ sets a crossover length
scale beyond which $J(\bm{x})$ varies approximately as
$J_2\,x^{-d-\sigma}$.

\subsection{Properties of the interaction potential}
\label{sec:Jprop}

In Appendix~\ref{app:FT} we show that the Fourier transform
\begin{equation}
  \label{eq:FT}
  \tilde{J}(\bm{q})\equiv\sum_{\bm{x}}J(\bm{x})\,e^{-i\bm{q}\cdot\bm{x}}
\end{equation}
 of this interaction can be written as
\begin{equation}
  \label{eq:Jtilde}
  \tilde{J}(\bm{q})=\tilde{J}(\bm{0})-K k_BT\,\Omega(\bm{q})
\end{equation}
with
\begin{equation}
  \label{eq:Kdef}
  K\equiv -\frac{1}{k_BT}\,\frac{\partial \tilde{J}(\bm{q})}{\partial
      q^2} \bigg|_{\bm{q}=\bm{0}} \;,
\end{equation}
where $\Omega(\bm{q})$ behaves as
\begin{equation}
  \label{eq:Omega}
  \Omega(\bm{q})=q^2-b q^\sigma+b_4q^4+b_{4,1}
    \sum_{\alpha=1}^d q_\alpha^4+ o(q^4)
\end{equation}
for small $q$, and $K>0$, $b>0$, $b_4>0$, and
$b_4+b_{4,1}>0$. The term $\propto b_{4,1}$ is anisotropic in
$\bm{q}$-space. It is a consequence of the fact that the hypercubic
lattice breaks the Euclidean symmetry down to the symmetry of a
hypercube. For other, less symmetric lattices more than two
fourth-order invariants and hence additional anisotropic $q^4$ terms
would appear.

Owing to our choice~(\ref{eq:sint}) of interaction constants, we have
$J(\bm{x})>0$ for all lattice displacements $\bm{x}$. A
straightforward consequence is that the Hamiltonian~(\ref{eq:spmH})
has a unique ground state whose energy for $h=0$ is given by
$\tilde{J}(\bm{0})$.  Furthermore, $\tilde{J}(\bm{0})>
\tilde{J}(\bm{q})$ for all nontrivial wave-vectors $\bm{q}$ in the
first Brillouin $\text{BZ}_1$. It follows that the resulting values of
$b, \ldots,b_{4,1}$ must be such that the equation $1-b\,
q^{\sigma-2}+q^2[b_4+b_{4,1}\sum_\alpha (q_\alpha/q)^4]=0$ has no
real-valued solutions $\bm{q}$.

Since the nonanalytic contribution $\sim q^\sigma$ arises from the
large-distance tail of $J(\bm{x})$, its coefficient
$k_BT Kb$ should not dependent on the details of how $J(\bm{x})$
behaves at small distances and hence be independent of
$\rho_0$. Our result
\begin{equation}\label{eq:b}
    b\,\frac{k_BTK}{J_2}
    =\frac{\pi^{d/2}\Gamma(-\sigma/2)}{2^\sigma \,\Gamma[(d+\sigma)/2]}\;,
\end{equation}
derived in Appendix \ref{app:FT}, confirms this expectation. On the
other hand, the coefficients of the analytic terms of orders $q^2$ and
$q^4$ of $\tilde{J}(\bm{q})$ depend, of course, on $J_1$ and $\rho_0$.

Note that the Fourier transform of the second term on the right-hand
side of Eq.~(\ref{eq:sint}) yields a contribution to the
nearest-neighbor coupling $J(\bm{x})|_{x=1}$ that depends on
$\kappa$. This dependence can be utilized to modify this contribution and
hence the nearest-neighbor coupling for a given value of $J_2$ by varying $\kappa$. If
we choose, for simplicity, the value zero for the coupling constant
$J_1$ in Eq.~(\ref{eq:sint}), then the parameter $b$ becomes
(cf.\ Appendix~\ref{app:FT})
\begin{equation}
  \label{eq:bfinal}
    b=-\frac{1}{\pi}\,\Gamma(-\sigma/2)\,\Gamma(2-\sigma/2)\,\sin(\pi
  \sigma/2)\,(\rho_0/2)^{\sigma-2}\;,
\end{equation}
which reduces to
\begin{equation}
  \label{eq:b33}
  b=\frac{2}{3}\,\rho_0\;,\;\;d=\sigma=3\,,
\end{equation}
for the case of nonretarded van-der-Waals interactions in
three dimensions.

\subsection{Solution of the model, free energy, and constraint equation}
\label{sec:modsol}

Defining
\begin{equation}
  \label{eq:v2}
  v_2\equiv -\frac{\partial}{\partial
    q^2}\ln\tilde{J}(\bm{q})\big|_{q=0}=\frac{Kk_BT}{\tilde{J}(\bm{0})}\;,
\end{equation}
we introduce the parameter
\begin{equation}
  \label{eq:rdef}
 r\equiv \frac{1}{v_2}\left[\frac{2sk_BT}{\tilde{J}(\bm{0})}-1\right]
\end{equation}
and the mode sum
\begin{equation}
\label{eq:Uds}
U_{d,\Omega}(r|\bm{L})=\frac{1}{2|{\mathfrak{L}}|}\sum_{\bm{q}\in \text{BZ}_1}
\ln[r+\Omega(\bm{q})]\;,
\end{equation}
where $\bm{L}\equiv(L_1,\ldots,L_d)$. As is shown in
Appendix~\ref{app:FT}, the coefficient $v_2$ for our
choice~(\ref{eq:sint}) of interaction constants takes the value
\begin{equation}
  \label{eq:v2res}
  v_2=\frac{\rho_0^2}{2(\sigma-2)}
\end{equation}
when $J_1=0$.

Expressed in terms of the above quantities, the total free energy
$F_{\bm{L}}(K,h)$  of our
model is given by \cite{BDT00}
\begin{equation}
  \label{eq:fed}
\frac{F_{\bm{L}}(K,h)}{k_BT|{\mathfrak{L}}|}=f^{(0)}(K)+\frac{1}{2}\sup_{r>0}
\left\{
 2 U_{d,\Omega}(r|\bm{L})-K r-\frac{h^2}{Kr}
\right\}
\end{equation}
with
\begin{equation}
  \label{eq:f0}
  f^{(0)}(K)=
\frac{1}{2}\bigg[ \ln\frac{K}{2\pi}-\frac{K}{v_2}\bigg]\;.
\end{equation}

To determine the required supremum, we differentiate
Eq.~(\ref{eq:fed}) with respect to $r$. This yields as condition from
which $r\equiv r_{\bm{L}}(K,h)$---or, equivalently, the spherical field
$s$ of Eq.~(\ref{eq:rdef})---is to be determined, the constraint
equation
\begin{equation}
  \label{eq:ceq}
K= \frac{h^2}{Kr_{\bm{L}}^2}+W_{d,\Omega}(r_{\bm{L}}|\bm{L})
\end{equation}
 with
\begin{equation}
\label{eq:Wds}
W_{d,\Omega}(r_{\bm{L}}|\bm{L})=\frac{1}{|{\mathfrak{L}}|}\sum_{\bm{q}\in\text{BZ}_1}
\frac{1}{r_{\bm{L}}+\Omega(\bm{q})}\;.
\end{equation}
The latter quantity is obviously related to $U_{d,\Omega}$ via
\begin{equation}
\label{eq:rel}
U_{d,\Omega}(r_{\bm{L}}|\bm{L})=U_{d,\Omega}(0|\bm{L})
+\frac{1}{2}\int_0^{r_{\bm{L}}} W_{d,\Omega}(x|\bm{L})\ dx.
\end{equation}

Let us recall that the constraint equation~(\ref{eq:ceq}) can be recast
in the form of an equation of state \cite{BDT00}. To see this, note
that Eq.~(\ref{eq:fed}) yields for the magnetization density
$m_{\bm{L}}$ the result
\begin{equation}
  \label{eq:minfty}
  m_{\bm{L}}(K,h)=-\frac{\partial}{\partial
    h}\,\frac{F_{\bm{L}}(K,h)/|\mathfrak{L}|}{k_BT}=\frac{h}{Kr_{\bm{L}}(K,h)} \;,
\end{equation}
whenever the supremum is attained for the solution $r_{\bm{L}}$ of
Eq.~(\ref{eq:ceq}). Using this to eliminate $r_{\bm{L}}$ in favor of
$m_{\bm{L}}$ and $h$ gives us the equation of state
\begin{equation}
  \label{eq:eqs}
  \left(1-m_{\bm{L}}^2\right)K=W_{d,\Omega} \bigg(\frac{h}{m_{\bm{L}}K}\bigg
  |\bm{L}\bigg)\;.
\end{equation}
In view of this correspondence between the constraint
equation~(\ref{eq:ceq}) and the equation of state~(\ref{eq:eqs}), we
will take the liberty of referring to the former henceforth as the
equation of state.

From Eq.~(\ref{eq:minfty}) one can easily read off that $r_{\bm{L}}$ for
${h=0}$ has the familiar meaning of an inverse susceptibility. Let us
define the susceptibility by
\begin{equation}
  \label{eq:chidef}
  \chi_{\bm{L}}(K,h)\equiv\frac{\partial m_{\bm{L}}(K,h)}{\partial h}\;.
\end{equation}
Taking the derivative of the above-mentioned equation with respect to
$h$ at $h=0$ then gives the desired relation
\begin{equation}
  \label{eq:rident}
  [r_{\bm{L}}(K,0)]^{-1}=\chi_{\bm{L}}(K,0)K\;.
\end{equation}

We are interested in the limit where all linear dimensions
$L_2,\ldots,L_d\to\infty$ while $L_1$ remains fixed at the finite
value $L_1\equiv L$. Let us employ the following convenient
convention: Whenever the bold symbol $\bm{L}$ in quantities such as
$W_{d,\Omega}(r|\bm{L})$ or $r_{\bm{L}}$ has been replaced by $L$, it
is understood that the so specified thermodynamic limit has been
taken. For instance, $U_{d,\Omega}(r|L)$ stands for
\begin{equation}
  \label{eq:tdlim}
  U_{d,\Omega}(r|L) \equiv
  \lim_{L_2,\ldots,L_d\to\infty}U_{d,\Omega}(r|\bm{L})\;,
\end{equation}
and $U_{d,\Omega}(r_L|L)$ means this function, taken at the
corresponding limiting value $r_L\equiv
\lim_{L_2,\ldots,L_d\to\infty}r_{\bm{L}}$ of the supremum
$r_{\bm{L}}$, i.e., of the solution to Eq.~(\ref{eq:ceq}).

\section{Finite size behavior of the equation of state~(\ref{eq:ceq})}
\label{sec:ES}

\subsection{Decomposition of mode sums into bulk and size-dependent
  contributions}
\label{sec:dec}

In order to determine the finite size behavior of the excess free
energy and its consequences for the Casimir force, we must investigate
the $L$~dependence of the mode sums $U_{d,\Omega}(r_L|L)$ and
$W_{d,\Omega}(r_L|L)$ for large $L$. Let us first focus our
attention on the explicit $L$~dependence of these quantities by
considering them at an arbitrary $L$-independent value of $r$. Writing
\begin{equation}
\label{eq:Udec}
U_{d,\Omega}(r|L)=U_{d,\Omega}(r|\infty)+\Delta U_{d,\Omega}(r|L)
\end{equation}
and
\begin{equation}
 \label{eq:Wdec}
W_{d,\Omega}(r|L)=W_{d,\Omega}(r|\infty)+\Delta W_{d,\Omega}(r|L)\;,
\end{equation}
we split off their $L$-independent bulk parts
\begin{equation}
  \label{eq:Ubulk}
    U_{d,\Omega}(r|\infty)=\frac{1}{2}
    \int^{(d)}_{\bm{q}\in\text{BZ}_1}\ln[r+\Omega(\bm{q})]
\end{equation}
and
\begin{equation}
 \label{eq:Wbulk}
 W_{d,\Omega}(r|\infty)=\int^{(d)}_{\bm{q}\in\text{BZ}_1}
  \frac{1}{r+\Omega(\bm{q})}\;,
\end{equation}
where
\begin{equation}
  \label{eq:intq}
  \int^{(d)}_{\bm{q}\in\text{BZ}_1}\equiv
  \prod_{\alpha=1}^d\int_{-\pi}^\pi\frac{dq_\alpha}{2\pi}
\end{equation}
is a convenient short-hand, from their $L$-dependent rests $\Delta
U_{d,\Omega}(r|L)$ and $\Delta W_{d,\Omega}(r|L)$. Using Poisson's
summation formula~(\ref{eq:Poisson}) (see Appendix~\ref{app:aUd0}),
the latter can be written as
\begin{equation}\label{eq:DUdLintro}
\Delta U_{d,\Omega}(r|L)=\sum_{k=1}^{\infty} \int^{(d)}_{\bm{q}\in\text{BZ}_1}
\cos\left(q_1kL\right)\ln[r+\Omega(\bm{q})]
\end{equation}
and
\begin{equation}
  \label{eq:Wfs}
\Delta W_{d,\Omega}(r|L)=\sum_{k=1}^\infty
\int^{(d)}_{\bm{q}\in\text{BZ}_1} \frac{2\cos(q_1kL)}{r+\Omega(\bm{q})}\;,
\end{equation}
respectively.

\subsection{Bulk equation of state}
\label{sec:bulkeqs}

Next, we consider the equation of state~(\ref{eq:ceq}) in the bulk
limit ${L\to\infty}$. At the bulk critical point ${K=K_{c,L=\infty}}$,
$h=0$, its solution $r\equiv r_{L=\infty}$ must vanish. Hence the
critical coupling $K_{c,\infty}$ is given by
\begin{equation}
  \label{eq:Kcinfty}
  K_{c,\infty}(b)=W_{d,\Omega}(0|\infty)\;.
\end{equation}
As indicated, this quantity depends on the interaction parameter $b$
as well as on all other interaction parameters $b_4$, $b_{4,1}$, etc
appearing in $\Omega(\bm{q})$. Defining the scaling fields $g_t$ and
$g_h$ as
\begin{equation}
  \label{eq:ghgt}
  g_t= K_{c,\infty}-K\;,\quad g_h=h/\sqrt{K}\;,
\end{equation}
we find from Eq.~(\ref{eq:ceq}) that the bulk quantity $r_\infty$ is
to be determined from
\begin{equation}
  \label{eq:bulkeqst}
 -g_t=(g_h/r_\infty)^2+W_{d,\Omega}(r_\infty|\infty)-W_{d,\Omega}(0|\infty)\;,
\end{equation}
the ``bulk equation of state''.

\subsubsection{The case $d+\sigma<6$ with $2<d<4$ and $2<\sigma<4$}

To study its solutions near the bulk critical point, we must know how
$W_{d,\Omega}$ behaves for small $r$. Since the long-ranged
interaction ${\propto b}$ does not modify the leading infrared
behavior, it is justified to expand in $b$. A straightforward
calculation (see Appendix~\ref{app:FT}) shows that provided $2<d<4$,
$2<\sigma<4$, and $d+\sigma<6$,
\begin{eqnarray}
  \label{eq:Wbulkres}
 \lefteqn{W_{d,\Omega}(r|\infty)-W_{d,\Omega}(0|\infty)}&&\nonumber\\
&\mathop{\approx}\limits_{r\to 0}&-A_d\,r^{d/2-1}+(w_d+b\,w_{d,\sigma})\,r+O(r^2)\nonumber\\
  &&\strut -b\,B_{d,\sigma}\,r^{(d+\sigma)/2-2}[1+o(r)]
  +O(b^2)\,,\quad
\end{eqnarray}
where
\begin{equation}
\label{eq:Ad}
A_d=-\frac{\Gamma(1-d/2)}{(4\pi)^{d/2}}>0
\end{equation}
and
\begin{equation}
  \label{eq:Bds}
B_{d,\sigma}=\frac{\pi\,(d+\sigma-2)}{2\,(4\pi)^{d/2}
\Gamma(d/2)\sin[\pi(d+\sigma)/2]}>0\;.
\end{equation}

We insert the above result into the bulk equation of state
(\ref{eq:bulkeqst}), keeping only the explicitly shown contributions.
The resulting equation for the scaled inverse susceptibility $r_\infty
g_t^{-\gamma}$ is expected to take a scaling form.  To the linear
order of our analysis in $b$ and the irrelevant scaling fields
$g_\sigma$ and $g_\omega$, this is the case provided a term linear in
$b$ is included in $g_\omega$. Such a contribution is anticipated on
general grounds because in a $\phi^4$ theory with coupling constant
$u$, the RG flow of the running variable $\bar{u}(\ell)$ should be
affected by terms linear in $b$; technically, this may be attributed
to the fact that single insertions of the long-ranged operator
(\ref{eq:Osigma}) require contributions linear in $b$ of the $\phi^4$
counterterm.

On the other hand, the scaling field $g_\sigma$ should have \emph{no
contribution of zeroth order in $b$} because, given an
initial Hamiltonian without long-range interactions ($b=0$), no
long-range interaction can be generated under a RG transformation.

In conformity with these ideas, the choices
\begin{equation}
  \label{eq:gomega}
  g_\omega(b)=w_d+b\,w_{d,\sigma}\;,\qquad g_\sigma(b)=b\;,
\end{equation}
(up to nonlinear contributions and a redefinition of the scales of
these fields) turn out to be appropriate. They entail that the
resulting bulk equation of state scales, so that solutions $r_\infty$
to it can be written as
\begin{equation}
  \label{eq:rinftyscf}
  r_\infty \approx g_t^{\gamma}\,{\mathcal R}^{(\infty)}\big(g_hg_t^{-\Delta},
  g_\omega g_t^{\nu\omega},g_\sigma g_t^{\nu\omega_\sigma}\big)\;,
\end{equation}
where the critical exponents  $\gamma=\nu\,(2-\eta)$, $\nu$, $\eta$,
$\Delta=(\nu/2)(d+2-\eta)$, $\omega_\sigma$, and $\omega$ take the
spherical-model values
\begin{eqnarray}
  \label{eq:nusm}
  \gamma_{\text{SM}}&=&2\nu_{\text{SM}}=\frac{2}{d-2}\;, \nonumber\\
  \eta_{\text{SM}}&=&0\;,\quad\Delta_{\text{SM}}=\frac{d+2}{2(d-2)}\;,
  \nonumber\\
   \omega_{\sigma,\text{SM}}&=&\sigma-2\;,
\end{eqnarray}
and (\ref{eq:omegainfty}), respectively. The function
${\mathcal R}^{(\infty)}$ is given by
\begin{eqnarray}
  \label{eq:Rexp}
\lefteqn{
{\mathcal R}^{(\infty)}(x_h,x_\omega,x_\sigma)
}&&\nonumber\\
&=&{\mathcal R}^{(\infty)}_0(x_h)
  +x_\omega\,{\mathcal R}^{(\infty)}_\omega(x_h)
  +x_\sigma\,{\mathcal R}^{(\infty)}_\sigma(x_h)
\nonumber\\&&\strut
+o(x_\omega,x_\sigma)\,,
 \;\;\qquad
\end{eqnarray}
where ${\mathcal R}^{(\infty)}_0(x_h)$ is the solution to the
asymptotic scaled bulk equation of state
\begin{equation}
  \label{eq:asscbulkeqst}
  1+x_h^2{\big[{\mathcal R}^{(\infty)}_0(x_h)\big]}^{-2}=
  A_d\,\big[{\mathcal R}^{(\infty)}_0(x_h)\big]^{(d-2)/2} \;,
\end{equation}
while the remaining two functions are given by
\begin{equation}
  \label{eq:Romega}
  {\mathcal R}^{(\infty)}_\omega(x_h)=
  \frac{2\,{\big[{\mathcal R}_0^{(\infty)}(x_h)\big]}^4}{(d-2) A_d
    \,{\big[{\mathcal R}_0^{(\infty)}(x_h)\big]}^{d/2+1}+4x_h^2}
\end{equation}
and
\begin{equation}
  \label{eq:Rsigma}
  {\mathcal R}^{(\infty)}_\sigma(x_h)= \frac{-2B_{d,\sigma}\,{\big[{\mathcal R}_0^{(\infty)}(x_h)\big]}^{(d+\sigma+2)/2}}{(d-2) A_d
    \,{\big[{\mathcal R}_0^{(\infty)}(x_h)\big]}^{d/2+1}+4x_h^2}\;.
\end{equation}

For zero magnetic field, the above findings simplify considerably, giving
\begin{eqnarray}
  \label{eq:rinftyhzero}
  \left.r_\infty\right|_{h=0}&\approx&
  \Big(\frac{g_t}{A_d}\Big)^{\gamma}
  \bigg[1+\frac{2 g_\omega(b)}{(d-2)A_d}
  \Big(\frac{g_t}{A_d}\Big)^{\nu\omega}
\nonumber\\ &&
\strut-\frac{2g_\sigma(b)\,B_{d,\sigma}}{(d-2)A_d}\,
\Big(\frac{g_t}{A_d}\Big)^{\nu\omega_\sigma}\bigg] \;,
\end{eqnarray}
where again the spherical-model values (\ref{eq:nusm}) and
(\ref{eq:omegainfty}) must be substituted for the critical exponents
$\gamma$, $\nu$, $\omega_\sigma$, and $\omega$.

\subsubsection{Logarithmic anomalies and the case $d+\sigma=6$ with
  $2<d<4$}
  \label{sec:loganom}

The above results get modified by the appearance of logarithmic
anomalies when $d=4$ or $d+\sigma=6$. Our ultimate interest is to
understand the consequences this has for finite-size scaling and the
Casimir force in the latter case. Since logarithmic anomalies occur
already in the bulk theory, it will be helpful to clarify their origin
first in this simpler context.

That the finite-size behavior gets modified by the presence of
logarithmic anomalies when $d+\sigma=6$ was recognized already in a
paper by Chamati and one of us \cite{CD02a}. However, no explanation
of their cause within the general context of RG theory was given
there. Here we wish to fill this gap. As we shall see, despite some
similarities with the situation at the upper critical dimension $d=4$,
the mechanisms by which they are produced in the case $d=4$ and
$d+\sigma=6$ with $2<d<4$ are different.

Let us begin by recalling the well understood case {$d=4$}
\cite{WR73}.  The coefficients $A_d$ and $w_d$ both become singular as
$d\to 4$ (see, e.g., Ref.~\cite{MZ03} and Appendix~\ref{app:FT}).
Although $w_d$ is nonuniversal, its pole part at $d=4$ (a single pole)
is universal and equal to that of $A_d$, so that the sum of these two
terms in Eq.~(\ref{eq:Wbulkres}) produces a finite $r\ln r$
contribution in the limit $d\to 4$. As a consequence, the leading
thermal singularity of $r_\infty$ takes the form
\begin{equation}
  \label{eq:rbulklogsing}
  \left.r_\infty\right|_{h=0}\sim g_t/|\ln g_t|\;.
\end{equation}

In the framework of RG theory the appearance of logarithmic anomalies
means that the Hamiltonian ${\mathcal H}$ transforms under a change of
momentum scale $\mu\to\mu \ell$ into a transformed one
${\mathcal H}(\{\bar{g}_j(\ell)\},\ell)$ whose $\ell$-dependence cannot fully
be absorbed through scale dependent scaling fields $\bar{g}_j(\ell)$ but
has an additional explicit dependence on $\ell$. We follow here the
notational conventions of Wegner \cite{Weg76,Weg72a,WR73}: The
$\bar{g}_j(\ell)$ are nonlinear scaling fields with initial values
$\bar{g}_j(1)=g_j$ and eigenexponents $y_j$; i.e.,
\begin{equation}
  \label{eq:gbar}
 \bar{g}_j(\ell)=\ell^{-y_j}\,g_j\;.
\end{equation}
We denote their linear counterparts as $\bar{\mu}_j(\ell)$, and let
$\bar{\mu}_0$ with $y_0=d$ be the special field associated with the
volume. If the linearized RG operator is diagonal in the variables
$\mu_i$, then these fields usually satisfy flow equations, which to
quadratic order can be written as
\begin{equation}
  \label{eq:RGmuquad}
  -\ell\frac{d\bar{\mu}_i(\ell)}{d \ell}=y_i\,\bar{\mu}_i+\frac{1}{2}\sum_{j,k}a_{ijk}\,\bar{\mu}_j\,\bar{\mu}_k\;,
\end{equation}
where $a_{ijk}=a_{ikj}$ and $a_{ij0}=0$ \cite{Weg76}. Provided the
conditions
\begin{equation}
  \label{eq:ycond}
  y_i\neq y_j+y_k
\end{equation}
are fulfilled, one arrives at an expansion of the form
\begin{eqnarray}
  \label{eq:muexp}
  \bar{\mu}_i=\bar{g}_i+\frac{1}{2}\sum_{j,k}
 b_{ijk}\,\bar{g}_j\,\bar{g}_k +\ldots
\end{eqnarray}
with
\begin{equation}
  \label{eq:bijk}
  b_{ijk}= \frac{a_{ijk}}{y_j+y_k-y_i}\;.
\end{equation}

Similar conditions involving sums of more than two eigenexponents,
e.g., $y_i\neq y_j+y_k+y_l$, must hold in order that the contributions
of third and higher orders have a corresponding form with
scale-independent expansion coefficients.

When conditions such as Eq.~(\ref{eq:ycond}) are violated so that
$y_i$ equals a sum of other RG eigenvalues, the coefficients of the
expansion of the linear fields $\bar{\mu}_i$ in the nonlinear ones
$\bar{g}_i$ become scale-dependent, involving logarithms of $\ell$ or
even powers of such logarithms. For example, when $y_i=y_j+y_k$ for a
single triple $(i,j,k)$ with $a_{ijk}\neq 0$, then $b_{ijk}$ gets
replaced by \cite{Weg76,WR73}
\begin{equation}
  \label{eq:coeffrep}
    b_{ijk}(\ell)= -a_{ijk}\ln \ell\;.
\end{equation}
Upon making the usual choice $\ell_t=\ell(g_t)$ such that
$|\bar{g}_t(\ell_t)|=1$, logarithms of $t$ result.

Let us first consider the case $d+\sigma<6$ with $d,\sigma\in(2,4)$,
and ignore the contributions from all irrelevant fields. Then the
inequalities (\ref{eq:ycond}) as well as the condition that the
linearized RG operator be diagonal at the critical fixed point in the
(relevant) fields are satisfied.

The logarithmic singularities one encounters at the upper critical
dimension $d=4$ have two sources \cite{WR73}: (i) The exponent
$\mu_0=d$ is equal to twice the thermal RG eigenexponent $y_t=1/\nu$;
(ii) a marginal operator ($\phi^4$) must be taken into account, so
that an infinite number of eigenexponent inequalities (\ref{eq:ycond})
and its analogs involving more than three RG eigenexponents are
violated. The known consequences are that the leading thermal
singularities have logarithmic anomalies which for general values of
$n$ consist of nontrivial powers of $\ln g_t$.

Next, we turn to the case $d+\sigma=6$ with $d,\sigma\in(2,4)$. A
similarity with the case $d=4$ is that the coefficient $w_{d,\sigma}$
(which is again universal) has a single pole at $d+\sigma=6$ that
cancels with the pole of $B_{d,\sigma}$ such that a contribution
$\propto r\ln r$ is produced in the limit $\sigma\to 6-d$ of
Eq.~(\ref{eq:Wbulkres}). Thus the analog of this equation for
$d+\sigma=6$ becomes
\begin{eqnarray}
  \label{eq:Wbulkres6}
\lefteqn{W_{d,\Omega}(r|\infty)-W_{d,\Omega}(0|\infty)}&&\nonumber\\
&\mathop{\approx}\limits_{r\to 0}&- A_d\, r^{d/2-1}+(w_d+\tilde{w}_d\,b)\,
r\nonumber\\
&&\strut +b \,K_{d} \,r \ln r+O(b^2)
\end{eqnarray}
with
\begin{equation}  \label{eq:wdtilde}
  \tilde{w}_{d}=\frac{K_d}{2}+w^{\text{reg}}_{d,6-d}\;,
\end{equation}
where $K_d$ denotes the conventional factor
\begin{equation}\label{Bds6}
K_d\equiv\int_{\bm{q}}\delta(|\bm{q}|-1)=
\frac{2}{(4\pi)^{d/2}\,\Gamma(d/2)}>0 \;,
\end{equation}
while $w^{\text{reg}}_{d,6-d}$  means the regular part
\begin{equation}
  \label{eq:wdreg}
  w^{\text{reg}}_{d,6-d}=\lim_{\sigma\to 6-d}\left[\frac{2
      K_d}{\sigma+d-6}+w_{d,\sigma}\right]
\end{equation}
of $w_{d,\sigma}$ at ${\sigma=6-d}$.

Upon substituting the above result into the bulk equation of
state~(\ref{eq:bulkeqst}), we see that instead of
Eqs.~(\ref{eq:rinftyscf})--(\ref{eq:Rsigma}) we now have
\begin{eqnarray}
  \label{eq:rinftydsig6}
  r_\infty&\approx& g_t^{\gamma}\Big[{\mathcal R}^{(\infty)}_0(x_h)
+ g_t^{\nu\omega}{\mathcal R}^{(\infty)}_\omega(x_h)\times
\nonumber\\&&\times
\Big\{w_d+b\,\tilde{w}_d  +
b\,K_d\ln\big[g_t^{\gamma}{\mathcal R}^{(\infty)}_0(x_h)\big]
\Big\}\Big],\qquad
\end{eqnarray}
which simplifies to
\begin{eqnarray}
  \label{eq:rinftyhzerodsig6}
  \left.r_\infty\right|_{h=0}&\approx&
  \Big(\frac{g_t}{A_d}\Big)^\gamma \bigg\{1 +\frac{2}{(d-2)A_d}
  \Big(\frac{g_t}{A_d}\Big)^{\nu\omega} \times
\nonumber\\ &&\strut\times
  \Big[w_d +b\,\tilde{w}_d+b\,\frac{2K_d}{d-2}\ln\frac{g_t}{A_d}\Big]
  \bigg\}\qquad
\end{eqnarray}
when $h=0$.

The origin of the logarithmic corrections $\propto b$ is due to the
previously mentioned mixing of the $b$-independent linear part of the
irrelevant scaling field $g_\omega$ (which we denote as $\mu_\sigma$)
with $\mu_\sigma\propto b$, which led us to conclude that the scaling
fields $g_\omega$ and $g_\sigma$ can be chosen as in
Eq.~(\ref{eq:gomega}) up to nonlinear contributions. Recalling that
$\mu_\sigma$ is expected to contribute to the change of $\mu_\omega$
under RG transformations, but $\mu_\sigma$ cannot be generated when
$2<\sigma<4$ if the initial Hamiltonian does not involve any
long-range interactions, one concludes that this translates into flow
equations of the form
\begin{eqnarray}
  \label{eq:muomsigfe}
  -\ell\frac{d}{d\ell}\,\bar{\mu}_\omega&=&y_\omega\,\bar{\mu}_\omega
    +a_{\omega\sigma}\,\bar{\mu}_\sigma+\ldots\;,\nonumber\\
-\ell\frac{d}{d\ell}\,\bar{\mu}_\sigma&=&y_\sigma\,\bar{\mu}_\sigma+\ldots\;,
\end{eqnarray}
with $a_{\omega\sigma}\neq 0$. As long as $d+\sigma<6$, the
eigenexponents $y_\sigma$ and $y_\omega$ differ. In that case these
flow equations yield
\begin{eqnarray}
  \label{eq:muomegaexp}
  \bar{\mu}_\omega(\ell)&=&\bar{g}_\omega(\ell)
  +\frac{a_{\omega\sigma}}{y_\sigma-y_\omega}\,\bar{g}_\sigma(\ell) + \ldots
  \;, \\ \label{eq:musigexp}
  \bar{\mu}_\sigma(\ell)&=&\bar{g}_\sigma(\ell)+\ldots\;,
\end{eqnarray}
which in turn implies that $g_\omega$ involves a linear combination of
$\mu_\omega$ and $\mu_\sigma$, in conformity with
Eq.~(\ref{eq:gomega}).

For $d+\sigma=6$, the spherical model yields $y_\sigma=y_\omega=d-4$
[cf.\ Eq.~(\ref{eq:ysigma})]. Owing to this degeneracy, the
expansion~(\ref{eq:muomegaexp}) gets replaced by
\begin{equation}
  \label{eq:muomdsig6}
  \bar{\mu}_\omega(\ell)=\bar{g}_\omega(\ell)
  -a_{\omega\sigma}\,\bar{g}_\sigma(\ell)\,\ln\ell \;,
\end{equation}
which in turns leads to the logarithmic temperature anomaly in
Eq.~(\ref{eq:rinftyhzerodsig6}).

The general mechanism we have identified here as producing the
logarithmic anomalies in the case $d+\sigma=6$ is, of course, not new;
a brief discussion of it may be found in Sec.~V.E.1 of Ref.~\cite{Weg76}.

\subsection{Finite-size scaling form of equation of state}
\label{sec:Ldep}

We now proceed with our analysis of the finite-size behavior.  To this
end we must work out the large-$L$ dependence of the functions
$\Delta W_{d,\Omega}$ and $\Delta U_{d,\Omega}$. Expanding again to
linear order in $b$ gives
\begin{equation}
  \label{eq:DWexp}
 \Delta W_{d,\Omega}(r|L)=\Delta W^{(0)}_{d,\Omega}(r|L) +b\,\Delta
  W^{(1)}_{d,\Omega}(r|L)+O(b^2) \;,
\end{equation}
where the superscripts $(0)$ and $(1)$ on the right-hand side indicate
respectively the function $\Delta W_{d,\Omega}(r|L)$ and its first
derivative with respect to $b$, taken at $b=0$. From Eqs.~(\ref{eq:Omega})
and (\ref{eq:Wbulk}) we obtain
\begin{equation}
  \label{eq:DW0}
\Delta W_{d,\Omega}^{(0)}(r|L)=\sum_{k=1}^\infty
\int^{(d)}_{\bm{q}\in\text{BZ}_1}
\frac{2\cos(q_1kL)}{r+\Omega(\bm{q})}
\end{equation}
and
\begin{equation}
  \label{eq:DW1}
\Delta W_{d,\Omega}^{(1)}(r|L)=\sum_{k=1}^\infty
\int^{(d)}_{\bm{q}\in\text{BZ}_1}
\frac{2\,q^\sigma\cos(q_1kL)}{{[r+\Omega(\bm{q})]}^2}\;.
\end{equation}

The $\bm{q}$-integrations (cosine transforms) appearing in these
equations are well-defined as long as $r>0$ and $L>0$. In order to
obtain the asymptotic behavior of the functions~(\ref{eq:DW0}) and
(\ref{eq:DW1}) for $r\to 0$, we extend the $\bm{q}$-integrations to
the full $\bm{q}$-space $\mathbb{R}^d$ and make the replacement
$\Omega(\bm{q})\to q^2$ in their denominators. This amounts to the
omission of contributions that are regular in $r$ or less singular
than those retained. The resulting expression for the right-hand side
of Eq.~(\ref{eq:DW0}) is easily evaluated by noting that it is nothing
else than the difference between the free propagator
$G_L^{\text{(pbc)}}$ of Eq.~(\ref{eq:GLpbc}) and its bulk counterpart
$G_\infty$, given by
\begin{eqnarray}
  \label{eq:Ginfty}
  G_\infty(d|r;x)&\equiv&
  \int^{(d)}_{\bm{q}}\frac{e^{i\bm{q}\cdot\bm{x}}}{r+q^2}
 \nonumber\\ &=&
  r^{(d-2)/2}\,
  \frac{K_{d/2-1}\big(x\sqrt{r}\big)}{(2\pi)^{d/2}\,
    \big(x\sqrt{r}\big)^{d/2-1}}\;,\qquad
\end{eqnarray}
with $x\equiv |\bm{x}|$.

One thus arrives at
\begin{eqnarray}
  \label{eq:DW0app}
  \Delta W_{d,\Omega}^{(0)}(r|L)\approx \frac{2}{L^{d-2}} \sum_{k=1}^\infty
  G_\infty(d|rL^2; k) \qquad &&\nonumber\\
=\frac{K_{d-1}}{L^{d-2}} \int_0^\infty
\frac{p^{d-2}}{\sqrt{rL^2+p^2}}\,\frac{dp}{e^{\sqrt{rL^2+p^2}}-1}\;,&&
\end{eqnarray}
where the second line follows from the first one with the aid of the
representation
\begin{equation}
  \label{eq:Ginftymixed}
  G_\infty(d|r;x)=\int_{\bm{q}_\|}^{(d-1)}\,
\frac{e^{-|x_1|(r+q_\|^2)^{1/2}}}{2(r+q_\|^2)^{1/2}}\,
e^{i\bm{x}_\parallel\cdot\bm{q}_\|}
\end{equation}
upon interchanging the summation over $k$ with the integration over
the $d-1$~dimensional wave-vector $\bm{q}_\|$ conjugate to
$\bm{x}_\parallel=(x_2,\ldots,x_d)$, the component of $\bm{x}$
perpendicular to $\hat{\bm{e}}_1$.

In order to compute the analogous approximation for $\Delta
W^{(1)}_{d,\Omega}(r|L)$, we proceed as follows. Applying the identity
\begin{equation}
  \label{eq:rewrite}
  \frac{q^\sigma}{(r+q^2)^2}
  =\partial_r\left(\frac{r\,q^{\sigma-2}}{r+q^2}\right)
\end{equation}
to the integrand of Eq.~(\ref{eq:DW1}), we see that the right-hand
side of this equation is the derivative $\partial_r$ of an expression
that differs from the right-hand side of Eq.~(\ref{eq:DW0}) merely
through an extra power of $q^{\sigma-2}$ in the integrand. This tells
us that the roles of the film propagator $G_L(d|r;\bm{x})$ and its
bulk counterpart $G_\infty(d|r;x)$ in Eq.~(\ref{eq:DW0app}) now are
taken over by the modified film propagator
\begin{equation}
  \label{eq:Gsiginfty}
  G_{L}(d,\sigma|r;\bm{x})=\frac{1}{L}\sum_{q_1\in
    \frac{2\pi}{L}\,\mathbb{Z}}\,
  \int_{\bm{q}_\|}^{(d-1)}\,\frac{q^{\sigma-2}\,
    e^{i\bm{q}\cdot\bm{x}}}{r+q^2}
\end{equation}
and its $L=\infty$ analog, respectively, which obviously reduce to the
former two when $\sigma=2$.

Let us define the function
\begin{eqnarray}
  \label{eq:Qdsigdef}
  Q_{d,\sigma}(y)&\equiv&
  \frac{y}{2}\big[G_{L=1}(d,\sigma|y;\bm{0})-
  G_\infty(d,\sigma|y;\bm{0})\big] \nonumber  \\
&=&y\,\sum_{k=1}^\infty\int^{(d)}_{\bm{q}}\frac{q^{\sigma-2}\,
   \cos(q_1k)}{y+q^2}\;.
\end{eqnarray}
Here the second representation follows again by Poisson's summation
formula~(\ref{eq:Poisson}).

In terms of this function, the analog of Eq.~(\ref{eq:DW0app}) becomes
\begin{equation}
  \label{eq:DW1app}
   \Delta W^{(1)}_{d,\Omega}(r|L)\approx \frac{2}{L^{d+\sigma-4}}\,
   Q_{d,\sigma}'(rL^2) \;,
\end{equation}
where the prime indicates a derivative, i.e., $Q_{d,\sigma}'(y)\equiv
\partial Q_{d,\sigma}(y)/\partial y$. Furthermore, our
result~(\ref{eq:DW0app}) for $\Delta W^{(0)}_{d,\Omega}(r|L)$ can be
written as
\begin{equation}
  \label{eq:DW0appQ}
   \Delta W^{(0)}_{d,\Omega}(r|L)\approx L^{-(d-2)}
   \,\frac{2}{rL^2}\,Q_{d,2}(rL^2)\;.
\end{equation}

Explicit results for the propagator $G_\infty(d,\sigma|r;\bm{x})$ and
the functions $Q_{d,\sigma}(y)$ are derived in Appendix~\ref{app:Qas}.
As is shown there, $G_\infty(d,\sigma|r;\bm{x})$ can be calculated for
general values of $\sigma\in(2,4)$ and expressed in terms of
generalized hypergeometric functions. From these results the
asymptotic behavior of the functions $Q_{d,\sigma}(y)$ for large and
small values of $y$ can be inferred in a straightforward manner (see
Appendixes~\ref{app:smalllargey}). We managed to express
$Q_{d,\sigma}(y)$ for general values of $(d,\sigma)$ in terms of
elementary and special functions up to a series of the form
$\sum_{j=1}^\infty(.)$, but have not been able to obtain closed-form
analytic results for these series in general. However, for a variety
of special choices $(d,\sigma)$, we succeeded in deriving explicit
analytic expressions for the functions $Q_{d,\sigma}$. In particular,
all functions $Q_{d,\sigma}$ required for the analysis of the case
${d=\sigma=3}$ of nonretarded van-der-Waals interactions in three
dimensions are determined analytically in Appendix~\ref{app:Qas}.

From the above results the finite-size scaling form of the equation of
state near the bulk critical point follows in a straightforward
fashion. Let us choose the scaling variables $\check{t}$ and
$\check{h}$ in Eq.~(\ref{eq:runvar}) as
\begin{eqnarray}
  \label{eq:scthgsigmahat}
  \check{t}&=&(K_{c,\infty}-K)L^{d-2}\;,\\
   \check{h}&=&hK^{-1/2}L^{(d+2)/2}\;,
\end{eqnarray}
$\check{g}_\omega$ and $\check{g}_\sigma$ in accordance with
Eq.~(\ref{eq:gomega}),
and introduce the scaled inverse susceptibility
\begin{equation}
  \label{eq:scr}
  \check{r}_L\equiv r_LL^{\gamma/\nu}=r_LL^2\;,
\end{equation}
where again the spherical-model values (\ref{eq:nusm}) were utilized for the
exponents $\Delta/ \nu$ and $\gamma/\nu$.

Upon subtracting from the equation of state~(\ref{eq:eqs}) its bulk
analog at the critical point and inserting Eqs.~(\ref{eq:Wdec}),
(\ref{eq:Wbulkres}), (\ref{eq:DW1app}), and (\ref{eq:DW0appQ}), we
obtain for the case ${2<d<4}$, {$2<\sigma<4$}, and $d+\sigma<6$:
\begin{eqnarray}
  \label{eq:sceqofst}
\check{t} &\approx& -\left(\check{h}/\check{r}_L\right)^2+
 A_d\,\check{r}_L^{d/2-1}-2\,\check{r}_L^{-1} Q_{d,2}(\check{r}_L)-
 \check{g}_\omega\check{r}_L
\nonumber\\ &&\strut
 +\check{g}_\sigma{\left[B_{d,\sigma} \check{r}_L^{(d+\sigma-4)/2}
  -2\,Q_{d,\sigma}'(\check{r}_L)\right]} \;.
\end{eqnarray}
The result has the expected scaling form. We can solve for $\check{r}_L$ (at
least in principle) to determine it as a function ${\mathcal R}$ of the
other scaled variables. Hence we have shown, to linear order in
$g_\omega$, $g_\sigma$, and $b$, that the inverse susceptibility $r_L$
can be written as
\begin{equation}
  \label{eq:rLscf}
  r_L=L^{-2}\,{\mathcal
    R}(\check{t},\check{h},\check{g}_\omega,\check{g}_\sigma)
\end{equation}
in the appropriate finite-size scaling regime. By analogy with the
expansion~(\ref{eq:Rexp}) made in the bulk case, we write
\begin{eqnarray}
  \label{eq:RLexp}
\lefteqn{
{\mathcal R}(\check{t},\check{h},\check{g}_\omega,\check{g}_\sigma)
}&&\nonumber\\
&=&{\mathcal R}_0(\check{t},\check{h})
  +\check{g}_\omega\,{\mathcal R}_\omega(\check{t},\check{h})
  +\check{g}_\sigma\,{\mathcal R}_\sigma(\check{t},\check{h})
+o(\check{g}_\omega,\check{g}_\sigma)\,.\nonumber\\ &&\strut
 \qquad
\end{eqnarray}
Here ${\mathcal R}_0(\check{t},\check{h})$ is the solution to
Eq.~(\ref{eq:sceqofst}) with $\check{g}_\omega$ and
$\check{g}_\sigma=0$ set to zero. The other two functions are found to
be given by
 \begin{equation}
  \label{eq:RLomega}
 {\mathcal R}_\omega(\check{t},\check{h})=\frac{2
 {[{\mathcal R}_0(\check{t},\check{h})]}^4}{{\mathcal N}(\check{t},\check{h})}
\end{equation}
and
\begin{eqnarray}
  \label{eq:RLsigma}
  {\mathcal R}_\sigma(\check{t},\check{h}) =\frac{
4{\mathcal R}_0^3\,Q_{d,\sigma}'({\mathcal R}_0)
-2B_{d,\sigma}\,{\mathcal R}_0^{(d+\sigma+2)/2}
}{{\mathcal N}(\check{t},\check{h})}
\end{eqnarray}
with
\begin{eqnarray}
  \label{eq:Nh}
{\mathcal N}(\check{t},\check{h})&=&
  (d-2)A_d
    \,{\mathcal R}_0^{d/2+1}+4\check{h}^2\nonumber\\ &&\strut
    +4{\mathcal R}_0\big[Q_{d,2}({\mathcal R}_0)
    -{\mathcal R}_0\,Q_{d,2}'({\mathcal R}_0)\big]\;,\quad
\end{eqnarray}
where ${\mathcal R}_0$ stands for ${\mathcal R}_0(\check{t},\check{h})$.

In the large-$L$ limit the foregoing results must reduce to our above
ones for the bulk, Eqs.~(\ref{eq:rinftyscf}) and
(\ref{eq:Rexp})--(\ref{eq:rinftyhzero}). This implies the limiting behavior
\begin{eqnarray}
  \label{eq:RlargeL}
  \lefteqn{{\mathcal R}[\check{t}(L),\check{h}(L),\check{g}_\omega(L),
  \check{g}_\sigma(L)]}&&\nonumber\\
  &\mathop{\approx}\limits_{L\to \infty}& \check{t}^{\,2\nu} \,
  {\mathcal R}^{(\infty)}(g_h g_t^{-\Delta},g_\omega
  g_t^{\nu\omega},g_\sigma g_t^{\nu\omega_\sigma} )
\end{eqnarray}
and corresponding relations between the other ${\mathcal R}$-functions
and their bulk counterparts, namely
\begin{equation}
  \label{eq:Raas}
  {\mathcal R}_{a}[\check{t}(L),\check{h}(L)]\mathop{\approx}\limits_{L\to \infty}  \check{t}^{\,2\nu} \,
  {\mathcal R}^{(\infty)}_{a} (g_h g_t^{-\Delta})\;,\quad
  a=0,\,\omega,\,\sigma\,.
\end{equation}

For $d+\sigma=6$ with $2<d<4$ and $2<\sigma<4$, a logarithmic
anomalies appear again in the equation of state and its solution. A
simple way to obtain these is to take the limits $\sigma\to 6-d$ of
Eqs.~(\ref{eq:sceqofst})--(\ref{eq:Nh}). This yields
\begin{eqnarray}
  \label{eq:sceqofst6}
\check{t} &\approx& -\left(\check{h}/\check{r}_L\right)^2+
 A_d\,\check{r}_L^{d/2-1}-
 2\,\check{r}_L^{-1}\,Q_{d,2}(\check{r}_L)\nonumber\\
&&\strut
-(w_d+\tilde{w}_d\,b)\,L^{d-4}\check{r}_L
\nonumber\\ &&\strut
 -\check{g}_\sigma\bigg[
  2\,Q_{d,6-d}'(\check{r}_L)
+K_d\,\check{r}_L
 \ln\frac{\check{r}_L}{L^2}\bigg]
\end{eqnarray}
and
\begin{eqnarray}
  \label{eq:rLscfsig6}
  r_L&\approx&L^{-2}\Big[{\mathcal R}_0(\check{t},\check{h})
  +L^{-(4-d)}{\mathcal
    R}_\omega(\check{t},\check{h}) \Big\{w_d+\tilde{w}_d\,b \quad\nonumber\\
  &&\qquad\strut
  +2b\,Q'_{d,6-d}[\mathcal{R}_0(\check{t},\check{h})]\,
    [\mathcal{R}_0(\check{t},\check{h})]^{-1}\nonumber\\ &&\qquad\strut  +b\,K_d\ln{\big[L^{-2}{\mathcal R}_0(\check{t},\check{h})\big]}
  \Big\}\Big]\;.
\end{eqnarray}

The logarithmic anomalies manifest themselves through the contributions that
depend explicitly on $\ln L^2$ (rather than merely on scaled
variables).

\subsection{Relation between finite-size and bulk inverse susceptibility}
\label{sec:rLrinfty}

The results of the previous section can be combined with those for the
bulk equation of state to express the inverse scaled finite-size
susceptibility $\check{r}_L$ in terms of its bulk counterpart
\begin{equation}
  \label{eq:rcheckinfty}
  \check{r}_\infty\equiv r_\infty\, L^{\gamma/\nu}=r_\infty\, L^2\;,
\end{equation}
rather than the scaled temperature field $\check{t}$. The relationship
between $\check{r}_L$ and $\check{r}_\infty$ will be needed in the
next section to determine the excess free energy as a function of the
inverse bulk susceptibility $r_\infty$. Since the second-moment
correlation length $\xi_\infty$ of the spherical model is given by
$r_\infty^{-1/2}$ (up to a normalization factor), this gives us $r_L$
and the excess free energy expressed in terms of $\xi_\infty$.

\subsubsection{The case $d+\sigma<6$ with $2<d<4$ and $2<\sigma<4$}
\label{sec:rLinftyless6}

We equate the finite-size equation of state~(\ref{eq:sceqofst}) with
its analog for $\check{r}_\infty$,
\begin{eqnarray}
  \label{eq:bulkscesless6}
\check{t}&\approx&
-\check{h}^2\check{r}_\infty^{-2}
+A_d\,\check{r}_{\infty}^{(d-2)/2}\nonumber\\ &&\strut -\check{g}_{\omega}\check{r}_{\infty}
+\check{g}_{\sigma}B_{d,\sigma} \check{r}_{\infty}^{(d+\sigma-4)/2}\;,
\end{eqnarray}
and substitute for $\check{r}_L$ the ansatz
\begin{equation}
\label{eq:ansatzcheckrLless6}
\check{r}_{L} \approx \mathsf{R}_0(\check{r}_{\infty},\check{h})
+\check{g}_{\sigma} \mathsf{R}_{\sigma}(\check{r}_{\infty},\check{h})
+\check{g}_{\omega} \mathsf{R}_{\omega}(\check{r}_{\infty},\check{h})\;.
\end{equation}
This yields for
$\mathsf{R}_0=\mathsf{R}_0(\check{r}_\infty,\check{h})$ the equation
\begin{eqnarray}
\label{eq:R0}
2\,\mathsf{R}_0^{-1}Q_{d,2}(\mathsf{R}_0)&=&A_{d}\big(\mathsf{R}_0^{(d-2)/2}
  -\check{r}_{\infty}^{(d-2)/2}\big)\nonumber\\ && \strut
-\check{h}^{2}\!\left(\mathsf{R}_0^{-2}
  -\check{r}_{\infty}^{-2}\right)
\end{eqnarray}
and for the other functions the solutions
\begin{equation}
  \label{eq:Romless6}
  \mathsf{R}_\omega(\check{r}_\infty,\check{h})=
  \frac{2\mathsf{R}_0}{\mathsf{N}}\,(\mathsf{R}_0
    -\check{r}_{\infty})
\end{equation}
and
\begin{eqnarray}
\label{eq:Rsigless6}
\mathsf{R}_{\sigma}(\check{r}_{\infty},\check{h}) &=&
\frac{2\mathsf{R}_0}{\mathsf{N}}\Big[B_{d,\sigma}
\Big(\check{r}_{\infty}^{(d+\sigma-4)/2}\nonumber\\ &&\strut
      -\mathsf{R}_0^{(d+\sigma-4)/2}\Big)
+2Q_{d,\sigma}^{\prime}(\mathsf{R}_0)\Big]\,,\quad
\end{eqnarray}
where $\mathsf{N}$ means the function
\begin{eqnarray}
  \label{eq:Nth}
  \mathsf{N}(\check{r}_\infty,\check{h})&=&4\check{h}^{2}\,
  {\/\mathsf{R}}_0^{-2}
 +(d-2)A_{d}\,\mathsf{R}_0^{(d-2)/2}
 \nonumber \\ &&\strut
    +4{\/\mathsf{R}}_0^{-1}Q_{d,2}(\mathsf{R}_0)
\strut -4Q_{d,2}^{\prime}(\mathsf{R}_0)\,.\quad
\end{eqnarray}

\subsubsection{The case $d+\sigma=6$ with $2<d<4$}
\label{sec:rLinfty6}

The analog of Eq.~(\ref{eq:bulkscesless6}) is given by
Eq.~(\ref{eq:sceqofst6}) with $\check{r}_L$ replaced by
$\check{r}_\infty$ and the terms involving $Q_{d,2}$ and
$Q'_{d,6-d}$ dropped.
Owing to the presence of the logarithmic anomaly $\propto b$, the
ansatz~(\ref{eq:ansatzcheckrLless6}) must be modified so as to allow
for an explicit $L$-dependence of $\mathsf{R}_{\sigma}$:
\begin{eqnarray}
\label{eq:ansatzcheckrL6}
\check{r}_{L}&\approx&\mathsf{R}_0(\check{r}_{\infty},\check{h})
+\check{g}_{\sigma}
\mathsf{R}_{\sigma}(\check{r}_{\infty},\check{h};L)
\nonumber\\ &&\strut
+(w_{d}+\tilde{w}_{d}b)L^{d-4}\,
\mathsf{R}_{\omega}(\check{r}_{\infty},\check{h})\;.
\end{eqnarray}
Instead of Eq.~(\ref{eq:Rsigless6}), we now have
\begin{eqnarray}
  \label{eq:Rsig6}
\mathsf{R}_\sigma(\check{r}_\infty,\check{h};L) &=&
  \frac{2\mathsf{R}_0}{\mathsf{N}}\Big\{
  K_d\Big[\mathsf{R}_0\ln(\mathsf{R}_0/L^2)
\nonumber\\&&\strut
  -\check{r}_\infty\ln\frac{\check{r}_\infty}{L^2} \Big]
  +2Q'_{d,6-d}(\mathsf{R}_0)\Big\},\;\qquad
\end{eqnarray}
where the function $\mathsf{N}$ continues to be given by
Eq.~(\ref{eq:Nth}). Likewise, Eqs.~(\ref{eq:R0}) and
(\ref{eq:Romless6}) for $\mathsf{R}_0$ and $\mathsf{R}_\omega$ remain
valid.

\begin{figure}[htbp]
  \centering
  \includegraphics[clip,scale=1]{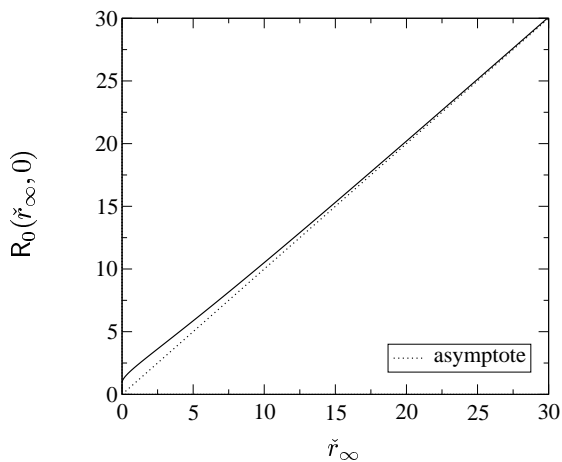}
  \caption{Scaling function $\mathsf{R}_0(\check{r}_\infty,0)$ for
    ${d=3}$, as given by Eq.~(\ref{eq:R0r3}). The dotted line represents
    the asymptote
    $\mathsf{R}_{0,\text{as}}(\check{r}_\infty,0)=\check{r}_\infty$
    that this function approaches for large values of
    $\check{r}_\infty$ in an exponential manner.}
  \label{fig:R0r3}
\end{figure}

\begin{figure}[htbp]
  \centering
  \includegraphics[clip,scale=1]{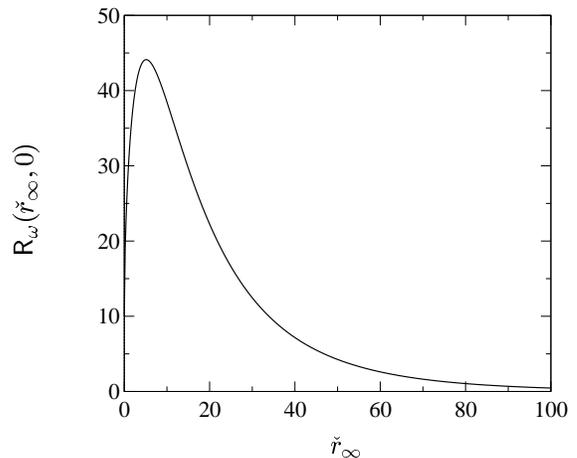}
  \caption{Scaling function $\mathsf{R}_\omega(\check{r}_\infty,0)$ for
    ${d=3}$ one obtains by inserting Eq.~(\ref{eq:R0r3}) into
    (\ref{eq:Romless6}).}
  \label{fig:Romr3}
\end{figure}

\begin{figure}[htbp]
  \centering
  \includegraphics[clip,scale=1]{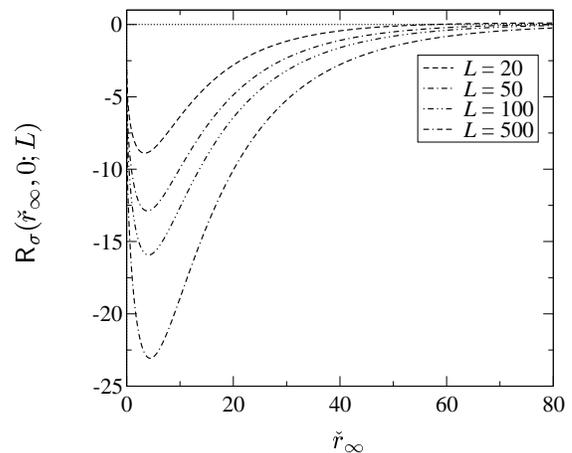}
  \caption{Function $\mathsf{R}_\sigma(\check{r}_\infty,0;L)$ for
    ${d=3}$ and the indicated values of $L$, as obtained by
    insertion of Eq.~(\ref{eq:R0r3}) into (\ref{eq:Rsig6}).}
  \label{fig:Rsigr6}
\end{figure}

In the case of primary interest, $d=3$, these results can be augmented
by determining the explicit solution to Eq.~(\ref{eq:R0}) for $h=0$.
To this end, we substitute the result~(\ref{eq:Q32}) derived in
Appendix~\ref{app:Gsig} for the function $Q_{3,2}$. Straightforward
algebraic manipulations then lead to
\begin{eqnarray}
  \label{eq:R0r3}
  \mathsf{R}_0(\check{r}_\infty,0) &=& 4\,\mathrm{arccsch}^{2}
  \big[2\,\exp\big(-\sqrt{\check{r}_\infty}/2\big)\big] \nonumber\\ &=&
  4\ln^2\Big[\frac{1}{2}\Big(e^{\sqrt{\check{r}_\infty}/2}
  +\sqrt{4+e^{\sqrt{\check{r}_\infty}/2}}\Big) \Big].\; \qquad
\end{eqnarray}
This and the associated scaling function that follows from it via
Eq.~(\ref{eq:Romless6}) are depicted in Figs.~\ref{fig:R0r3} and
\ref{fig:Romr3}), respectively. Figure~\ref{fig:Rsigr6} shows the
function $\mathsf{R}_\sigma(\check{r}_\infty,0;L)$ defined in
Eq.~(\ref{eq:Rsig6}).


In conjunction with Eqs.~(\ref{eq:Romless6}), (\ref{eq:Rsigless6}),
and (\ref{eq:ansatzcheckrL6}), the result~(\ref{eq:R0r3}) gives us the
asymptotic behavior of $r_L$ for $h=0$ in three dimensions including
corrections-to-scaling, in an explicit analytic form.

\section{Finite-Size Behavior of Free Energy and Casimir Force}
\label{sec:fsfe}

We now turn to the computation of the finite-size free energy
(\ref{eq:fed}), beginning again with the case $d+\sigma<6$.

\subsection{The case $d+\sigma<6$ with $2<d<4$ and $2<\sigma<4$}
\label{sec:dsigless6}

In order to use Eq.~(\ref{eq:rel}), we need the $L$-dependent part of
$U_{d,\Omega}(0|L)$. The calculation is performed in
Appendix~\ref{app:aUd0}, giving
\begin{equation}
  \label{eq:DelUdOmrzero}
  \Delta U_{d,\Omega}(0|L)\mathop{\approx}\limits_{L\to \infty}
  L^{-d}\big[\Delta^{\text{GM}}_C(d) +
  \check{g}_\sigma\,\Delta^{\text{GM}}_{\sigma,C}(d,\sigma)+O(b^2)\big] \;.
\end{equation}
Here
\begin{equation}
  \label{eq:DeltaCGM}
  \Delta^{\text{GM}}_{C}(d)=-\pi^{-d/2}\,\Gamma(d/2)\,\zeta(d)
\end{equation}
and
\begin{equation}
  \label{eq:DeltaCsigmaGM}
  \Delta^{\text{GM}}_{\sigma,C}(d,\sigma)=-\frac{2^{\sigma-2}\,\zeta(d+\sigma-2)\,
  \Gamma[(d+\sigma-2)/2]}{\pi^{d/2}\,\Gamma(1-\sigma/2)}
\end{equation}
are the values of the Casimir amplitudes (\ref{eq:X000}) and
(\ref{eq:X0sigma}) for our Gaussian model, where $\zeta(d)$ is the
Riemann zeta function.

Upon exploiting the relation~(\ref{eq:Qdp2rel}) between the
  derivative of $Q_{d+2,2}(r)/r$ and $Q_{d,2}(r)/r$ derived in
  Appendix~\ref{app:Qas}, one can readily integrate
  Eq.~(\ref{eq:DW0appQ}) to obtain
\begin{equation}
  \label{eq:DU0app}
  \Delta U^{(0)}_{d,\Omega}(r|L)\approx
  -L^{-d}\Big[\Delta_C^{\text{GM}}(d)
  +\frac{4\pi}{rL^2}\,Q_{d+2,2}(r L^2)\Big].
\end{equation}
Likewise, $\Delta U^{(1)}_{d,\Omega}(r|L)$ follows by integration of Eq.~(\ref{eq:DW1app}).
A simple integration by parts yields
\begin{equation}
  \label{eq:DU1app}
  \Delta U^{(1)}_{d,\Omega}(r|L)
\approx L^{-(d+\sigma-2)}\,Q_{d,\sigma}(r L^2) \;.
\end{equation}

The above results can now be combined in a straightforward fashion to
determine the scaled free-energy density $f_L L^{d-1}$. One gets
\begin{equation}
  \label{eq:fLdsigless6}
 L^{d-1}(f_L-Lf^{(0)})\approx
 \left.\Upsilon(\check{t},\check{r},\check{h},\check{g}_\omega,
   \check{g}_\sigma)\right|_{\check{r}=\check{r}_L}
\end{equation}
with
\begin{eqnarray}
\label{eq:Y}
\lefteqn{\Upsilon(\check{t},\check{r},\check{h}, \check{g}_\omega,
  \check{g}_\sigma)}&&
\nonumber\\
&\equiv&\frac{1}{2}\,\check{r}\,\check{t}
-\frac{\check{h}^2}{2\check{r}}- \frac{A_d}{d}\, \check{r}^{d/2}
-\frac{4\pi}{\check{
      r}}\,Q_{d+2,2}(\check{r})  +\frac{\check{g}_\omega}{4}\check{r}^2
\nonumber\\ &&\strut
+\check{g}_\sigma\Big[\Delta^{\text{GM}}_{\sigma,C}(d,\sigma)+Q_{d,\sigma}(\check{
  r})
-\frac{B_{d,\sigma}}{d+\sigma-2}\,\check{r}^{(d+\sigma-2)/2}
\Big],\nonumber\\
\end{eqnarray}
where $f^{(0)}\equiv f^{(0)}(K)$ denotes the smooth background term
(\ref{eq:f0}). As indicated, the function $Y$ in the first equation
must be taken at the solution $\check{r}_L$ of the scaled equation of
state $\partial Y(\check{t},\check{r},\check{h},\check{g}_\omega,
\check{g}_\sigma)/\partial\check{r}=0$, Eq.~(\ref{eq:sceqofst}).

The bulk free-energy density (per volume) $f_{\text{bk}}$ follows
from this in a straightforward manner. As is shown in
Appendix~\ref{app:Qas}, the functions $Q_{d,2}(r)$ and $Q_{d,\sigma\neq
2}(r)$
behave for large values of $r$ as
\begin{equation}
  \label{eq:Qd22}
  Q_{d,2}(r)\mathop{=}\limits_{r\to
    \infty}\frac{r^{(d+1)/4}}{2\,(2\pi)^{(d-1)/2}}\, e^{-\sqrt{r}} \,\big[1+
  O\big(r^{-1/2}\big)\big]
\end{equation}
and
\begin{equation}
  \label{eq:Qdsig}
  Q_{d,\sigma}(r)\mathop{=}\limits_{r\to
    \infty}-\Delta^{\text{GM}}_{\sigma,C}(d,\sigma)-\frac{D_\sigma(d)}{r}
+O\big(r^{-2}\big)\;,
\end{equation}
respectively, where
\begin{equation}
  \label{eq:Dsigd}
  D_\sigma(d)=\frac{2^\sigma\, \Gamma[(d+\sigma)/2]}{\pi^{d/2}\,
    \Gamma(-\sigma/2)}\,\zeta(d+\sigma)\;,
\end{equation}
according to Eq.~(\ref{eq:Qd2largey}).  Though not needed here, the
value of this coefficient appears in our subsequent analysis; it is
positive for $2<\sigma<4$ and vanishes both at $\sigma=2$ and $4$.
Since the same applies to $\Delta^{\text{GM}}_{\sigma,C}(d+2,\sigma)$, the
results~(\ref{eq:Qd22}) and (\ref{eq:Qdsig}) are in conformity with
each other.

Hence neither the term $\propto Q_{d+2,2}$ in
Eq.~(\ref{eq:fLdsigless6}) nor the sum of $Q_{d,\sigma}$ and
$\Delta^{\text{GM}}_{\sigma,C}(d,\sigma)$ contribute in the thermodynamic
bulk limit. The remaining terms yield
\begin{eqnarray}
  \label{eq:fbdsigless6}
  f_{\text{bk}}-f^{(0)}&\approx &\frac{r_\infty}{2}\,
  g_t-\frac{g_h^2}{2r_\infty}- \frac{A_d}{d}\, r_\infty^{d/2}
  +\frac{g_\omega}{4}r_\infty^2\nonumber\\
&&\strut -g_\sigma\,\frac{B_{d,\sigma}}{d+\sigma-2}\,
r_\infty^{(d+\sigma-2)/2} \;.
\end{eqnarray}

The difference of the right-hand sides of Eqs.~(\ref{eq:fLdsigless6})
and (\ref{eq:fbdsigless6}) gives us the scaled excess free-energy
density $L^{d-1}f^{\text{sing}}_{\text{ex}}$. In the result, the
scaling function ${\mathcal R}$ of Eq.~(\ref{eq:rLscf}) must be
substituted for $\check{r}$, and for $r_\infty$, we have the scaling
form (\ref{eq:rinftyscf}) and the relationship (\ref{eq:RlargeL})
between the scaling functions ${\mathcal R}$ and ${\mathcal
  R}_\infty$.  Obviously, the resulting expression for
$L^{d-1}f^{\text{sing}}_{\text{ex}}$ therefore complies with the
scaling form (\ref{eq:freeenergyper}).

To derive and describe what this means in terms of explicit results
for scaling functions, it is advantageous to eliminate the temperature
field $g_t$ in favor of the inverse bulk susceptibility $r_\infty$
(which in the spherical model is related to the bulk correlation
length $\xi_\infty$ via $r_\infty\propto \xi_\infty^{-2}$). The
advantage originates from the explicit results we have been able to
get for the dependence of $\check{r}_L$ on
$\check{r}_\infty$. Denoting the corresponding analogs of the scaling
functions $X,\ldots,X_\omega$ in Eqs.~(\ref{eq:freeenergyper}) and
(\ref{eq:Xexp}) by $Y,\ldots,Y_\omega$, we write
\begin{equation}
  \label{eq:fexrinftyless6}
  f^{\text{sing}}_{\text{ex}}\approx L^{-(d-1)}\,Y(r_\infty L^2, h
  L^{\Delta/\nu},g_\omega L^{-\omega},g_\sigma L^{-\omega_\sigma})
\end{equation}
with
\begin{eqnarray}
  \label{eq:Yexexpless6}
  Y(\check{r}_\infty,\check{h},\check{g}_\omega,
  \check{g}_\sigma) &=&
  Y_0(\check{r}_\infty,\check{h}) +\check{g}_\omega\,
  Y_\omega(\check{r}_\infty,\check{h}) \nonumber\\&&
  \strut +\check{g}_\sigma\,
  Y_\sigma(\check{r}_\infty,\check{h})+\ldots\;.
\end{eqnarray}

The above results in conjunction with those of Sec.~\ref{sec:Ldep} and
\ref{sec:rLrinfty} yield the scaling functions
\begin{eqnarray}
  \label{eq:Yex0}
  Y_0(\check{r}_\infty,\check{h})&=&
  -\frac{A_d}{2}\,\check{r}_\infty^{(d-2)/2}\,
  [\check{r}_\infty-\mathsf{R}_0(\check{r}_\infty,\check{h})]
\nonumber\\ &&\strut
+\frac{A_d}{d}\,
\big[\check{r}_\infty^{d/2}-\mathsf{R}_0^{d/2}(\check{r}_\infty,\check{h})\big]
\nonumber\\&&\strut
-\frac{4\pi\, Q_{d+2,2}[\mathsf{R}_0(\check{r}_{\infty},\check{h})]
}{\mathsf{R}_0 (\check{r}_{\infty},\check{h})}
\nonumber\\&&\strut
-\frac{\check{h}^{2} \left[\check{r}_{\infty}
    -\mathsf{R}_0(\check{r}_{\infty},\check{h})\right]^{2}
}{2\check{r}_{\infty}^{2}\, \mathsf{R}_0(\check{r}_{\infty},\check{h})}\;,
\end{eqnarray}
\begin{equation}
  \label{eq:Yexomega}
   Y_\omega(\check{r}_\infty,\check{h})= \frac{1}{4}\,
   [\check{r}_\infty -\mathsf{R}_0(\check{r}_\infty,\check{h})]^2\;,
\end{equation}
and
\begin{eqnarray}
  \label{eq:Yexsigmaless6}
  Y_\sigma(\check{r}_\infty,\check{h})&=&
  \Delta_{\sigma,C}^{\text{GM}}(d,\sigma)
  +Q_{d,\sigma}[\mathsf{R}_0(\check{r}_{\infty},\check{h})]
\nonumber\\ && \strut
 +B_{d,\sigma}\,\bigg\{\frac{\check{r}_{\infty}^{(d+\sigma-2)/2}-\mathsf{R}_0^{(d+\sigma-2)/2}(\check{r}_\infty,\check{h})}{d+\sigma-2}
\nonumber\\ &&\strut
-\frac{\check{r}_\infty -\mathsf{R}_0(\check{r}_\infty,\check{h})}{2}
\, \check{r}_{\infty}^{(d+\sigma-4)/2}
\bigg\}\;.
\end{eqnarray}
Note that $\check{r}_{\infty}$ is the full inverse bulk
susceptibility, which itself has corrections to scaling $\sim
g_\omega$ and $g_\sigma$ according to Eq.~(\ref{eq:rLscf}). Expanding
it in powers of $g_\omega$ and $g_\sigma$ to express
$f_{\text{ex}}^{\text{sing}}$ in terms of
$\check{r}|_{g_\omega=g_\sigma=0}$ would produce contributions linear
in $g_\omega$ and $g_\sigma$, in addition to those involving $Y_\omega$
and $Y_\sigma$.

\subsubsection{Behavior at bulk criticality}

At the bulk critical point (bcp) $T=T_{c,\infty}$, $h=0$, the above
results reduce to
\begin{eqnarray}
  \label{eq:fexrinftyless6bc}
  f^{\text{sing}}_{\text{ex,bcp}}&\mathop{\approx}\limits_{L\to\infty}&
 L^{-(d-1)} \Big[ \Delta_{C}^{\text{SM}}(d)+
  \Delta_{\omega,C}^{\text{SM}}(d)\,\frac{g_\omega(b)}{L^{4-d}} \nonumber\\
&&\strut +
  \Delta_{\sigma,C}^{\text{SM}}(d,\sigma)\,
  \frac{g_\sigma(b)}{L^{\sigma-2}}+\ldots\Big]\;,
\end{eqnarray}
where $\Delta_{C}^{\text{SM}}$, $\Delta_{\omega,C}^{\text{SM}}$, and
$\Delta_{\sigma,C}^{\text{SM}}$, the spherical-model values of the
amplitudes (\ref{eq:X000})--(\ref{eq:X0omega}), are given
by \cite{rem:exDel3,Dan98}
\begin{eqnarray}
 \label{eq:DelCid}
\Delta_{C}^{\text{SM}}(d)&\equiv& Y_{0}(0,0)\nonumber\\
&=& -\frac{A_{d}}{d}\,\mathsf{R}_{0,\text{bcp}}^{d/2} -\frac{4\pi \,
    Q_{d+2,2}(\mathsf{R}_{0,\text{bcp}})}{\mathsf{R}_{0,\text{bcp}}}\;,\qquad\quad
\end{eqnarray}
\begin{eqnarray}
  \label{eq:Delomid}
\Delta_{\omega,C}^{\text{SM}}(d)&\equiv&Y_{\omega}(0,0)=
{\mathsf{R}_{0,\text{bcp}}^{2}}/{4} \;,
\end{eqnarray}
and
\begin{eqnarray}
 \label{eq:Delsigid}
\Delta_{\sigma,C}^{\text{SM}}(d,\sigma)&\equiv&Y_{\sigma}(0,0)\nonumber\\
&=&\Delta_{\sigma,C}^{\text{GM}}(d,\sigma)
+Q_{d,\sigma}(\mathsf{R}_{0,\text{bcp}})\nonumber \\ &&\strut
 -\frac{B_{d,\sigma}}{d+\sigma-2}\,
 \mathsf{R}_{0,\text{bcp}}^{(d+\sigma-2)/2}\,. \qquad
\end{eqnarray}
Here $\mathsf{R}_{0,\text{bcp}}\equiv\mathsf{R}_0(0,0)$ is the
$d$-dependent solution to Eq.~(\ref{eq:R0}) at the bulk critical point.

Thus at bulk criticality, the scaling fields $g_\omega$ and $g_\sigma$
indeed give leading and next-to-leading corrections to the familiar first
term involving the Casimir amplitude $\Delta_{C}^{\text{SM}}(d)$ of
the spherical model with short-range interactions.

\subsubsection{Behavior for $T>T_{c,\infty}$ and $h=0$}

Next, we consider the case $T>T_{c,\infty}$ and $h=0$. As $L\to
\infty$, the scaled inverse finite-size and bulk susceptibilities
$\check{r}_L$ and $\check{r}_\infty$ both tend towards $+\infty$.
Hence, to obtain the asymptotic large-$L$ behavior, we must study the
behavior of the functions $\mathsf{R}_0$, $\mathsf{R}_\omega$, and
$\mathsf{R}_\sigma$ in the limit $\check{r}_\infty\to\infty$. Clearly,
$\mathsf{R}_0(\check{r}_\infty,\check{h})\to \check{r}_\infty$ as
$\check{r}_\infty\to\infty$. To determine the asymptotic
large-$\check{r}_\infty$ behavior of
$\mathsf{R}_0(\check{r}_\infty,0)$, we choose $\check{r}_\infty$ so
large that the function $Q_{d,2}(\mathsf{R}_{0})$ in
Eq.~(\ref{eq:R0})) can safely be replaced by the first term of its
asymptotic expansion~(\ref{eq:Qd22}). Solving for
$\mathsf{R}_0(\check{r}_\infty,0)$ then yields
\begin{eqnarray}
 \label{eq:R0aslt6}
\mathsf{R}_{0}(\check{r}_{\infty},0)&
  \mathop{=}\limits_{\check{r}_{\infty}\to\infty}&
  \check{r}_{\infty} +\frac{2\,(2\pi)^{(1-d)/2}}{(d-2)\,A_{d}}\,
\check{r}_{\infty}^{(5-d)/4}\,
    e^{-\check{r}_\infty^{1/2}}
  \nonumber \\
  & & \strut\times  \big[1+O\big(\check{r}_{\infty}^{-1/2}\big)\big]\;,
\end{eqnarray}
where $\check{r}_\infty$ now also is to be taken at $h=0$.

Using this result together with Eqs.~(\ref{eq:Qd22}) and
(\ref{eq:Qdsig}), one can derive the large-$\check{r}_\infty$ behavior
of the scaling functions $Y_0$, $Y_\omega$ and $Y_\sigma$ in a
straightforward fashion. One obtains
\begin{equation}
 \label{eq:Y0aslt6}
  Y_{0}(\check{r}_{\infty},0)
  \mathop{=}\limits_{\check{r}_{\infty}\to\infty} - \frac{1
  +O(\check{r}_{\infty}^{-1/2})}{(2\pi)^{(d-1)/2}}\,
  \check{r}_{\infty}^{(d-1)/4} e^{-\check{r}_{\infty}^{1/2}}\;,
\end{equation}
\begin{eqnarray}
 \label{eq:Yomaslt6}
Y_{\omega}(\check{r}_{\infty},0)
&\mathop{=}\limits_{\check{r}_{\infty}\to\infty}&
\frac{(2\pi)^{1-d}}{(d-2)^2\,A_d^2} \,
  \check{r}_{\infty}^{(5-d)/2}\,
e^{-2\check{r}_{\infty}^{1/2}}\nonumber\\&&\times
\big[1+O\big(\check{r}_{\infty}^{-1/2}\big)\big]\,,
\end{eqnarray}
and
\begin{equation}
 \label{eq:Ysigaslt6}
Y_{\sigma}(\check{r}_{\infty},0)
\mathop{=}\limits_{\check{r}_{\infty}\to\infty}
-D_{\sigma}(d)\,\check{r}_{\infty}^{-1} +
O(\check{r}_{\infty}^{-2})\;,
\end{equation}
where $D_\sigma(d)$ is the constant introduced in
Eq.~(\ref{eq:Dsigd}). Unlike $Y_{0}(\check{r}_{\infty},0)$ and
$Y_{\omega}(\check{r}_{\infty},0)$, which decay exponentially, the
scaling function $Y_{\sigma}(\check{r}_{\infty},0)$ decays in an
algebraic manner.

Thus the contribution due to this latter slowly decaying term governs
the large-$L$ behavior of the excess free energy
$f_{\text{ex}}^{\text{sing}}$ for $T>T_{c,\infty}$ whenever the
coupling constant $b$ of the long-range potential does not vanish.
One has
\begin{eqnarray}
 \label{eq:asymptfexbgt0}
 f_{\text{ex}}^{\text{sing}}
  &\mathop{\approx}\limits_{L\to\infty}&
 -g_{\sigma}(b)\,\frac{D_{\sigma}(d)}{r_{\infty}}\,L^{-(d+\sigma-1)}
 \nonumber\\ &\approx &
  -g_\sigma(b)\,L^{-(d+\sigma-1)} D_{\sigma}(d)
  \,\Big(\frac{g_{t}}{A_{d}}\Big)^{-\gamma},\qquad
\end{eqnarray}
where we substituted Eq.~(\ref{eq:rinftyhzero}) for $r_\infty$ to
obtain the second line. This strongly contrasts with the asymptotic
form that applies in the absence of long-range interactions:
\begin{equation}
 \label{eq:asymptfexbeq0}
  \left.f_{\text{ex}}^{\text{sing}}\right|_{b=0}
  \mathop{\approx}\limits _{L\to\infty}  -(2\pi L)^{-\frac{d-1}{2}}
  \Big(\frac{g_{t}}{A_{d}}\Big)^{\frac{d-1}{4}\,\gamma} \,
  e^{-(g_t/A_d)^{\nu}L}\;.
\end{equation}

\subsection{The case $d+\sigma=6$ with $2<d<4$}
\label{sec:dsig6}

Proceeding along similar lines as in the foregoing subsection, one can
derive the analogs of Eqs.~(\ref{eq:fLdsigless6}),
(\ref{eq:fbdsigless6}), (\ref{eq:fexrinftyless6}), and
(\ref{eq:Yexexpless6}). They read
\begin{eqnarray}
  \label{eq:fLdsig6}
 \lefteqn{L^{d-1}(f_L-Lf^{(0)})}&&
\nonumber\\
&\approx&\frac{1}{2}\,\check{r}_L\,\check{t} -\frac{\check{
      h}^2}{2\check{r}_L}- \frac{A_d}{d}\,
  \check{r}_L^{d/2}-\frac{4\pi}{\check{r}_L}\,Q_{d+2,2}(\check{r}_L)
\nonumber\\ &&\strut
+\frac{1}{4}(w_d+\tilde{w}_d\,b)L^{d-4}\,\check{r}_L^2
\nonumber\\ &&\strut
+\check{g}_\sigma\Big[\Delta^{\text{GM}}_{6-d,C}(d)+Q_{d,6-d}(\check{
  r}_L)
+\frac{K_d}{4}\,\check{r}_L^2\,\ln\frac{\check{r}_L}{L^2}
\Big]\;,
\nonumber\\
\end{eqnarray}
\begin{eqnarray}
  \label{eq:fbdsig6}
  f_{\text{bk}}-f^{(0)}&\approx &\frac{r_\infty}{2}\,
  g_t-\frac{g_h^2}{2r_\infty}- \frac{A_d}{d}\, r_\infty^{d/2}
\nonumber\\ &&\strut
  +\frac{w_d+b\,\tilde{w}_d}{4}\,r_\infty^2
+g_\sigma\,\frac{K_d}{4}\,r_\infty^2\ln r_\infty
\,,\nonumber\\
\end{eqnarray}
and
\begin{eqnarray}
  \label{eq:Yexexp6}
  L^{d-1}\,f^{\text{sing}}_{\text{ex}} &\approx&
  Y_0(\check{r}_\infty,\check{h})  +(w_d+\tilde{w}_d\,b)L^{d-4}\,
  Y_\omega(\check{r}_\infty,\check{h})\nonumber\\&&
  \strut+\check{g}_\sigma\,
  Y_\sigma(\check{r}_\infty,\check{h};L)+\ldots\;,
\end{eqnarray}
where
\begin{eqnarray}
  \label{eq:Yexsigma6}
   Y_\sigma(\check{r}_\infty,\check{h};L)&=&
   \Delta_{\sigma,C}^{\text{GM}}(d,6-d)
   +Q_{d,6-d}[\mathsf{R}_0(\check{r}_{\infty},\check{h})]
\nonumber\\ &&\strut
+\frac{K_{d}}{4}\bigg\{
\big[\check{r}_{\infty}
-2\,\mathsf{R}_0(\check{r}_{\infty},\check{h})\big]\,\check{r}_\infty
\ln\frac{\check{r}_\infty}{L^2}
 \nonumber \\ &&\strut
+ \mathsf{R}_0^2(\check{r}_{\infty},\check{h})\,
\ln\frac{\mathsf{R}_0(\check{r}_\infty,\check{h})}{L^2}\bigg\}\;.
\end{eqnarray}
while the functions $Y_0$ and $Y_\omega$ remain given by
Eqs.~(\ref{eq:Yex0}) and (\ref{eq:Yexomega}), respectively. Owing to
the presence of logarithmic anomalies, the analogs of the scaling
functions $Y$ and $Y_\sigma$ in Eq.~(\ref{eq:Yexexpless6}) have an
additional explicit dependence on $L$. It should also be remembered
that logarithmic anomalies reside also in the temperature-dependence
of the $b$-dependent corrections to scaling of $\check{r}_\infty$.

\begin{figure}[htbp]
  \centering
  \includegraphics[clip,scale=1]{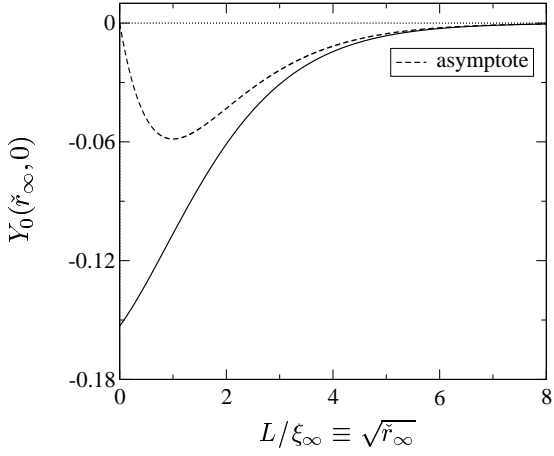}
  \caption{Scaling function $Y_0(\check{r}_\infty,0)$ for $d=3$ (full
    line). The dashed line represents the
    asymptote~(\ref{eq:Y0aslt6}). The value
    $Y_0(0,0)=\Delta^{\text{SM}}_C(d=3)$ which
    $Y_0(\check{r}_\infty,0)$ approaches as $\check{r}_\infty\to 0$ is
    known exactly: According to Ref.~\cite{Dan98}, it is given by
    $\Delta^{\text{SM}}_C(3)=-2\zeta(3)/(5\pi)=-0.15305\ldots$.}
  \label{fig:Y03}
\end{figure}

\begin{figure}[htbp]
  \centering
  \includegraphics[clip,scale=1]{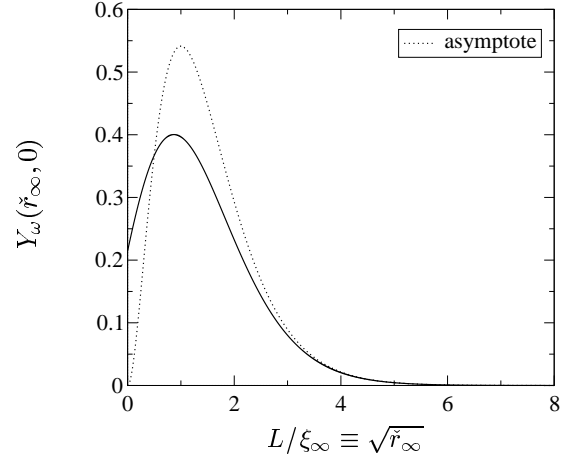}
  \caption{Scaling function $Y_\omega(\check{r}_\infty,0)$ for $d=3$
    (full lined).  The dashed line represents the
    asymptote~(\ref{eq:Yomaslt6}).}
  \label{fig:Yom3}
\end{figure}

\begin{figure}[htbp]
  \centering
  \includegraphics[clip,scale=1]{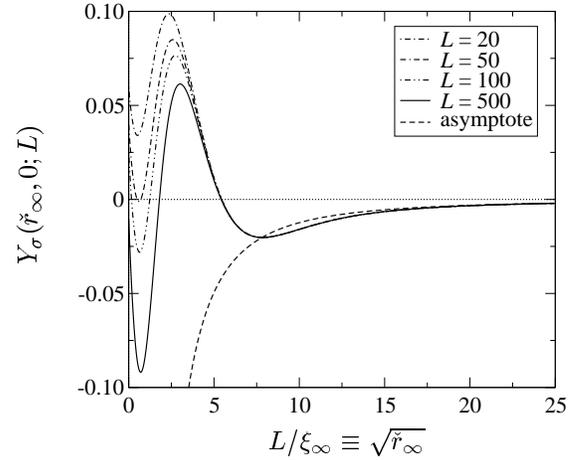}
  \caption{Function $Y_\sigma(\check{r}_\infty,0;L)$ for $d=\sigma=3$
    and the indicated values of $L$ (dashed-dotted and full
    lines). The dashed line  represents the corresponding
    asymptote~(\ref{eq:Ysigaslt6}) with $d=\sigma=3$.}
  \label{fig:Ysig3}
\end{figure}

The scaling functions $Y_0(\check{r}_\infty,0)$ and
$Y_\omega(\check{r}_\infty,0)$ for the three-dimensional case are
plotted in Figs.~\ref{fig:Y03} and \ref{fig:Yom3}, respectively.
In Fig.~\ref{fig:Ysig3}, the function
$Y_\sigma(\check{r}_\infty,h=0;L)$ is displayed for the case
$d=\sigma=3$ and some values of $L$.

\subsubsection{Behavior at bulk criticality}

Let us again see how these results simplify at the bulk critical
point. From Eqs.~(\ref{eq:Yexexp6})--(\ref{eq:Yexsigma6}) one easily
deduces the asymptotic behavior
\begin{eqnarray}
  \label{eq:fexrinftyeq6bc}
  f^{\text{sing}}_{\text{ex,bcp}}&\mathop{\approx}\limits_{L\to\infty}&
 L^{-(d-1)} \bigg\{ \Delta_{C}^{\text{SM}}(d) \nonumber\\
&&\strut +
  \frac{g_\sigma(b)}{L^{4-d}}\bigg[
 \frac{K_{d}}{4}\,
\mathsf{R}_{0,\text{bcp}}^{2}\,\ln\frac{\mathsf{R}_{0,\text{bcp}}}{L^{2}}
\nonumber\\&&\strut+
  \Delta_{\sigma,C}^{\text{GM}}(d,6-d)
  +Q_{d,6-d}\big(\mathsf{R}_{0,\text{bcp}}\big)
 \bigg]
  \nonumber\\ &&\strut +
  \Delta_{\omega,C}^{\text{SM}}(d)\,
  \frac{w_{d}+\tilde{w}_{d}\,b}{L^{4-d}}+\ldots\bigg\}\;.
\end{eqnarray}

The leading corrections to scaling now results from the $b$-dependent
contribution involving the logarithmic anomaly.

\subsubsection{Behavior for $T>T_{c,\infty}$ and $h=0$}

Turning to the case of $T>T_{c,\infty}$ and $h=0$, let us again
consider the asymptotic behavior of $f^{\text{sing}}_{\text{ex}}$ for
$L\to \infty$ at fixed $T>T_{c,\infty}$. Upon inserting the
large-$\check{r}_\infty$ form (\ref{eq:R0aslt6}) of $\mathsf{R}_0$
into the result~(\ref{eq:Yexsigma6}) for $Y_\sigma$, one sees that the
contribution in curly brackets decays exponentially and hence
asymptotically negligible compared to the algebraically decaying
contribution from the sum of the two terms in the first line of this
equation. This means that the limiting form~(\ref{eq:Ysigaslt6})
carries over to the present case, except that we must set
$\sigma=6-d$.  Since the expressions~(\ref{eq:Yex0}) and
(\ref{eq:Yexomega}) for the scaling functions $Y_0$ and
$Y_\omega$---and hence their limiting forms~(\ref{eq:Y0aslt6}) and
(\ref{eq:Yomaslt6})---continue to hold, the
results~(\ref{eq:asymptfexbgt0}) and (\ref{eq:asymptfexbeq0}) for the
leading asymptotic behavior of $f_{\mathrm{ex}}^{\mathrm{sing}}$ when
$b\neq 0 $ or $b=0$, respectively, also remain valid.

\subsubsection{The case $d=\sigma=3$}

\begin{figure*}[htb]
\begin{center}\hspace*{1.5em}\includegraphics[%
  clip,scale=1]{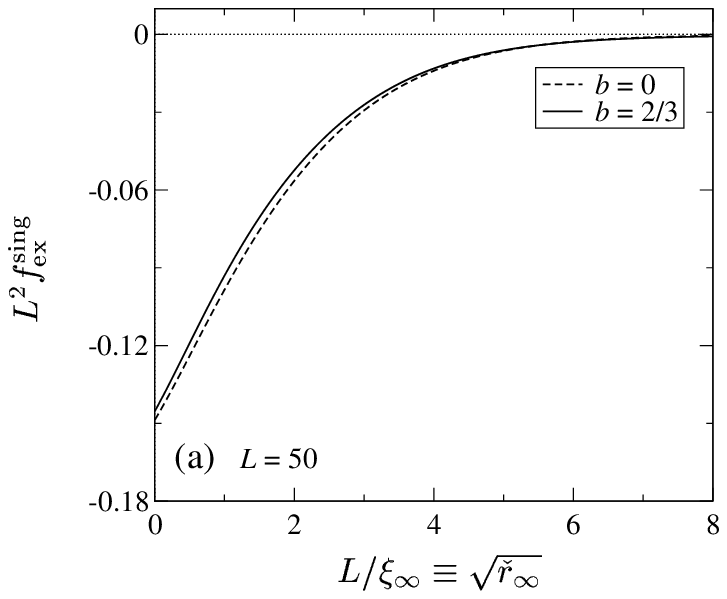}\;\;\hfill{}\;\;\includegraphics[clip,%
  scale=1]{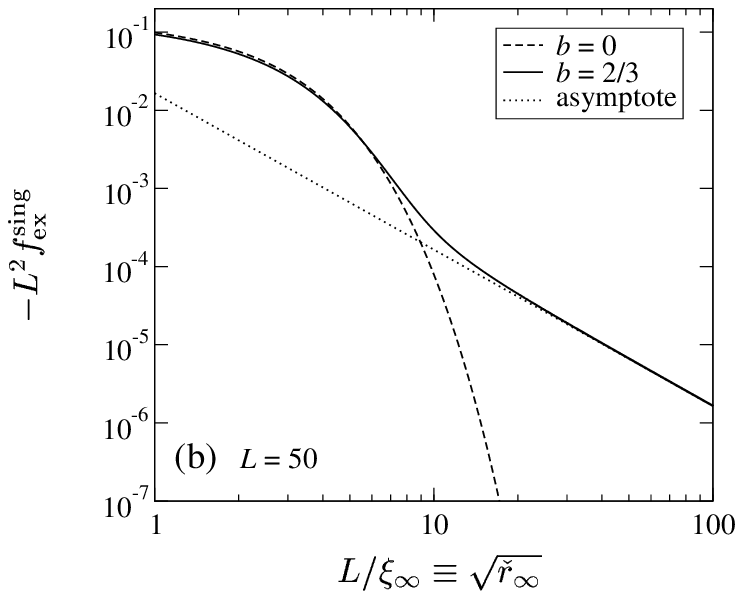}\hspace*{1.5em}~\end{center}
\caption{Scaled excess free-energy density~(\ref{eq:Yexexp6}) of the
  three-dimensional spherical model in a $L\times\infty^{2}$~slab with
  periodic boundary conditions for $L=50$, plotted versus the
  finite-size scaling variable
  $L/\xi_{\infty}\equiv\sqrt{\check{r}_\infty}$ (a). The solid line
  corresponds to the case $\sigma=3$ of nonretarded van-der-Waals-type
  interactions with $g_\sigma(b)\equiv b=2/3$; the dashed line shows
  results for the short-range case $b=0$ for comparison. In (b) the
  graphs displayed in (a) are plotted in a double-logarithmic manner.
  In this representation the asymptote~(\ref{eq:Ysigaslt6}) (dotted
  line) becomes a straight line with the slope $-2$. The nonuniversal
  constants $w_{d}$ and $\tilde{w}_{d}$ both have been set to unity.}
\label{fig:plotsfexab}
\end{figure*}

In Fig.~\ref{fig:plotsfexab} the scaled excess free-energy
densities~(\ref{eq:Yexexp6}) of the three-dimensional case with
nonretarded van-der-Waals-type interactions ($\sigma=3$) and without
those are compared for the chosen value $L=50$ of the slab thickness
$L$. For simplicity, we have set the nonuniversal constants $w_{d}$
and $\tilde{w}_{d}$ to unity. As can clearly be seen from the
double-logarithmic plot (b), the asymptotic behavior for large
$L/\xi_\infty$ when $b\neq 0$ is characterized by the
asymptote~(\ref{eq:Ysigaslt6}) and differs strongly from its
counterpart for the short-range case $b=0$.

\begin{figure*}[htb]
\begin{center}\hspace*{1.5em}\includegraphics[%
  clip,
  scale=1]{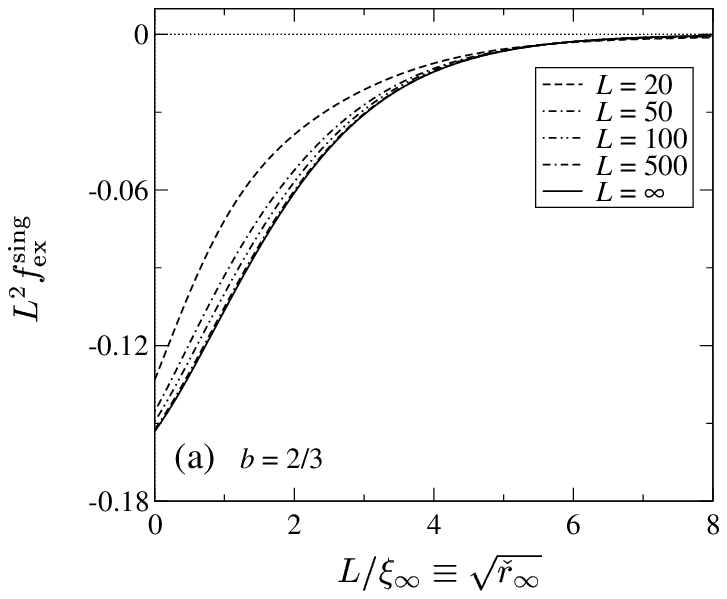}\;\;\hfill{}\;\;\includegraphics[%
  clip,
  scale=1]{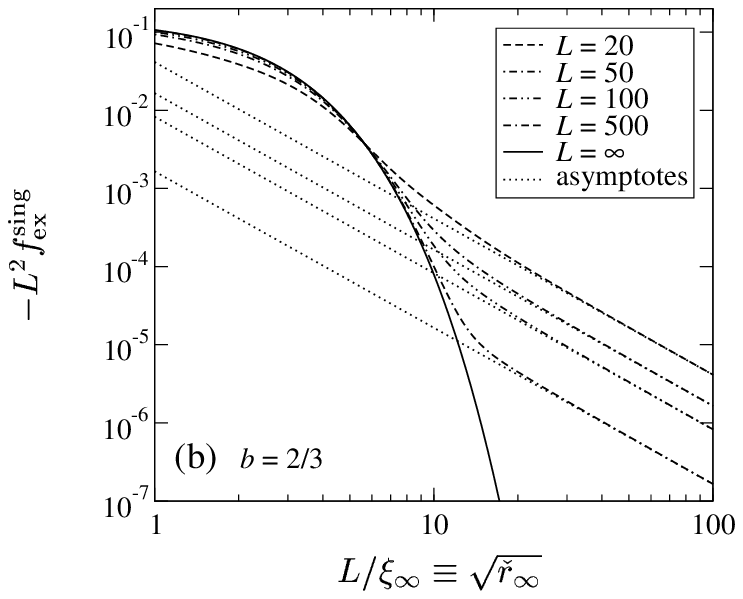}\hspace*{1.5em}~\end{center}
\caption{Scaled excess free-energy densities~(\ref{eq:Yexexp6}) of the
  three-dimensional spherical model in a $L\times\infty^{2}$~slab with
  periodic boundary conditions and nonretarded van-der-Waals-type
  interactions $(\sigma=3$). The results for various choices of $L$
  including $L=\infty$ are shown as linear (a) and double-logarithmic
  plots (b). The asymptotes (dotted lines) correspond to the power-law
  behavior~(\ref{eq:Ysigaslt6}). The nonuniversal constants $w_{d}$
  and $\tilde{w}_{d}$ both have been set to unity.}
\label{fig:plotsfexcd}
\end{figure*}

In order to illustrate the effect of the explicit dependence of the
scaled excess free-energy density~(\ref{eq:Yexexp6}) for nonvanishing
interaction constant $b$ on $L$, we display in
Fig.~\ref{fig:plotsfexcd} linear and double-logarithmic plots of
$L^2\,f^{\text{sing}}_{\text{ex}}$ for a variety of values of $L$,
including $L=\infty$. For the sake of simplicity, we have set the
nonuniversal constants $w_d$ and $\tilde{w}_d$ to unity.

\subsection{The Casimir force}
 \label{sec:Casi}

Using the results of the foregoing subsection for the excess free
energy, the large-scale behavior of the Casimir force~(\ref{eq:FCdef})
can be derived in a straightforward fashion. Depending on whether
$d+\sigma<6$ or $d+\sigma=6$, we have
\begin{eqnarray}
 \label{eq:FClt6}
\frac{\mathcal{F}_{C}^{\mathrm{sing}}}{k_{B}T} & \approx & L^{-d}\big[
\Xi_{0}(\check{r}_{\infty},\check{h})
+g_{\omega}(b)\,L^{-\omega}\,\Xi_{\omega}(\check{r}_{\infty},\check{h})
\nonumber\\ &&\strut
+ g_{\sigma}(b)\,L^{-\omega_\sigma}\,
\Xi_{\sigma}(\check{r}_{\infty},\check{h}) +\ldots \big]
\end{eqnarray}
or
\begin{eqnarray}
 \label{eq:FCeq6}
\frac{\mathcal{F}_{C}^{\mathrm{sing}}}{k_{B}T} & \approx & L^{-d}\big[
\Xi_{0}(\check{r}_{\infty},\check{h})+(w_{d}+\tilde{w}_{d}b)\,L^{-\omega}
\,\Xi_{\omega}(\check{r}_{\infty},\check{h})\nonumber \\&&\strut
+g_{\sigma}(b)\,L^{-\omega}\,
\Xi_{\sigma}(\check{r}_{\infty},\check{h};L)+\ldots \big]\;,
\end{eqnarray}
where $\omega$ and $\omega_\sigma$ take their spherical-model
values~(\ref{eq:omegainfty}) and (\ref{eq:nusm}), respectively. The
scaling form~(\ref{eq:FClt6}) should hold more generally for the
$n$-vector model with $2<d<4$ even when $d+\sigma=6$, as long as
$\omega$ and $\omega_\sigma$ are not degenerate. This applies, in
particular, to the case $d=\sigma=3$ of nonretarded van-der-Waals
interactions, albeit with the appropriate (different) values of
$\omega$ and $\omega_\sigma$, and different scaling functions.

By taking the derivatives of Eqs.~(\ref{eq:fexrinftyless6}) and
(\ref{eq:Yexexp6}) with respect to $L$, one can express the above
functions $\Xi_0$,\ldots, $\Xi_\sigma$ in terms of the
functions $Y_0$,\ldots, $Y_\sigma$. One finds
\begin{equation}
 \label{eq:CasimirXi0}
\Xi_0(\check{r}_{\infty},\check{h})= \bigg[d-1-
  2\check{r}_\infty\partial_{\check{r}_{\infty}}-
  \frac{\Delta}{\nu}\,\check{h}\,\partial_{\check{h}}
  \bigg]Y_0(\check{r}_\infty,\check{h})\;,
\end{equation}
\begin{equation}
 \label{eq:CasimirXom}
\Xi_\omega(\check{r}_\infty,\check{h})= \bigg[d+\omega-1
-2\check{r}_\infty\partial_{\check{r}_\infty}
- \frac{\Delta}{\nu}\,\check{h}\,\partial_{\check{h}}\bigg]
    Y_\omega(\check{r}_\infty,\check{h}) \,,
\end{equation}
\begin{equation}
\Xi_{\sigma}(\check{r}_{\infty},\check{h})=\bigg[d+\omega_{\sigma}-1
-2\check{r}_\infty\partial_{\check{r}_\infty}
- \frac{\Delta}{\nu}\,\check{h}\,\partial_{\check{h}}\bigg]
Y_{\sigma}(\check{r}_{\infty},\check{h}) \;,
\end{equation}
and
\begin{eqnarray}
 \label{eq:CasimirXisigma2}
\Xi_{\sigma}(\check{r}_{\infty},\check{h};L) &=&
\bigg[d+\omega_{\sigma}-1
-2\check{r}_\infty\partial_{\check{r}_\infty}
- \frac{\Delta}{\nu}\,\check{h}\,\partial_{\check{h}}
\nonumber\\ &&\strut
-L\partial_L\bigg] Y_{\sigma}(\check{r}_{\infty},\check{h};L)\;,
\end{eqnarray}
where again the spherical-model values (\ref{eq:omegainfty}) and
(\ref{eq:nusm}) must be substituted for $\omega$, $\omega_\sigma$, and
$\Delta/\nu$ \cite{rem:gamnu}.

Noting that the limit of $\mathsf{R}_0$ as $\check{r}_\infty$ and
$\check{h}$ approach the bulk critical point exists,
\begin{equation}
  \label{eq:R0lim}
  \mathsf{R}_0(\check{r}_\infty,\check{h})
  \mathop{=}\limits_{\check{r}_\infty,\check{h}\to 0}
  \mathsf{R}_{0,\text{bcp}}+o(\check{r}_\infty,\check{h})\;,
\end{equation}
one sees that the same applies to the scaling functions
$Y_i(\check{r}_\infty,\check{h})$:
\begin{equation}
  \label{eq:Yibcp}
  Y_i(\check{r}_\infty,\check{h})
  \mathop{=}\limits_{\check{r}_\infty,\check{h}\to 0} Y_i(0,0)
  +o(\check{r}_\infty,\check{h})\;,\quad i=0,\omega,\sigma.
\end{equation}
Hence the terms in
Eqs.~(\ref{eq:CasimirXi0})--(\ref{eq:CasimirXisigma2}) involving the
derivatives with respect to $\check{r}_\infty$ and $\check{h}$ yield
vanishing contributions as the bulk critical point is approached.
Using this in conjunction with
Eqs.~(\ref{eq:DelCid})--(\ref{eq:Delsigid}) and
(\ref{eq:fexrinftyeq6bc}), one finds that the values of these
functions at the bulk critical point become
\begin{eqnarray}
 \label{eq:X0bcp}
\Xi_{0}(0,0) &=&  (d-1)\,\Delta_{C}^{\text{SM}}(d)\;,\\[\medskipamount]
 \label{eq:Xombcp}
\Xi_{\omega}(0,0) &=&
(d+\omega-1)\,\Delta_{\omega,C}^{\text{SM}}(d)\;, \\[\medskipamount]
 \label{eq:Xsigbcp}
\Xi_{\sigma}(0,0) &=&
(d+\omega_{\sigma}-1)\Delta_{\sigma,C}^{\text{SM}}(d,\sigma)\;,
\end{eqnarray}
and
\begin{eqnarray}
 \label{eq:Xsig2crit}
\Xi_{\sigma}(0,0;L)&=&3\big[\Delta_{\sigma,C}^{\text{GM}}(d,6-d)
  +Q_{d,6-d}\big(\mathsf{R}_{0,\text{bcp}}\big)\big] \nonumber\\
  &&\strut +\frac{K_{d}}{4}\,
\mathsf{R}_{0,\text{bcp}}^{2}\bigg(2
+3\ln\frac{\mathsf{R}_{0,\text{bcp}}}{L^{2}}\bigg) \;.\qquad
\end{eqnarray}

To obtain the critical Casimir forces in the cases $d+\sigma<6$ and
$d+\sigma=6$, we must simply substitute the scaling functions
$\Xi_{i}$ in Eqs.~(\ref{eq:FClt6}) and (\ref{eq:FCeq6}), respectively,
by their above values at the bulk critical point.

The asymptotic forms of the Casimir force as ${L\to \infty}$ at fixed
temperature $T>T_{c,\infty}$ and zero magnetic field can be inferred
in a straightforward fashion from the corresponding
results~(\ref{eq:asymptfexbgt0}) and (\ref{eq:asymptfexbeq0}) for the
excess free-energy density. Depending on whether a long-range
interaction $\propto b$ is present or absent, one has
\begin{equation}
 \label{eq:FCTgrTcasb}
\left.\frac{\mathcal{F}_{C}^{\mathrm{sing}}}{k_BT}\right|_{b\neq 0}
\mathop{\approx}\limits _{L\to\infty}
-g_\sigma(b)\,(d+\sigma-1)\,\frac{D_{\sigma}(d)}{L^{d+\sigma}}
\,\bigg(\frac{g_{t}}{A_{d}}\bigg)^{-\gamma}
\end{equation}
or the exponential decay
\begin{equation}
 \label{eq:FCTgrTcas0}
\left.\frac{\mathcal{F}_{C}^{\mathrm{sing}}}{k_BT}\right|_{b=0}
\mathop{\approx}\limits _{L\to\infty} -
\frac{\left({g_{t}}/{A_{d}}\right)^{(d+1)\gamma/4}}{(2\pi L)^{(d-1)/2}}\,
e^{-L\,\left({g_{t}}/{A_{d}}\right)^{\nu}}.
\end{equation}
Here again the spherical-model values~(\ref{eq:nusm}) must be
substituted for $\gamma$ and $\nu$.

\begin{figure*}[htb]
\begin{center}\hspace*{1.5em}\includegraphics[%
  clip,
  scale=1]{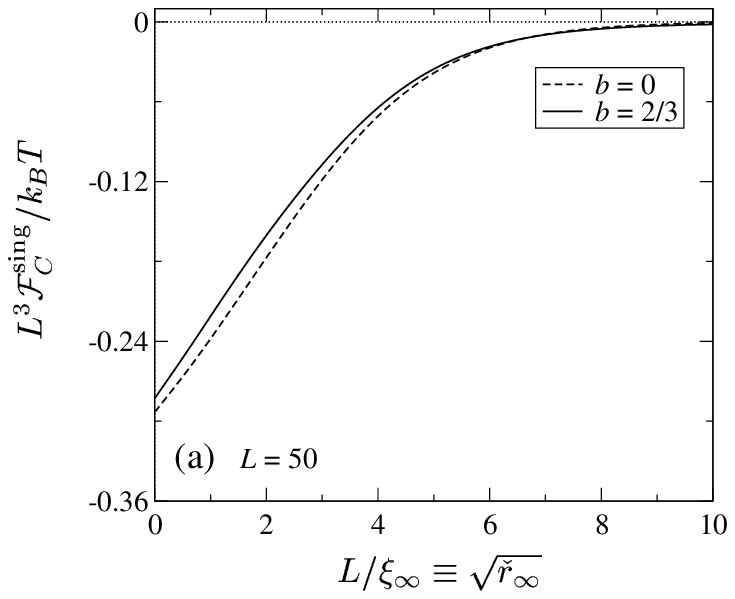}\;\;\hfill{}\;\;\includegraphics[%
  clip,
  scale=1]{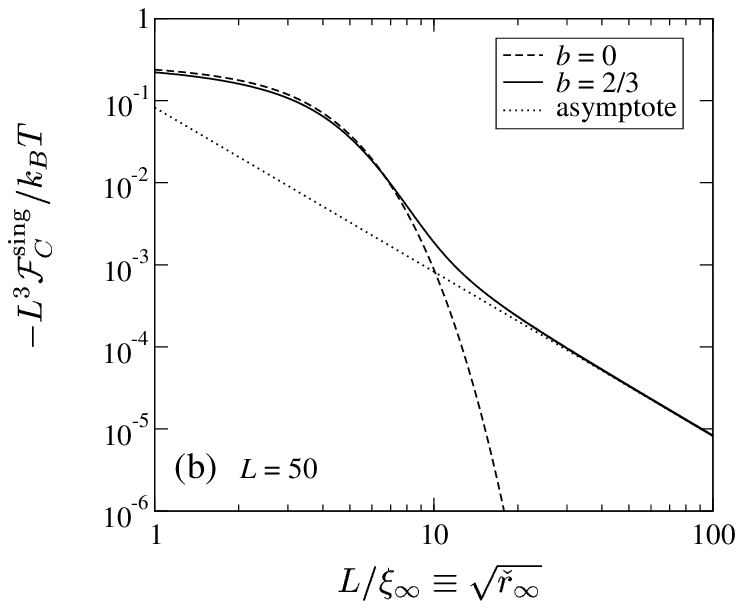}\hspace*{1.5em}~\end{center}
\caption{Scaled Casimir force~(\ref{eq:FCeq6}) of the
  three-dimensional spherical model in a $L\times\infty^{2}$~slab with
  periodic boundary conditions for $L=50$, plotted versus the
  finite-size scaling variable
  $L/\xi_{\infty}\equiv\sqrt{\check{r}_\infty}$ (a). The solid line
  corresponds to the case $\sigma=3$ of nonretarded van-der-Waals-type
  interactions with $g_\sigma(b)\equiv b=2/3$; the dashed line
  represents results for the short-range case $b=0$ for comparison.
  In (b) the graphs displayed in (a) are plotted in a
  double-logarithmic manner.  In this representation the
  asymptote~(\ref{eq:FCTgrTcasb}) $\sim\check{r}_\infty^{-1}$ (dotted line)
  becomes a straight line with the slope $-2$. The nonuniversal
  constants $w_{d}$ and $\tilde{w}_{d}$ both have been set to unity.}
 \label{fig:FCab}
\end{figure*}

Figure~\ref{fig:FCab} shows a comparison of the Casimir
forces~(\ref{eq:FCeq6}) of a $\infty^2\times L$ slab of thickness
$L=50$ with and without van-der-Waals-type interactions ($\sigma=3$),
where we have again set the nonuniversal constants $w_{d}$ and
$\tilde{w}_{d}$ to unity. The double-logarithmic plot (b) again nicely
demonstrates the approach to the asymptote~$\sim \check{r}_\infty^{-1}$
and the qualitatively different behavior in the short-range case.

\begin{figure*}
\begin{center}\hspace*{1.5em}\includegraphics[%
  clip,
  scale=1]{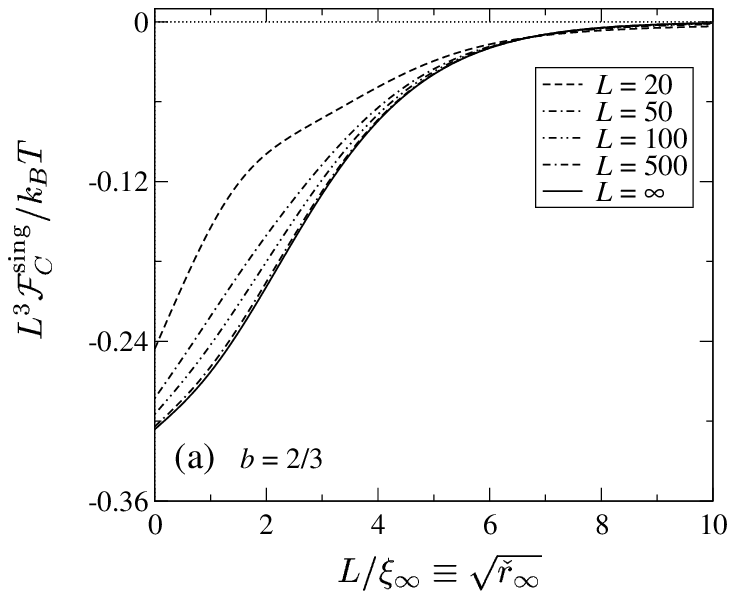}\;\;\hfill{}\;\;\includegraphics[%
  clip,
  scale=1]{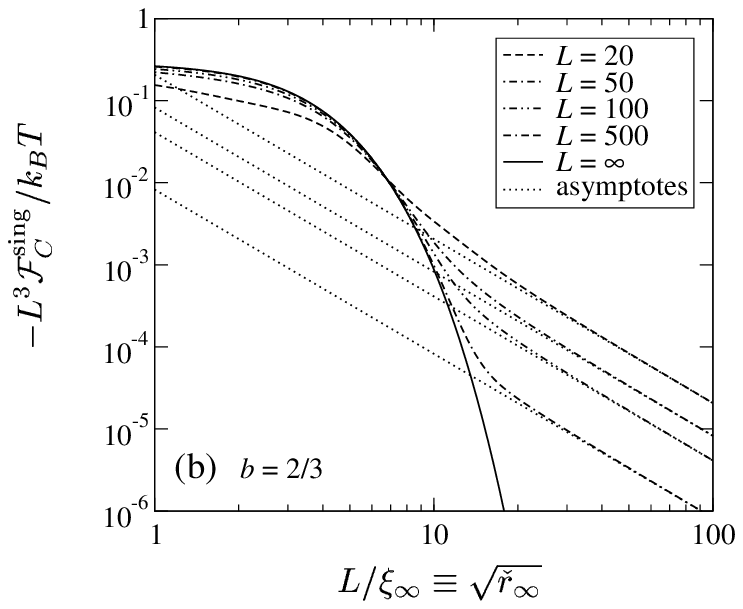}\hspace*{1.5em}~\end{center}
\caption{Scaled Casimir force~(\ref{eq:FCeq6}) of the
  three-dimensional spherical model in a $L\times\infty^{2}$~slab with
  periodic boundary conditions and nonretarded van-der-Waals-type
  interactions $(\sigma=3$). The results for various choices of $L$
  including $L=\infty$ are shown as linear (a) and double-logarithmic
  plots (b). The asymptotes (dotted lines) correspond to the power-law
  behavior $\sim\check{r}_\infty^{-1}$ of Eqs.~(\ref{eq:Ysigaslt6})
  and (\ref{eq:FCTgrTcasb}). The nonuniversal constants $w_{d}$
  and $\tilde{w}_{d}$ both have been set to unity.}
\label{fig:FCcd}
\end{figure*}

In Fig.~\ref{fig:FCcd} we illustrate how the scaled Casimir force for
the case with van-der-Waals-type interactions ($\sigma=3$) varies
under changes of the slab thickness $L$.

\section{Summary and Conclusions}
\label{sec:concl}

In this paper we have studied the effects of long-range interactions
whose pair potential decays at large distances as $x^{-d-\sigma}$ with
$2<\sigma<4$. Prominent examples of such interactions are nonretarded
and retarded van-der-Waals forces. The latter are ubiquitous in
nature; in particular, they are present in fluids.

Application of the phenomenological theory of finite-size scaling
revealed that such long-range interactions are of the kind termed
``subleading long-range interactions'' \cite{DR01} and hence should
yield corrections to scaling in the critical regime near the bulk
critical point.

For systems belonging to the universality class of the $n$-component
$\phi^4$ model in $d$ dimensions, the associated correction-to-scaling
exponent $\omega_\sigma$, given in Eq.~(\ref{eq:ysigma}), has a larger
value than its counterpart $\omega$ associated with the conventional
leading corrections-to-scaling of $d\geq 3$ dimensional systems with
short-range interactions. Hence the corrections-to-scaling governed by
$\omega_\sigma$ are next to leading.

However, irrespective of whether $\omega$ is smaller or larger than
$\omega_\sigma$, the subleading long-range interactions yield a
contribution to the Casimir force that decays in a power-law fashion
as a function of the film thickness $L$, both at and away from the
bulk critical temperature $T_{c,b}$. Since the fluctuation-induced
Casimir force one has even in the absence of these long-range
interactions, for $T>T_{c,b}$ decays exponentially on the scale of the
correlation length, the contribution due to the long-range
interactions becomes dominant for sufficiently large $L$.

To corroborate these findings we solved a mean spherical model with
such long-range interactions---and hence the limit $n\to\infty$ of the
corresponding $n$-vector model---exactly. For general values of
$\sigma,\,d\in(2,4)$, we confirmed the anticipated finite-size scaling
behavior, and determined the scaling functions to first order in the
irrelevant scaling fields $g_\omega$ and $g_\sigma$.

A crucial, though not unexpected, discovery was that the scaling field
associated with the conventional leading corrections to scaling,
$g_\omega$, depends on the strength $b$ of the long-range
interactions.  This dependence plays a role in the mechanism producing
the logarithmic anomalies by which the finite-size scaling behavior of
our model turned out to be modified when $d+\sigma=6$. In these
special cases---which include, in particular, the physically important
one of nonretarded van-der-Waals interactions in three
dimensions---anomalies of this kind showed up in $b$-dependent
(leading) corrections to scaling. We were able to clarify their origin
(see Sec.~\ref{sec:loganom}): They are caused by the degeneracy
$\omega=\omega_\sigma$ of the two correction-to-scaling exponents in
conjunction with the $b$-dependence of $g_\omega$.

Thus, three-dimensional systems belonging to the universality classes
of the scalar $\phi^4$ model and its $O(n)$ counterparts with
$n<\infty$ should not exhibit such logarithmic anomalies because their
correction-to-scaling exponents $\omega$ and $\omega_\sigma$ are not
degenerate.

It would be worthwhile to extend the present work in a number of
different directions. We have focused our attention here on the case
of temperatures $T\geq T_{c,\infty}$. An obvious next step is a
detailed investigation of the model for temperatures below the bulk
critical temperature $T_{c,\infty}$. For the spherical model with
periodic boundary conditions considered here, such an extension, which
we defer to a future publication, is relatively straightforward.

The scaling forms derived in this paper on the basis of
phenomenological scaling ideas involved nontrivial critical indices,
such as $\eta$ and $\gamma/\nu=2-\eta$, and the correction-to-scaling
exponent $\omega$. Although the exact results for the spherical model
we were able to present are in conformity with the predicted more
general finite-size scaling forms, they neither permit us to
corroborate the appearance of a nontrivial value of $\eta$ nor to
verify the $n$-dependence of $\omega$ for the $n$-vector model. A
desirable complementary check of the phenomenological predictions that
is capable of identifying nontrivial values of $\eta$ as well as the
$n$-dependence of it and other exponents can be made by performing a
two-loop RG analysis for small $\epsilon=4-d$. We have performed such
an analysis; its results will be published elsewhere \cite{GDD05}.

Valuable alternative checks of our phenomenological predictions should
be possible by means of Monte Carlo simulations. Although it is quite
a challenge to perform accurate Monte Carlo simulations of
near-critical systems with long-range interactions, suitable
algorithms were developed and demonstrated to be quite efficient
recently \cite{LB95,LB97,GH04}. We therefore believe that accurate
tests of our predictions via such simulations are feasible.

An obviously important direction for further research is the extension
of our work to other than periodic boundary conditions, namely, those
of a kind giving a better representation of typical experimental
situations. Important examples are slabs with Dirichlet boundary
conditions on both boundary planes, or more generally, Robin boundary
conditions. Although some aspects of our above findings should carry
over to such boundary conditions---e.g., the form of the scaled
variables encountered here and the power-law decrease of the Casimir
away from the bulk critical temperature---it is clear that any
quantitative comparison between theoretical predictions and results of
a given experiment requires that appropriate boundary conditions have
been chosen in the calculations. We leave such extensions to future
work.

\acknowledgments

D.~D.\ would like to thank H.W.\ Diehl's group and Fachbereich Physik
of the Universit{\"a}t Duisburg-Essen for their hospitality at Campus
Essen.

We gratefully acknowledge the financial support of this work by the
Deutsche Forschungsgemeinschaft via Grant No.~Di-378/5, the Bulgarian
NSF Academy of Sciences (436 BUL 113/118/0-1), and the Bulgarian NSF
(Project F-1402).

\appendix

\section{Fourier transform of the interaction potential}
\label{app:FT}

In this appendix we wish to derive the small-momentum behavior of the
Fourier transform (\ref{eq:FT}) of the pair interaction $J(\bm{x})$
introduced in Eq.~(\ref{eq:sint}). To this end we introduce a lattice
constant $a_\alpha$ for each of the principal directions of the simple
hypercubic lattice in $d$ dimension we are concerned with. We assume
that the lattice has an odd number $2N_\alpha+1$ (with $N_\alpha\in
\mathbb{N}$) of layers perpendicular to the $x_\alpha$ axis, so that
its linear extension along the $x_\alpha$-direction is $L_\alpha=
(2N_\alpha+1) a_\alpha$.

With these conventions the lattice Fourier transform~(\ref{eq:FT}) of
the pair interaction (\ref{eq:sint}) becomes
\begin{equation}
  \label{eq:lattFT}
  \tilde{J}(\bm{q})=\sum_{j_1=-N_1}^{N_1}\cdots
   \sum_{j_d=-N_d}^{N_d} J\big(\bm{x}^{(\bm{j})}\big)\prod_{\alpha=1}^d
   e^{-i\hat{q}_\alpha
     j_\alpha} \;,
\end{equation}
where $\bm{x}^{(\bm{j})}=(j_\alpha a_\alpha)$ and we have introduced the
dimensionless momentum components $\hat{q}_\alpha=q_\alpha a_\alpha$. The momentum
$\bm{q}$ takes values in the first Brillouin zone, i.e., $q_\alpha
=2\pi\nu_\alpha/L_\alpha$ with $\nu_\alpha =-N_\alpha, -N_\alpha+1, \ldots,N_\alpha$.

The contribution proportional to $J_1$ of Eq.~(\ref{eq:sint}) gives
the usual result for nearest-neighbor interactions on a hypercubic
lattice:
\begin{eqnarray}
  \label{eq:J1q}
  \tilde{J}_1(\bm{q})&=& 2J_1\sum_{\alpha=0}^d\cos(\hat{q}_\alpha)\\
&=&2J_1{\bigg\{d-\frac{1}{2}\,\hat{q}^2
+\frac{1}{24}\sum_{\alpha=1}^d
\left[\hat{q}_\alpha^4+O{\big(\hat{q}_\alpha^6\big)}\right]\bigg\}}
.\qquad
\end{eqnarray}

To compute the Fourier transform of the remaining part of the
interaction~(\ref{eq:sint}), which we denote as $J_2(x)$, it is
useful to recall Poisson's summation formula \cite[p.~31]{GS64}
\begin{equation}
  \label{eq:Poisson}
  \sum_{j=-\infty}^{\infty}\delta(t-j a)
 =\frac{1}{a}\sum_{m=-\infty}^{\infty}
  e^{i2\pi m t/a} \;.
\end{equation}
Applying the generalized functions on both sides to a test function
$f(t)$ whose support is restricted to $[-L/2,L/2]$ gives
\begin{equation}
  \label{eq:Poisapp}
  \sum_{j=-N}^{N}f(ja)=\sum_{m=-\infty}^\infty \int_{-L/2}^{L/2}
  \frac{dt}{a} \,e^{i2\pi mt/a}\,f(t)\;.
\end{equation}

Since we are interested in a system of macroscopic lateral
extent, we take the limits ${N_\alpha\to \infty}$ for all
$\alpha>1$, keeping the associated lattice constants $a_\alpha>0$
fixed. We thus obtain
\begin{equation}
  \label{eq:J2q}
 \tilde{J}_2(\bm{q})
=\sum_{\bm{m}\in\mathbb{Z}^d}\,
 \int_{\mathfrak{V}_L}\frac{d^dx}{v_{\bm{a}}}
  J_2(x) \prod_{\alpha=1}^d e^{-i\,(q_\alpha-2\pi
  m_\alpha/a_\alpha)x_\alpha} \;,
\end{equation}
where $v_{\bm{a}}=\prod_{\alpha=1}^d a_\alpha$ is the volume of the
unit cell, and the integration is over a slab
$\mathfrak{V}_L=[-L/2,L/2]\times\mathbb{R}^{d-1}$ of thickness
$L\equiv L_1$.

The terms with $\bm{m}\neq\bm{0}$ reflect the lattice structure of the
model and give contributions anisotropic in $\bm{q}$ space (as well as
isotropic ones). Owing to the restricted integration regime, the
$\bm{m}=\bm{0}$ term also yields $\bm{q}$-dependent contributions
(which, however, are small for large $L$). These anisotropies add to
those originating from the short-range contribution~(\ref{eq:J1q}) and
produce, in particular, a nonzero value of the coefficient $b_{4,1}$ of
the anisotropic $q_\alpha^4$ terms in Eq.~(\ref{eq:Omega}).

We have emphasized the importance of long-range van-der-Waals type
interactions for fluids before. Let us therefore consider the case of
simple isotropic fluids. For such systems it is appropriate to take
the continuum limit $a_\alpha\to 0$. Then the contributions of the
$\bm{m}\neq\bm{0}$ terms vanish by the Riemann-Lebesgue lemma. We have
\begin{equation}
  \label{eq:Jiso}
  v_{\bm{a}}\tilde{J}_2(\bm{q})\;
  \mathop{\longrightarrow}\limits_{\{a_\alpha\to 0\}}
  \;\tilde{J}_{2,L}^{(\text{cont})}(\bm{q})\equiv
  \int_{\mathfrak{V}_L}d^dx \,J_2(x)\,e^{-i\bm{q}\cdot\bm{x}}\;.
\end{equation}
In order that $\tilde{J}_2(\bm{q})$ have a nontrivial continuum limit,
the coupling constant $J_2$ must be scaled such that $J_2/v_{\bm{a}}$
approaches a finite value $J_2^{\text{(cont})}>0$.

The Fourier transform has an explicit $L$-dependence due to the
restriction of the $x_1$-integration to a finite interval. However,
the deviation from its bulk analog is small, unless $L$ is very small:
The integration over the parallel coordinates $\bm{x}_\|$ yields a
function of $x_1$ that varies $\sim x_1^{-\sigma-1}$ for large $x_1$.
The error resulting from $\int_{L}^\infty dx_1$ therefore decreases as
$L^{-\sigma}$, i.e., decays $\sim N_1^{-\sigma}$ when $a_1>0$. Let us
ignore this $L$-dependence and determine the behavior of its bulk
counterpart $\tilde{J}_{2,\infty}^{(\text{cont})}(\bm{q})$ for small
$q$.

The calculation of the latter is straightforward. The required
angular integral is
\begin{eqnarray}
  \label{eq:Eav}
  \int
  d\Omega_d\,e^{i\bm{q}\cdot\bm{x}}=
  (2\pi)^{d/2}(qx)^{1-d/2}\,J_{\frac{d-2}{2}}(qx)\;.
\end{eqnarray}
Performing the remaining radial integration gives
\begin{equation}
  \label{eq:J2qInfty}
  \tilde{J}_{2,\infty}^{(\text{cont})}(\bm{q})=J_2^{\text{(cont)}}
\,2\pi^{d/2}\,\Big(\frac{q}{2\rho_0}\Big)^{\sigma/2}
    \frac{K_{\sigma/2}(\rho_0q)}{\Gamma[(d+\sigma)/2]}\;,
\end{equation}
where $K_{\sigma/2}$ is a modified Bessel function. From its known
asymptotic behavior for small values of $q$ one easily derives the
limiting form
\begin{eqnarray}
  \label{eq:J2qexp}
  \frac{\rho_0^\sigma
    \tilde{J}_{2,\infty}^{(\text{cont})}(\bm{q})}{J_2^{\text{(cont})}}&
  \mathop{=}\limits_{q\to 0}&A_0+A_2\,
    (\rho_0 q)^2+A_4\,(\rho_0 q)^4 \nonumber\\
    &&\strut -A_\sigma\,(\rho_0q)^\sigma  +
    O\big(q^{\sigma+2},q^6\big)\,.\;\;\;
\end{eqnarray}
in which
\begin{eqnarray}
  \label{eq:A0}
  A_0&=&-2(\sigma-2)\,A_2=-8(4-\sigma)(\sigma-2)\,A_4\nonumber\\
  &=&\frac{\pi^{d/2}\,\Gamma(\sigma/2)}{\Gamma[(d + \sigma)/2])} \;,
\end{eqnarray}
while $A_\sigma$ is given by the right-hand side of Eq.~(\ref{eq:b}).

The ratios $-A_2/A_0$ and $A_\sigma/A_2$ yield the values
~(\ref{eq:v2}) and (\ref{eq:bfinal}) of the coefficients $v_2$ and
$b$, respectively, for the case of vanishing nearest-neighbor
interaction constant $J_1$.

\section{Calculation of the functions $Q_{d,\sigma}(y)$}
\label{app:Qas}

According to Eq.~(\ref{eq:Qdsigdef}) the function $Q_{d,\sigma}(y)$
can be represented as
\begin{eqnarray}
  \label{eq:Qrep1}
  Q_{d,\sigma}(y)&=&\frac{y}{2}\,\Bigg[\sum_{q_1\in
    2\pi\mathbb{Z}}\,
\int_{\bm{q}_\|}^{(d-1)} -\int^{(d)}_{\bm{q}}\Bigg]
\frac{q^{\sigma-2}}{y+q^2}\quad\\ &=&
\label{eq:Qrep2}
y^{(d+\sigma-2)/2} \,\sum_{k=1}^\infty
  G_{\infty}(d,\sigma|1;k\sqrt{y})\;.
\end{eqnarray}
To obtain the second form~(\ref{eq:Qrep1}), we have utilized the
property
 \begin{equation}
  \label{eq:scinv}
   G_{\infty}(d,\sigma|r;x)=r^{(d+\sigma-4)/2}\,
   G_{\infty}(d,\sigma|1;x\sqrt{r}) \;,
\end{equation}
of the bulk propagator.

The case $\sigma=2$ is special in that the summation and integration over
$q_1$ in Eq.~(\ref{eq:Qrep1}) can easily be performed to reduce
$Q_{d,2}$ to a single integral, namely
\begin{equation}
  \label{eq:Qd2}
  Q_{d,2}(y)=\frac{y\,K_{d-1}}{2}\,\int_0^\infty dp\,
  \frac{p^{d-2}}{\big[e^{\sqrt{y+p^2}}-1\big]\sqrt{y+p^2}} \;.
\end{equation}
Integrals of this kind were also encountered in Krech and Dietrich's
work \cite{KD92a,KD92b} on the Casimir effect in systems with
short-range interactions.

From Eq.~(\ref{eq:Qd2}) it is not difficult to derive a useful
relation between $Q_{d+2,2}$ and $Q_{d,2}$:
\begin{equation}
  \label{eq:Qdp2rel}
  \frac{\partial}{\partial y}\,\frac{Q_{d+2,2}(y)}{y}
  =-\frac{ Q_{d,2}(y)}{4\pi y}\;.
\end{equation}
To do this, one simply must interchange the differentiation of
$Q_{d+2,2}(y)/y$ with respect to $y$ with the integration over $p$,
replace the $y$-derivative of the integrand's $y$-dependent part by a
derivative with respect to $p^2$, and then integrate by parts.

Returning to the case of general $\sigma$, we note that the
representation~(\ref{eq:Qrep2}) has the advantage of linking the
asymptotic behavior of $Q_{d,\sigma}(y)$ for large values of $y$ to
that of $G_\infty(d,\sigma|1;x)$. In addition, there are some special
values of $(d,\sigma)$ for which it allows one to derive closed-form
analytical expressions for $Q_{d,\sigma}$ in a straightforward
fashion. We therefore begin by computing the bulk propagator.

\subsection{Calculation of the propagator $G_{\infty}(d,\sigma|r;\bm{x})$}
\label{app:Gsig}

We start from Eq.~(\ref{eq:Gsiginfty}) and perform the angular
integrations using our previous result~(\ref{eq:Eav}). This gives
\begin{equation}
  \label{eq:Gsigstart}
  G_{\infty}(d,\sigma|1;x)=\frac{x^{1-d/2}}{(2\pi)^{d/2}}\int_0^\infty
  dq\,\frac{q^{\sigma-2+d/2}}{1+q^2}\,J_{\frac{d-2}{2}}(qx)\;.
\end{equation}
The required integral can be evaluated with the aid of Eq.~(6.565.8)
of Ref.~\cite{GR80} or {\sc Mathematica} \cite{Mat}. One obtains
\begin{eqnarray}
  \label{eq:Gsiginftyres}
\lefteqn{G_{\infty}(d,\sigma|1;x)}&&\nonumber\\
&=&\frac{\pi\,\csc[\pi(d+\sigma)/2]}{(4\pi)^{d/2}} \bigg[-
{_{0}F}_1^{(\text{reg})}{\Big(;\frac{d}{2};\frac{x^2}{4}\Big)}
\nonumber\\
&&\strut +
  \Big(\frac{x}{2}\Big)^{4-d-\sigma}\;
  {_{1}F}_2^{(\text{reg})}\Big(1;2-\frac{\sigma}{2},3-
  \frac{d+\sigma}{2};\frac{x^2}{4}\Big)\bigg]\,,\nonumber\\
\end{eqnarray}
where ${_{0}F}_1^{(\text{reg})}$ and ${_{0}F}_1^{(\text{reg})}$ are
regularized generalized hypergeometric function which can be expressed
as
\begin{equation}
  \label{eq:0F1}
   {_{0}F}_1^{(\text{reg})}{\big(;d/2;x^2/4\big)}=
   {({x}/{2})}^{1-d/2}\,I_{\frac{d-2}{2}}(x)
\end{equation}
and
\begin{equation}
  \label{eq:hypgeomreg}
  {_{1}F}_2^{(\text{reg})}(\alpha;\beta,\gamma;z)
  =\frac{{_{1}F}_2(\alpha;\beta,\gamma;z)}{\Gamma(\beta)\,
    \Gamma(\gamma)}\;.
\end{equation}
in terms of the modified Bessel function of the first kind $I_\nu$ and
the generalized hypergeometric function
${_{1}F}_2(\alpha;\beta,\gamma;z)$, respectively. Their Taylor
expansions read
\begin{equation}
  \label{eq:0F1exp}
   {_{0}F}_1^{(\text{reg})}{(;d/2;z)}
   =\sum_{j=0}^\infty\frac{z^j}{j!\,\Gamma(j+d/2)}
\end{equation}
and
\begin{eqnarray}
  \label{eq:1F2exp}
   \lefteqn{{_{1}F}_2^{(\text{reg})}[1;2-\sigma/2,3-
  (d+\sigma)/{2};z]}&&\nonumber\\ &=&
   \sum_{j=0}^\infty\frac{z^j}{\Gamma(j+2-\sigma/2)
     \,\Gamma[j+3-(d+\sigma)/2]}\;.\quad
\end{eqnarray}

For $\sigma=2$, we recover the familiar result for the free propagator
of systems with short-range interactions:
\begin{equation}
  \label{eq:Gd2}
  G_\infty(d,2|1;x)=(2\pi)^{-d/2}\,x^{-(d-2)/2}
  \,K_{\frac{d-2}{2}}(x) \;.
\end{equation}
The latter is known to decay exponentially; the familiar
asymptotic expansion of the Bessel functions $K_\nu$ implies that
\begin{eqnarray}
  \label{eq:Ginfsrlarge}
\lefteqn{G_\infty(d,2|1;x)}&&\nonumber\\ &\mathop{=}\limits_{x\to\infty}&
  \frac{x^{-(d-1)/2}\,e^{-x}}{2(2\pi)^{(d-1)/2}}
  \bigg[\sum_{j=0}^{m-1}
  \frac{\Gamma[(d-1+2j)/2]}{j!\,\Gamma[(d-1-2j)/2]}\,(2x)^{-j}
\nonumber\\ &&\qquad\qquad\strut \qquad\qquad
+O\big(x^{-m}\big)\bigg]\;.
\end{eqnarray}

When $2<\sigma<4$, the Fourier transform of the propagator
$G_\infty(d,\sigma|1;x)$ is not regular in $q$ at $q=0$. This entails
that the propagator decays only as an inverse power of $x$. An easy
way to obtain its asymptotic expansion for this case is to start from
Eq.~(\ref{eq:Gsigstart}), do a rescaling $q\to Q=qx$, expand the
factor $(1+Q\,x^{-2})^{-1}$ of the resulting integrand in powers of
$x^{-2}$, and integrate the series termwise. This leads to the
asymptotic expansion
\begin{eqnarray}
  \label{eq:Ginfgenlarge}
\lefteqn{G_\infty(d,\sigma|1;x)}&&\nonumber\\ &\mathop{=}\limits_{x\to\infty}&
\frac{2^{\sigma-2}}{\pi^{d/2}\,x^{d+\sigma-2}}
\sum_{j=0}^{m-1}
\frac{\Gamma[j+(d+\sigma-2)/2]}{\Gamma(1-j-\sigma/2)}\,
\frac{(-4)^j}{x^{2j}}
\nonumber\\  &&\strut +O(x^{-2m})\;.
\end{eqnarray}

\begin{figure}[htb]
\begin{center}\includegraphics[%
  clip,scale=1]{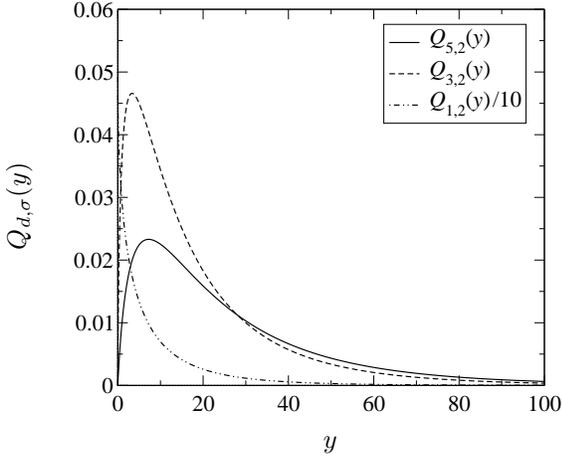}\end{center}
\caption{The functions $Q_{5,2}(y)$ (full line). $Q_{3,2}(y)$
  (dashed), and $Q_{1,2}(y)/10$ (dash-dotted), respectively.}
\label{fig:Qd2s}
\end{figure}

Let us see how the above results can be employed to compute the
required $Q_{d,\sigma}$. To treat the three-dimensional case, we need
$Q_{d,2}(y)$ for $d=3$ and $d=5$, as well as $Q_{3,3}$, and their
derivatives. Since $Q_{3,2}'(y)$ involves $Q_{1,2}(y)$, we also
determine the latter. For these choices of $d$, the
result~(\ref{eq:Gd2}) reduces to
\begin{equation}
  \label{eq:G12}
  G_\infty(1,2|1;x)=\frac{1}{2}\,e^{-x}\;,
\end{equation}
\begin{equation}
  \label{eq:G32}
  G_\infty(3,2|1;x)=\frac{1}{4\pi x}\,e^{-x}\;,
\end{equation}
and
\begin{equation}
  \label{eq:G52}
  G_\infty(5,2|1;x)=\frac{1+x}{8\pi^2 x^3}\,e^{-x}\;,
\end{equation}
respectively. Upon substituting these expressions into
Eq.~(\ref{eq:Qrep2}), the series can be summed, giving
\begin{equation}
  \label{eq:Q12}
  Q_{1,2}(y)=\frac{1}{2}\,\frac{\sqrt{y}}{\exp(\sqrt{y})-1}\;,
\end{equation}
\begin{equation}
  \label{eq:Q32}
  Q_{3,2}(y)=-\frac{y}{4\pi}\,\ln\big(1-e^{-\sqrt{y}\,}\big)\;,
\end{equation}
and
\begin{equation}
  \label{eq:Q52}
  Q_{5,2}(y)=\frac{y}{8\pi^2}\big[\mathrm{Li}_3\big(e^{-\sqrt{y}}\,\big)
  +\sqrt{y}\,\mathrm{Li}_2\big(e^{-\sqrt{y}}\,\big)\big]\;,
\end{equation}
where $\mathrm{Li}_p$ is the polylogarithmic function,
\begin{equation}
  \label{eq:Lidef}
 \mathrm{Li}_p(z)=\sum_{j=1}^\infty\frac{z^j}{j^p}\;.
\end{equation}
As can easily be checked, these results~(\ref{eq:Q12})--(\ref{eq:Q52})
are in conformity with Eq.~(\ref{eq:Qdp2rel}). Plots of the functions
are displayed in Fig.~\ref{fig:Qd2s}.

For other choices of $\sigma$ and $d$---including $\sigma=d=3$---the
series~(\ref{eq:Qrep2}) cannot in general be summed analytically. This
suggests that one has to resort to numerical means. To this end a
different representation of $Q_{d,\sigma}$, which we are now going to
derive from Eq.~(\ref{eq:Qrep1}), proved to be more effective.
Remarkably, this representation enabled us to derive even a
closed-form analytical expression for $Q_{3,3}(y)$.

Note, first, that the subtracted $\bm{q}$-integral in
Eq.~(\ref{eq:Qrep1}) is the bulk propagator at $x=0$. Both the limit
$x\to 0$ of Eq.~(\ref{eq:Gsiginftyres}) as well as the explicit
calculation of the integral $\int^{(d)}_{\bm{q}}$ yield
\begin{equation}
  \label{eq:Ginftyzero}
  G_\infty(d,\sigma|y;0)=-y^{(d+\sigma-4)/2}\,
  \frac{\pi}{2}\,K_d\,\csc[\pi\,(d+\sigma)/2]\;.
\end{equation}
When $d+\sigma\geq 4$, this result involves analytic continuation in
$d$ since the integral is ultraviolet (uv) divergent in this case. The
same $L$-independent uv divergences must occur in the first term of
Eq.~(\ref{eq:Qrep1}), so that they cancel in the difference. We find
it most convenient to handle uv divergences of this kind, which occur
at intermediate steps, by means of dimensional regularization. Readers
preferring to work with a large-momentum cutoff $\Lambda$ are
encouraged to utilize a smooth variant of it, since a sharp cutoff is
known to give unphysical results in treatments of finite-size effects
based on the small-momentum form of the inverse free propagator, i.e.,
of $\Omega(\bm{q})$ \cite{DR01}.

Next, consider a term of the series $\sum_{q_1}$ in
Eq.~(\ref{eq:Qrep1}). It is given by the
integral
\begin{equation}
  \label{eq:Iq1}
  I_{d,\sigma}(q_1,y)=K_{d-1}\int_0^\infty \!dq_\|\,q_\|^{d-2}\,
  \frac{(q_1^2+q_\|^2)^{(\sigma-2)/2}}{y+q_1^2+q_\|^2} \;.
\end{equation}
Its calculation for $q_1=0$ is straightforward, giving
\begin{equation}
  \label{eq:I0}
   I_{d,\sigma}(0,y)=K_{d-1}\,y^{(d+\sigma-5)/2}\,
   \frac{\pi}{2\cos[\pi\,(d+\sigma)/2]}\;.
\end{equation}

It can also be computed in closed form for $q_1\neq 0$; the result
involves a hypergeometric function ${_2}F_1$ and algebraic functions
of $y$ and $q_1$. Rather than working with this expression directly,
it is more convenient to split off appropriate terms containing the uv
singularities they contribute to the series $\sum_{q_1}$ in
Eq.~(\ref{eq:Qrep1}). Whether and what kind of subtractions are
necessary depends on the values of $d$ and $\sigma$ for which
$Q_{d,\sigma}$ is needed. The $Q_{d,\sigma}(y)$ with the largest
values of $d+\sigma$ encountered in our analysis of the
three-dimensional case with $\sigma=3$ are $Q_{5,2}$ and $Q_{3,3}$.
Thus the largest value of $d+\sigma$ for which $Q_{d,\sigma}$ is
required is $7$. Using power counting we see that the strongest
possible uv singularity of the bulk integral $\int_{\bm{q}}^{(d)}$ in
Eq.~(\ref{eq:Qrep1}) is $\sim \Lambda^{d+\sigma-4}$. Hence it is
sufficient to subtract from the integral~(\ref{eq:Iq1}) its Taylor
expansion to first order in $y$.  This ensures that the difference,
summed over $q_1$, produces a uv finite result. All poles must
originate from the subtracted terms and cancel with those of the bulk
contribution in Eq.~(\ref{eq:Qrep1}).

Accordingly, we decompose $I_{d,\sigma}(q_1,y)$ as
\begin{equation}
  \label{eq:Iq1dec}
  I_{d,\sigma}(q_1,y)=\sum_{k=0}^1I^{(0,k)}_{d,\sigma}(q_1,0)\,\frac{y^k}{k!}
  +Z_{d,\sigma}(q_1,y)\;,
\end{equation}
where $I_{d,\sigma}^{(0,k)}$ denotes the $k$th derivative of the
function $I_{d,\sigma}$ with respect to its second argument. Computing
the integrals of the Taylor coefficients and the remainder
$Z_{d,\sigma}$ yields
\begin{eqnarray}
  \label{eq:Iq1y0k}
I^{(0,k)}_{d,\sigma}(q_1,0)
&=&\frac{(-1)^k}{2}\,k!\,K_{d-1}\,|q_1|^{d+\sigma-5-2k}\nonumber\\&&\times
   B\Big(\frac{d-1}{2},\frac{5+2k-d-\sigma}{2}\Big)
\end{eqnarray}
and
\begin{eqnarray}
  \label{eq:Rdsigy}
\lefteqn{Z_{d,\sigma}(q_1,y)}\nonumber\\
&=& \frac{K_{d-1}\,\pi}{2\cos[\pi\,(d+\sigma)/2]}\Bigg[
\frac{y^{(\sigma-2)/2}}{(y+q_1^2)^{(3-d)/2}}
\nonumber\\ &&\strut
-\frac{y^2\,|q_1|^{d+\sigma-7}}{y+q_1^2}
\,\frac{\Gamma[(d-1)/2]}{\Gamma(3-\sigma/2)} \times
\nonumber\\ &&\strut
\times {_2}F^{(\text{reg})}_1\bigg(1,\frac{d-1}{2};
\frac{d+\sigma-5}{2};\frac{q_1^2}{y+q_1^2}\bigg)\Bigg]\,,
\end{eqnarray}
where $B(a,b)$ and ${_2}F^{(\text{reg})}_1$ are the Euler beta function
and the regularized hypergeometric function
\begin{equation}
  \label{eq:reg2F1}
   {_2}F^{(\text{reg})}_1(a,b;c;z)= {_2}F_1(a,b;c;z)/\Gamma(c)\;,
\end{equation}
respectively.

We now substitute the above results into the
representation~(\ref{eq:Qrep1}) of $Q_{d,\sigma}$, utilizing the fact
that series $\sum_{q_1\neq0}$ of pure powers of $q_1$ give zeta functions:
\begin{equation}
  \label{eq:sumzeta}
  \sum_{\substack{q_1 \in 2\pi\mathbb{Z}\\ \neq0}}
  |q_1|^{-s}=2\,\frac{\zeta(s)}{(2\pi)^s}\;.
\end{equation}
The result is
\begin{eqnarray}
  \label{eq:Qdsigfr}
Q_{d,\sigma}(y)
&=& y\bigg[\frac{1}{2}\,y^{(d+\sigma-5)/2}\, I_{d,\sigma}(0,1)
+\sum_{j=1}^\infty Z_{d,\sigma}(2\pi j,y)
\nonumber\\ &&\strut
+ \frac{\zeta(5-d-\sigma)}{(2\pi)^{5-d-\sigma}}\,
  I_{d,\sigma}(1,0) \nonumber\\ &&\strut
+y\,  \frac{\zeta(7-d-\sigma)}{(2\pi)^{7-d-\sigma}}\,
I^{(0,1)}_{d,\sigma}(1,0) \nonumber\\  &&\strut
+\frac{K_d\,\pi}{4}\,y^{(d+\sigma-4)/2}\,\csc[\pi\,(d+\sigma)/2]
\bigg].\;\;
\end{eqnarray}

Evaluating this expression for $(d,\sigma)=(3,2)$ and $(5,2)$ with the
aid of {\sc Mathematica} \cite{Mat}, we have checked that the previous
results~(\ref{eq:Q12})--(\ref{eq:Q52}) for $Q_{1,2}(y)$, $Q_{3,2}$, and
$Q_{5,2}(y)$ are recovered. It can also be utilized to determine
$Q_{3,3}(y)$ analytically. To this end one rewrites the series
coefficient $Z_{3,3}$ as
\begin{eqnarray}
  \label{eq:rewrsc}
Z_{3,3}(2\pi j,y)&=&\frac{y}{4\pi^2 j}-\frac{\sqrt{y}}{2\pi}\,
\arccos\big[\frac{2\pi j}{\sqrt{y+4\pi^2j^2}}\big]
\nonumber \\ &=&\sqrt{y}\int_0^y
dt\, \frac{\sqrt{t}}{8\pi^2 j\,(t+4\pi^2j^2)}
\end{eqnarray}
and interchanges the integration over $t$ with the summation over $j$.
In this manner the series $\sum_j Z_{3,3}$ can be computed, and one
obtains
\begin{eqnarray}
  \label{eq:Q33res}
  Q_{3,3}(y)&=&\frac{y}{12}+\frac{y^2}{4\pi^2}\bigg[ 1-
  \ln\bigg(\frac{\sqrt{y}}{2\pi}\bigg)\bigg]
  \nonumber\\ &&\strut
+\frac{y^{3/2}}{2\pi}\bigg\{\frac{\pi}{4} +\mathrm{Im\,}
\bigg[{\ln\Gamma\bigg(i\, \frac{\sqrt{y}}{2\pi}\bigg)}\bigg]
 \bigg\}\,.\qquad
\end{eqnarray}

\begin{figure}[htb]
\begin{center}\includegraphics[%
  clip,scale=1]{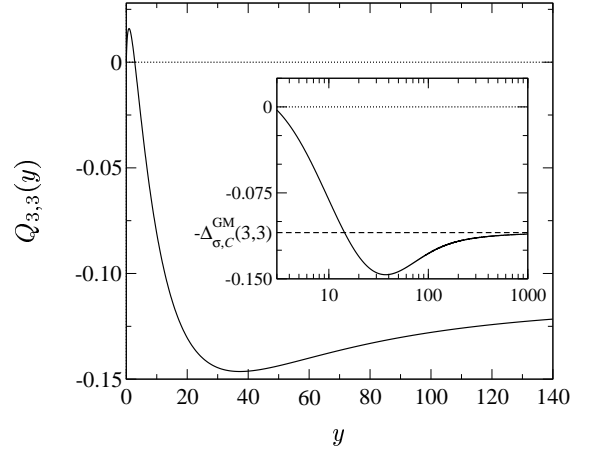}\end{center}
\caption{The function $Q_{3,3}(y)$. The in-set is a logarithmic-linear
  plot of this function, which illustrates the approach to the
  limiting value $-\Delta_{\sigma,C}^{\text{GM}}(3,3)=-\pi^2/90$ implied by
  Eqs.~(\ref{eq:Qdsig}) and (\ref{eq:DeltaCsigmaGM}).}
\label{fig:Q33}
\end{figure}

A plot of this function is shown in Fig.~\ref{fig:Q33}.

\subsection{Asymptotic behavior of $Q_{d,\sigma}(y)$ for small and large values of $y$}
\label{app:smalllargey}

The asymptotic behavior of the function $Q_{d,2}(y)$ for large values
of $y$ readily follows from the representation~(\ref{eq:Qrep2}) in
conjunction with the asymptotic expansion (\ref{eq:Ginfsrlarge}) of
the bulk propagator~(\ref{eq:Gd2}). The result one finds for general
values of $d$,
\begin{equation}
  \label{eq:Qd2largey}
  Q_{d,2}(y)\mathop{=}\limits_{y\to\infty}
  \frac{y^{(d+1)/4}}{2\,(2\pi)^{(d-1)/2}}\,e^{-\sqrt{y}}\,
  \left[1+O\left(y^{-1/2}\right)\right],
\end{equation}
can be verified to be in accordance with the large-$y$ behavior of the
explicit expressions (\ref{eq:Q12})--(\ref{eq:Q52}) of these functions
for $d=1,\,3$, and $5$.

To determine the large-$y$ behavior of $Q_{d,\sigma}(y)$ with
$2<\sigma<4$, we insert the asymptotic
expansion~(\ref{eq:Ginfgenlarge}) into Eq.~(\ref{eq:Qrep2}).  The
summations over $k$ can be performed for the expansion coefficients,
giving $\zeta$-functions.  In this way one arrives at the asymptotic
expansion
\begin{eqnarray}
  \label{eq:Qdsiglargey}
  Q_{d,\sigma}(y)&\mathop{=}\limits_{y\to\infty}&
  \frac{2^{\sigma-2}}{\pi^{d/2}}
  \sum_{j=0}^{m-1}\bigg[
  \frac{\Gamma[j+(d+\sigma-2)/2]}{\Gamma(1-j-\sigma/2)}\nonumber\\ &&\times
  \,\zeta(d+\sigma+2j-2)\,\frac{(-4)^j}{y^{j}}\bigg]+O(y^{-m})\,.\nonumber\\
\end{eqnarray}
Again, one can employ our explicit result~(\ref{eq:Q33res}) for
$Q_{3,3}$ to verify this asymptotic series. Note, that the
series~(\ref{eq:Qdsiglargey}) truncates when $\sigma$ is even, e.g.,
when $\sigma=2$. This ensures the consistency with the exponential
decay~(\ref{eq:Qd2largey}) one has for $\sigma=2$.

The asymptotic behavior of $Q_{d,2}(y)$ for small $y$ can be
conveniently obtained from Eq.~(\ref{eq:Qdsigfr}) for $1<d<7$. One
finds that
\begin{eqnarray}
  \label{eq:Qd2small}
  Q_{d,2}(y)&\mathop{=}\limits_{y\to 0}&
    \frac{\sqrt{\pi}\,\Gamma[(3-d)/2]}{(4\pi)^{d/2}}\,\Big[y^{(d-1)/2}
\nonumber\\ &&\strut
    -\frac{B(1-d/2,1/2)}{2\pi}\,y^{d/2}  +
    \frac{2\,\zeta(3-d)}{(2\pi)^{3-d}}\,y \nonumber\\ &&\strut
+\frac{(d-3)\,\zeta(5-d)}{(2\pi)^{5-d}}\,y^2+O(y^3)
\Big],
\end{eqnarray}
provided $d\neq 1,3,5$. The behavior in the latter cases follows by
expansion about these values of $d$. The poles that the $\Gamma$
function yields for individual terms at such odd integer values of $d$
cancel, and logarithms of $y$ emerge when $d=3$ or $5$. The expansions
one gets in this manner,
\begin{equation}
  \label{eq:Q12smexp}
  Q_{1,2}(y)\mathop{=}\limits_{y\to 0}
    \frac{1}{2}-\frac{\sqrt{y}}{4}+\frac{y}{24}-\frac{y^2}{1440}
    +O(y^3)\;,
\end{equation}
\begin{equation}
\label{eq:Q32smexp}
 Q_{3,2}(y)\mathop{=}\limits_{y\to 0}
-\frac{y}{8\pi}\ln y+\frac{y^{3/2}}{8\pi}-\frac{y^2}{96\pi}+O(y^3) \;,
\end{equation}
and
\begin{equation}
\label{eq:Q52smexp}
 Q_{5,2}(y)\mathop{=}\limits_{y\to 0}
 \frac{\zeta(3)\,y}{8\pi^2}-\frac{1-\ln
   y}{32\pi^2}\,y^2-\frac{y^{5/2}}{48\pi^2}+O(y^3) \;,
\end{equation}
agree with those of the analytic expressions~(\ref{eq:Q12}),
(\ref{eq:Q32}), and (\ref{eq:Q52}).

The small-$y$ behavior of $Q_{d,\sigma}$ with $2<\sigma<4$ can be
determined from Eq.~(\ref{eq:Qdsigfr}) in a similar fashion. One obtains
\begin{eqnarray}
\label{eq:Qdsigsmexp}
 \lefteqn{Q_{d,\sigma}(y)}&&\nonumber\\ &\mathop{=}\limits_{y\to 0}&
 \frac{K_{d-1}}{4}\,
 B\Big(\frac{d+\sigma+1}{2}},{\frac{1-d-\sigma}{2}\Big)
 \bigg\{y^{(d+\sigma-3)/2} \nonumber\\ &&\strut
 + B\Big(\frac{d-1}{2},
   \frac{1}{2}\Big)\,
   \frac{\cot[\pi(d+\sigma)/2]}{2\pi}\,
   y^{(d+\sigma-2)/2} \nonumber\\ &&\strut
+ B\Big(\frac{d-1}{2}},{\frac{5-d-\sigma}{2}\Big)
\,\frac{4\cos[(\pi(d+\sigma)/2]}{(2\pi)^{6-d-\sigma}}\nonumber\\ &&\times
\Big[\zeta(5-d-\sigma)\,y +\frac{(d+\sigma-5)\,
  \zeta(7-d-\sigma)\,y^2}{(4-\sigma)\,(2\pi)^2}\,
\Big]\nonumber\\
&&\strut +O(y^3)\bigg\}\,.
\end{eqnarray}

To deal with the case of $Q_{3,3}$, one can set $\sigma=3$ and expand
about $d=3$. This gives
\begin{eqnarray}
\label{eq:Q33smexp}
 Q_{3,3}(y)&\mathop{=}\limits_{y\to 0}&\frac{y}{12}-
 \frac{y^{3/2}}{8} +\frac{y^2}{8\pi^2}\,\Big(2-2C_E-\ln\frac{y}{4\pi^2}\Big)
   \nonumber\\ &&\strut+O(y^3),
\end{eqnarray}
where $C_E=0.7772156\ldots$ is the Euler-Mascheroni constant. The result
is consistent with what one obtains from the analytic
expression~({\ref{eq:Q33res}) for $Q_{3,3}$.

\section{Calculation of $\Delta U_{d,\omega}(0|L)$}
\label{app:aUd0}

We start from Eqs.~(\ref{eq:Uds}) and (\ref{eq:Udec}), take the
thermodynamic limit $L_\parallel\to\infty$, and utilize the continuum
approximation. Upon transforming the discrete sum over the momentum
component $q_1$ by means of Poisson's summation formula~(\ref{eq:Poisson}), we arrive at
\begin{equation}
  \Delta
  U_{d,\Omega}(0|L)=
  \sum_{j=1}^\infty\int_{\bm{q}}^{(d)}\cos(jq_1L)\,
  \ln\Omega_{\bm{q}}\;.
\end{equation}
We substitute for $\Omega_{\bm{q}}$ its small-momentum
form~(\ref{eq:Omega}), expand the logarithm as
\begin{equation}
  \label{eq:logOmexp}
  \ln\Omega_{\bm{q}}=\ln\big[q^2+O(q^4)\big] -b\,q^{\sigma-2}
 +O\big(b^2\big)\;,
\end{equation}
drop all suppressed terms, and extend the $\bm{q}$-integration to
$\mathbb{R}^d$.

The contribution from $\ln q^2$ is known from the short-range case
\cite{Sym81,KD92a,KD92b}, easily calculated, and given by the
$b$-independent term on the right-hand side of
Eq.~(\ref{eq:DelUdOmrzero}).  The $O(b)$~contribution involves a
difference of critical bulk and finite-size propagators at
$\bm{x}=\bm{0}$ for which one obtains, using
Eqs.~(\ref{eq:Gsiginftyres}), (\ref{eq:0F1exp}), and
(\ref{eq:1F2exp}),
\begin{eqnarray}
  \label{eq:Obcontr}
 \frac{1}{2}\,(G_L-G_\infty)(d,\sigma+2|0;\bm{0})= \sum_{j=1}^\infty
 G_\infty(d,\sigma+2|0;jL)\nonumber\\ =
L^{2 - d - \sigma}\,\frac{2^{\sigma-2 }\,
    \Gamma[(d + \sigma-2)/2]\,
    \zeta( d + \sigma-2 )}{\pi^{d/2}\,\Gamma(1 - \sigma/2)}\;.\qquad
\end{eqnarray}

Adding both contributions yields the result displayed in
Eq.~(\ref{eq:DelUdOmrzero}).


\begin{thebibliography}{88}
\bibitem[{cas({\natexlab{a}})}]{casiqmrev}
\bibinfo{note}{For reviews of the work and extensive lists of references on the
  Casimir effect, the interested reader might consult
  Refs.~\cite{MT97,KG99,BMM01,Mil01,PMG86}.}
\bibitem[{rem({\natexlab{b}})}]{rem:additivity}
\bibinfo{note}{If long-range interactions were included whose pair potential
  decay as $x^{-d-\sigma}$ with $\sigma<0$, then surface contributions would
  not be negligible compared to bulk contributions and the total bulk free
  energy of a finite system would not be additive. We never consider such
  extreme long-range interactions; since we presume that $\sigma>2$, the
  positivity of $\sigma$ and hence the additivity of the free energy are
  guaranteed. Likewise, thermodynamic limits such as the bulk free energy
  density $f_{\text{bulk}}$ and the surface excess free energy densities
  $f_{s,j}$ [see Eq.~(\ref{eq:fexinfty})] should exist.}
\bibitem[{rem({\natexlab{c}})}]{rem:corrlength}
\bibinfo{note}{At this stage of our discussion, the precise definition of the
  bulk correlation length $\xi_\infty$ does not matter; $\xi_\infty$ may be
  viewed here either as the true correlation length (defined via the
  exponential decay of the two-point bulk correlation function) or else as the
  usual second-moment correlation length.}
\bibitem[{rem({\natexlab{e}})}]{rem:lbound}
\bibinfo{note}{In Ref.~\cite{IS77} also rigorous lower bounds for the decay of
  the two-point cumulant are derived. However, these are not sufficiently
  strong to imply that this function decays asymptotically according to the
  same power law as the interaction potential. For the mean spherical model,
  this type of decay of the order-parameter two-point cumulant has been
  demonstrated in Ref.~\cite{Dan01}.}
\bibitem[{rem({\natexlab{f}})}]{rem:exDel3}
\bibinfo{note}{The value of $\Delta_C^{\text{SM}}(3)$ is known exactly.
  According to Ref.~\cite{Dan98}, one has
  $\Delta_C^{\text{SM}}(3)=-2\zeta(3)/(5\pi)$.}
\bibitem[{rem({\natexlab{g}})}]{rem:gamnu}
\bibinfo{note}{In order for these relations to hold for the $n$-vector model
  with $n<\infty$, one must interpret $r_\infty$ as $\xi_\infty^{-2}$ rather than
  the inverse bulk susceptibility because the latter scales as $L^{-(2-\eta)}$.
  Otherwise both the factor $2$ multiplying
  $\check{r}_\infty\partial_{\check{r}_\infty}$ as well as the exponent of $L$
  in $\check{r}_\infty=r_\infty L^2$ must be replaced by $\gamma/\nu=2-\eta$.}
\end{thebibliography}

\end{document}